\documentclass[letterpaper, 10 pt, conference]{ieeeconf}  % Comment this line out if you need a4paper
\IEEEoverridecommandlockouts                              
\usepackage[english]{babel}

\usepackage{amsmath, amssymb, amsthm}
\usepackage{enumerate}
\usepackage{graphicx}  % for pdf, bitmapped graphics files
\usepackage[dvipsnames]{xcolor}
\usepackage{booktabs} 
\usepackage{tabularx}
\usepackage{subcaption}

\usepackage{algorithm}
\usepackage{algpseudocode}

\usepackage[overload]{empheq}

\usepackage[hyperindex=true, breaklinks=true,colorlinks=true,bookmarks=false, pdftitle={Receding Horizon Control and Dissipativity -- Optimal Control, Games and Uncertainty},pdfauthor={Sophie Hall, Jonas Schie\ss{}l, Sergio Grammatico, Timm Faulwasser}, pagebackref=false, plainpages=false, pdfpagelabels, linkcolor=black, citecolor=black, filecolor=black, urlcolor=black]{hyperref}

\definecolor{lightgray}{gray}{0.9}
\definecolor{ggreen}{rgb}{0,0.5,0}
\definecolor{rred}{rgb}{0.5,0,0}

\newcommand{\mbf}[1]{\mathbf{#1}}
\newcommand{\bb}[1]{\mathbb{#1}}

\newcommand{\mc}[1]{\mathcal{#1}}
\newcommand{\col}[1]{\text{col}({#1})}
\newcommand{\R}{\mathbb{R}}
\newcommand{\x}{\mathbf{x}}
\newcommand{\N}{\mathbb{N}}
\newcommand{\xs}{{x}_s}
\newcommand{\us}{{u}_s}

\DeclareMathOperator*{\argmin}{argmin}

\DeclareMathOperator{\subjectto}{\ {s.}{t.}\ }
\DeclareMathOperator{\dist}{dist}
\DeclareMathOperator{\inte}{int}

\newtheorem{dfn}{Definition}

\newtheorem{rmk}{Remark}
\newtheorem{ex}{Example}
\newtheorem{pthm}{Insight}

\renewenvironment{proof}{\paragraph*{Proof}}{\hfill$\square$}
\newcommand{\sh}[1]{\textcolor{NavyBlue}{[#1]\raise 0.5ex \hbox{\footnotesize{SH}}}}
\newcommand{\js}[1]{\textcolor{Purple}{[#1]\raise 0.5ex \hbox{\footnotesize{JS}}}}
\newcommand{\sg}[1]{\textcolor{ForestGreen}{[#1]\raise 0.5ex \hbox{\footnotesize{SG}}}}
\newcommand{\tf}[1]{\textcolor{Orange}{[#1]\raise 0.5ex \hbox{\footnotesize{TF}}}}

\newcommand{\edit}[1]{\textcolor{black}{#1}}
\newcommand{\CHECK}[1]{\textcolor{Red}{[#1]\raise 0.5ex \hbox{\footnotesize{CHECK}}}}

\graphicspath{{figs/}}

\title{\LARGE \bf
Receding Horizon Control and Dissipativity -- Optimal Control, Games and Uncertainty
}

\author{Sophie Hall$^{1}$, Jonas Schie\ss{}l$^{2}$, Sergio Grammatico$^{3}$, Timm Faulwasser$^{4}$%
\thanks{$^{1}$SH is with the  Automatic Control Lab, ETH Z\"urich, Physikstrasse 3, 8092 Z\"urich, Switzerland, 
        {\tt\small shall@ethz.ch}}%
\thanks{$^{2}$JS is with the Department of Mathematics, University of Bayreuth, Universitätsstraße 30, 95447 Bayreuth, Germany,
        {\tt\small jonas.schiessl@uni-bayreuth.de}}%
\thanks{$^{3}$SG is with the Delft Center for Systems and Control, Delft University of Technology, Mekelweg 2, 2628 CD Delft, Netherlands,
        {\tt\small s.grammatico@tudelft.nl}}%
\thanks{$^{4}$ TF is with the Institute of Control Systems, Hamburg University of Technology, Harburger Schlo{\ss}stra{\ss}e 22a, 21079 Hamburg, Germany,
        {\tt\small timm.faulwasser@ieee.org}} %
\thanks{JS and TF acknowledge funding by the Deutsche Forschungsgemeinschaft (DFG, German Research Foundation) – project number 499435839.
}
}

\begin{document}

\maketitle
\thispagestyle{empty}
\pagestyle{empty}

\begin{abstract}
We provide a tutorial overview of receding-horizon control across deterministic, stochastic, multi-agent, and game-theoretic settings. We contrast the minimizer of an optimal control problem with the equilibrium of a game in terms of cost and constraint handling to motivate the difference of MPC schemes involving multiple agents/players. Interestingly, dissipativity theory and the turnpike property have turned out to be a unifying thread and system-theoretic backbone across all these schemes. This paper is the first to give an overview of results, summarizing fundamental insights, drawing parallels, and highlighting technical differences in dissipativity and turnpike analysis across economic, game-theoretic, and stochastic settings.  \edit{We also explore recent trends in numerical methods for games.} We conclude by pointing to a list of open challenges. 
\end{abstract}

\section{Introduction}It stands to reason that the optimal control twin breakthroughs, i.e., the maximum principle by Lev Pontryagin and co-workers and dynamic programming by Richard Bellman, as well as the system-theoretic dissipativity notion due to Jan C. Willems are fundamental pillars of systems and control.
Indeed, the close relation between optimal control and dissipativity is a bi-directional one:
In his foundational 1972 paper~\cite{Willems72a} Willems used optimal control concepts to provide equivalent characterizations of dissipativity through the available storage and the required supply.
Even earlier, he analyzed linear-quadratic optimal control and the underlying Riccati equations through the lens of dissipation inequalities linking value functions and stage costs~\cite{Willems71a}.
At the same time optimal control has seen tremendous research interest -- ranging from the maximum principle~\cite{PMP56}, dynamic programming \cite{Bellman54a} and early LQ approaches \cite{Kalman60a} to numerical optimal control and modern receding horizon concepts, a.k.a., model predictive control (MPC) for linear and nonlinear systems~\cite{Mayne00a}. In particular, MPC can be regarded as a success story of control research as it finds application in essentially all relevant branches of industry. A key driver for the success of MPC is that it is one of the few methods which can handle nonlinear systems with multiple inputs and subject to constraints. Moreover, through the repeated optimization, it also allows one to directly consider performance-related objectives. Finally, the availability of very efficient open-source tailored optimization codes for convex and nonconvex problems enables efficient MPC design and implementation for a wide range of hardware platforms, see, e.g., \cite{Andersson19a,Kapernick14a,Verschueren22a}.

However, it took from 1971/1972 until around 2008-2013 before the importance of dissipativity for the analysis of MPC schemes started to emerge. In 2008 Jim Rawlings coined the term \textit{economic MPC} to express that in many applications it may be of interest to consider performance functionals more general than convex quadratic settings~\cite{Rawlings09b}. In \cite{Diehl11a,Angeli12a} it has been observed that dissipation inequalities which consider the stage cost of the optimal control problem in the supply rate allow for tailored closed-loop analysis. In this context, the turnpike phenomenon became of interest to study  nonlinear MPC trajectories. In essence, this phenomenon states that in an optimal control problem (OCP) optimal solutions for varying initial conditions and varying horizon length often tend to cluster around a steady state which is optimal with respect to the considered stage cost. Despite the long history of turnpike results for optimal control problems---which can be traced back to Frank Ramsey~\cite{Ramsey28}, John von Neumann~\cite{vonNeumann38}, Dorfman et al. \cite{Dorfman58}, and many others such as \cite{Wilde72a}---it remains surprising that \cite{Rawlings09b,Wuerth09} appear to be the earliest papers recognizing its role for MPC analysis, while \cite{Gruene13a} formally linked turnpikes, dissipativity, and MPC.

Jumping back in time to the first half of the 20th century---largely in parallel to the foundational optimal control developments---game theory concepts for static and dynamic problems started to emerge. Once more John von Neumann played a pivotal role with his foundational works on zero-sum games~\cite{vonNeumann28} and the book \cite{vonNeumann44} before John Nash entered the stage, establishing non-cooperative equilibrium concepts~\cite{Nash50}. While both concepts---optimal control and game theory---allow the consideration of multi-agent settings, games are particularly well-suited for non-cooperative settings, i.e., settings in which agents are self-interested and thus primarily concerned with their own outcome. Zero-sum two-player games, in which one agent's cost is the negative of the other's, lead to $\min\max$ optimal control problems~\cite{Basar08}, a well-known link between both perspectives.

Interestingly, in dynamic games turnpike behavior has been observed and analyzed since the 1980s ~\cite{fershtman1986turnpike} with early theoretical results focusing on infinite-horizon open-loop games~\cite{carlson1995turnpike, carlson1996turnpike}. Turnpike analysis in games has recently received renewed attention in stochastic zero-sum~\cite{li2025turnpike} and non-zero-sum settings~\cite{cohen2025turnpike}, as well as under a mean field assumption~\cite{cirant2021long, carmona2024leveraging, ersland2025long, fedorov2025studying}.

In parallel active research around a receding-horizon implementation of games developed, see, e.g.~\cite{stephens2015game, lecleach2022algames, benenati2024probabilistic, hall2024receding}. So far the system-theoretic and closed-loop analysis of \textit{Receding Horizon Games} (RHG) has been limited~\cite{hall2026stability, benenati2026linear, hall2026towards}. \edit{Clearly, a deep conceptual understanding and a solid theoretical foundation are essential before deploying RHG on a real system, since receding horizon controllers already exhibit counter-intuitive closed-loop behavior on their own, which is even worse when pairing them with competitive agent interactions.} Especially, comparing it against the combined backdrop of 50+ years of dissipativity-based analysis of optimal control, 30+ years of closed-loop analysis for MPC, and its industrial success. 

In the context of system-theoretic and closed-loop analysis of games and RHG a multitude of open problems exist. In this paper, however, we narrow the scope to four main aspects:
\begin{itemize}
    \item Can dissipativity notions be leveraged for the analysis of dynamic games? Are there fundamental barriers, hidden commonalities, or a mix of both?
    \item What is the prospect of transferring analysis concepts from MPC to receding horizon games?
     \item How do dissipativity notions enable the analysis of stochastic decision problems beyond Gaussian uncertainty?
    \item Given the tremendous progress on tailored numerical algorithms for optimal control and MPC, what are promising research directions for solving open-loop dynamic games?
\end{itemize}

In approaching these questions a crucial hurdle is induced by the different domains touched upon: dissipativity, optimal control, game theory, stochastic control, and numerical optimization. These domains rely on specific technical concepts which tend to overshadow fundamental structures. Our approach of uncovering common features and fundamental obstacles thus avoids phrasing exact theorems, as this would require partially laborious, detailed technical conditions and assumptions. Instead, we will formulate \textit{insights} and will provide links to relevant literature which formalizes these statements. One may regard this as an attempt to partially filter the technical proof-oriented and apparatus from the underlying foundational concepts.

The remainder of this paper is structured as follows:
In Section~\ref{sec:openloop} we recap key aspects of open-loop optimal control and games, while Section~\ref{sec:recedinghorizon} turns towards receding-horizon concepts for OCPs and games. In Section~\ref{sec:diss} we explore dissipativity notions and their role for games and optimal control, before Section~\ref{sec:uncertainty} turns towards stochastic problems.
As any receding-horizon control approach relies on efficient numerical algorithms, \edit{Section~\ref{sec:rec_hor_clNE}} investigates recent progress on this end. The paper concludes with a discussion of open problems and bottlenecks in Section~\ref{sec:conclusion}.

\section{Open-Loop Optimal Control and Games} \label{sec:openloop}
One key advantage of optimization-based control design is its conceptual scalability to multi-dimensional settings. While for classic  control methods, the design procedure becomes more and more complex if one goes from SISO to MIMO settings (consider concepts like transmission zeros, relative gain arrays and static/dynamic decoupling designs \cite{Skogestad05}), optimization-based designs are usually based on solving optimality conditions (or suitable reformulations thereof). Consider, e.g., the ubiquitous linear quadratic regulator; the underlying Riccati equations do not change their structure if one goes from a SISO to a MIMO setting, see~\cite{Willems71a} for a dissipativity-based analysis. The same can be said about classic optimal control, e.g., based on the maximum principle~\cite{PMP56,Boltyanskii60a} or on dynamic programming~\cite{Bellman54a}. \edit{Conceptual scalability does not, however, mean that large scale optimization problems are free of numerical challenges.} Indeed, even in convex linear-quadratic settings with linear inequality, solving such quadratic programs will usually scale no better than $O(N(n_x\cdot n_u)^3)$, where $N$ is the time horizon and $n_x,n_u$ are, respectively, state and input dimensions of the considered dynamics.

In multi-agent settings, where $v \in \mc V$ distinguishes different agents, one often has
\begin{equation}\label{eq:sysv}
    x_{k+1} =  f(x_k,u_k^v,u_k^{-v}), \quad  x_0 = \x
\end{equation}
i.e., the agents $v\in\mc{V} := \{1,\dots, M\}$ all act on the same state variable. Here, as a shorthand, $-v$ indicates $\mc V\setminus \{v\}$, i.e., all agents except $v$.
In principle, the agents may act \textit{cooperatively} or \textit{competitively}. In the former case, one may formally rewrite \eqref{eq:sysv} as one centralized system
\begin{equation}\label{eq:sysu}
    x_{k+1} =  f(x_k,u_k), \quad  x_0 = \x, 
\end{equation}
where $u_k = \begin{bmatrix}u_k^1, \dots, u_k^M\end{bmatrix}^\top$. Here and henceforth, for the sake of readability, we suppress the transposition of the entries in $u_k$. 

From the analytic point of view one may treat a set of \textit{cooperative agents} as one holistic entity. Of course, whenever any kind of cooperative distributed design is of interest, it makes sense to bring back the distinction between agents~\cite{Scattolini09a, muller2017economic}. 
Yet, as we will discuss below, in any competitive or \textit{non-cooperative setting} the distinction between the single agents is of fundamental importance to define appropriate solution concepts for the arising open-loop dynamic games. 

Another major watershed of optimization-based control is the nature of the computed control actions. \edit{In both cooperative and competitive settings, the inputs are generated by an explicit feedback law $u_k = \mu(k, x_k)$, possibly time-varying or dynamic.} Alternatively, one may consider the application of a sequence of open-loop inputs: $u_0,\dots, u_{N-1}$ over some horizon $\bb{Z}_{N}:=\{0, \dots, N-1\}$.

These two viewpoints lead to \textit{closed-loop} and \textit{open-loop} control. Here, we first introduce the open-loop perspective on optimal control and dynamic games before we return to the closed-loop receding horizon point of view in Section~\ref{sec:recedinghorizon}. 

\subsection{Optimal Control}
In any cooperative setting, it is natural to optimize over all inputs at once. One therefore works with \eqref{eq:sysu} rather than \eqref{eq:sysv} to formulate an \textit{optimal control problem} (OCP) such as
\begin{subequations}
\label{eq:OCP}
\begin{align}
\label{eq:OCPRunningCost}
\displaystyle \min_{x,\,u}  J_N(x,u):= &\;\sum_{k= 0}^{N-1} \ell(x_k, u_k) \\
\textrm{s.t.} \quad &  x_{k+1} =  f(x_k,u_k)  \hspace{1.25em}  k \in \bb{Z}_{N} \label{eq:OCPConstr1}\\
&g(x_k,u_k) \leq 0, \hspace{2.5em} k \in \bb{Z}_{N} \label{eq:OCPConstr2}\\
 & h(u_k)\leq 0,      \hspace{3.9em} k \in \bb{Z}_{N} \label{eq:OCPConstr3}\\ 
&\; x_0 = \x. \label{eq:OCPConstr4}
\end{align}
\end{subequations}
Subsequently, we indicate optimal solutions by the superscript $\cdot^\diamond$ and pairs of optimal input and state trajectories (optimal pairs in short) are collected in the solution set $(x^\diamond, u^\diamond) \in \mc{S}^{\text{\tiny OCP}}_N(\x)$. Observe that this solution set and its elements are parametrized by the initial condition $\x$ of the dynamics \eqref{eq:OCPConstr4}. For the sake of simplified exposition, we have not included any terminal penalty/Mayer term  or terminal constraints in \eqref{eq:OCP} but both could be added without conceptual difficulties. 
\edit{If we assume a minimum exists}, we have that 
\begin{align*}
    (x^\diamond, u^\diamond) \in \argmin_{x, u} J_N(x,u) \text{ s.t. } \eqref{eq:OCPConstr1}-\eqref{eq:OCPConstr4}.
\end{align*}
Consider the feasible set 
\begin{equation}\label{eq:FeasibleSetOCP} \tag{\edit{FS}}
    \mc{Z}^{\text{\tiny OCP}}_N(\x) = \{(x, u) \in \R^{(N+1)n_x+N n_u}~|~  \eqref{eq:OCPConstr1} - \eqref{eq:OCPConstr4}\}.
\end{equation}
Then the minimization implies
\begin{equation}\label{eq:optimalityOCP}
     J_N(x^\diamond, u^\diamond) \leq  J_N(x,u),~~ \forall (x, u)\in  \mc{Z}^{\text{\tiny OCP}}_N(\x)
\end{equation}
i.e., the optimal input sequence $u^\diamond(\x) =\begin{pmatrix} u^\diamond_0, \dots, u^\diamond_{N-1}\end{pmatrix}$ will achieve the lowest objective among all feasible inputs sequences. This optimal sequence also depends on the initial condition $\x$. Moreover, it is well-known that under mild differentiability assumptions any $(x^\diamond, u^\diamond) \in \mc{S}^{\text{\tiny OCP}}_N(\x)$ also has to satisfy the KKT optimality system \edit{$\forall k\in \bb{Z}_N$}
\begin{subequations} \label{eq:KKT_OCP}
\begin{align} 
  x_{k+1} &= f(x_k,u_k) \label{eq:KKT_OCPx}\\
\lambda_{k} &= \ell_x +  g_{x}^\top \mu_{k} + f_x^\top \lambda_{k+1}, \quad \\ 
0 &=   \ell_{u}  +  g_{u}^\top \mu_{k}  + f_{u}^\top \lambda_{k+1} +  h_{u}^\top \eta_k, \label{eq:KKT_OCPu}
\end{align} 
including the split boundary conditions\footnote{In case of additional terminal constraints $g^\mathrm{f}(x_N)\leq 0$, the boundary condition will read $\lambda_N = {g^\mathrm{f}_{x}}^\top \mu_{N}$.}
\begin{equation} \label{eq:KKT_OCP_boundary}
    x_0 =\x\quad\text{and}\quad\lambda_N = 0%
\end{equation}
 \edit{subject to primal feasibility and complementarity slackness
\begin{align}
0 \leq \mu_k &\quad\perp\quad -g(x_k,u_k) \geq 0, \\
0 \leq \eta_k &\quad\perp\quad -h(u_k) \geq 0.
\end{align}}
 \end{subequations} 
We refer to \cite{Bryson69a,Bryson99a} for a detailed derivation. 
Using this approach, which is mirrored in continuous time by the maximum principle~\cite{PMP56} and the Euler-Lagrange equations~\cite{Sussmann97,Kalman63}, one conceptualizes the solution of \eqref{eq:OCP} as an open-loop problem: for each value of $\x$ the solution to \eqref{eq:KKT_OCP} will entail the primal trajectory pair $(x^\diamond, u^\diamond)$ and the trajectories of dual variables (a.k.a. multipliers) $(\lambda^\diamond(\x), \mu^\diamond(\x), \eta^\diamond(\x))$. 

Alternatively, one could rely on dynamic programming and split the determination of optimal feedback law into two parts: (i) compute the optimal value function  $V^\diamond_N: \R^{n_x} \to \R$\footnote{Often one extends the range of $V_N^\diamond$ to the extended real line. This \edit{is} done such that $V_N^\diamond(\x) = \infty$ can be used as an infeasibility certificate for an OCP and, conversely, $V_N^\diamond(\x) < \infty$ is equivalent to requiring feasibility of the OCP with initial condition $\x$.}
\begin{equation}\label{eq:VN}
   V^\diamond_N(\x):= J_N(u^\diamond(\x), x^\diamond(\x))
\end{equation} which has to satisfy the Bellman equation
\begin{equation}\label{eq:BE}
    V^\diamond_N(\x) = \min_{u}\, \ell(\x, u) + V^\diamond_{N-1}(f(\x, u))~\text{s.t. }  \mc{Z}^{\text{\tiny OCP}}_1(\x). \tag{BE}
\end{equation}
(ii) given $V^\diamond_N$ and $\x$, the optimal one-step input can be obtained by solving the one-step optimization on the right hand side of the Bellman equation above. Naturally, this points towards using modern reinforcement learning concepts in this process which are closely related to dynamic programming, see, e.g. \cite{Bertsekas19a}.

\subsection{Open-Loop Games}
In contrast to the OCP setting above, in open-loop games, we consider a set of self-interested agents $v\in\mc{V} := \{1,\dots, M\}$ which aim to minimize their specific cost functions $\ell^v$ over a finite horizon $N$. The agents can be coupled in three distinct ways: (i)~they  share and influence the same dynamics $x_{k+1} =  f(x_k,u_k^v,u_k^{-v})$ (e.g., shared battery storage, accessing the same aquifer); (ii)~the cost each agent incurs is influenced by the joint state and other agents' actions $u^{-v}$ (e.g., cost of energy or a product increase with higher demand); and (iii) their feasible set of actions depends on the joint state and the actions of others $g(x_k,u_k^v, u_k^{-v})$ (e.g., road space is limited, grids get overloaded). Examples include autonomous driving~\cite{lecleach2022algames}, wireless networks~\cite{pavel2012game}, smart grids~\cite{atzeni2013noncooperative}, and environmental games~\cite{krawczyk2005coupled, bahn2008class}. 

The resulting $M$ interdependent problems  constitute a non-cooperative dynamic generalized game as follows~\cite{facchinei2009generalized}
\begin{subequations}
\label{eq:GNEP}
\begin{empheq}[left=\hspace{-.5mm}\forall\hspace{-.25mm}v\hspace{-.25mm}\in\hspace{-.25mm}\mc{\hspace{-.25mm}V\hspace{-.5mm}}:\hspace{-1.25mm} \empheqlbrace]{align}
\label{eq:RunningCost}
\displaystyle \min_{x, u^v}  &\; \sum_{k= 0}^{N-1} \ell^v(x_k, u_k^v,u_k^{-v})\\%
\textrm{s.t.} \quad &  x_{k+1} =  f(x_k,u_k^v,u_k^{-v}),  \hspace{1.25em}  k \in \bb{Z}_{N} \label{eq:Constr1}\\
&g^v(x_k,u_k^v, u^{-v}_k) \leq 0, \hspace{2.5em} k \in \bb{Z}_{N} \label{eq:Constr2}\\
  & h^v(u_k^v)\leq 0,      \hspace{6.0em} k \in \bb{Z}_{N} \label{eq:Constr3}\\ 
&\; x_0 = \x, \label{eq:Constr4}
\end{empheq}
\end{subequations}
with initial condition $\x$ and agent-specific local input constraint $h^v$. 
Note that~\eqref{eq:GNEP} is presented in its most general form relevant for control applications. As required by the application, this setting can be specialized to a local state $x^v$ with coupled $x^v_{k+1} = f(x_k^v, u_k^v,u_k^{-v})$ or local dynamics $x^v_{k+1} = f(x_k^v, u_k^v)$, as well as input only or state only coupling in the cost and constraints.

The description of this problem is to be interpreted as follows: Each agent $v\in \mc V$ solves \eqref{eq:GNEP} supposing the actions of the other agents $-v \in \mc V^{-v} := \mc V\setminus\{v\}$. That is, the generalized open-loop game \eqref{eq:GNEP} consists of $M$ interdependent OCPs. \textit{However, this does not imply that the solutions of \eqref{eq:GNEP} and \eqref{eq:OCP} will necessarily coincide}. Clarifying this point requires to specify a suitable solution concept for \eqref{eq:GNEP}. 

We define the following per-agent and global action sets
\begin{subequations} 
\begin{align}
&\mc{Z}_N^v( \x, u^{-v}) = \{(x,u^v)  ~|~ \eqref{eq:Constr1} - \eqref{eq:Constr4}\} \\[5pt]
&\mc{Z}_N(\x) = \{(x, u) \in \R^{(N+1)n_x+N n_u}~|~  \eqref{eq:Constr1} - \eqref{eq:Constr4}\} \label{eq:GlobalFeasibleGNESet} .
\end{align}
\end{subequations}
and the per-agent cumulative cost 
\[
J_N^v(x,u^v, u^{-v}) :=\sum_{k= 0}^{N-1} \ell^v(x_k, u_k^v, u^{-v}_k).
\]
With these abstractions we can write~\eqref{eq:GNEP} as follows:
\begin{equation}\label{eq:GNEPcompact}
\forall v\in \mc{V}: \; 
\left\{\begin{aligned}\;\min_{u^v, x}\;& J^v_N(x,u^{v},u^{-v}) \\
\subjectto &(x,u^{v})\in \mathcal{Z}_N^v( \x, u^{-v}).
\end{aligned}\right.
\end{equation}

A set of strategies  $ u = [(u^1), \ldots, (u^M)]^\top$ which jointly solves~\eqref{eq:GNEP} is a \textit{Generalized Nash equilibrium} (GNE).

\begin{dfn}[Generalized Nash equilibrium] \label{dfn:GNE} ~\\A joint decision $(x^*, u^*)\in \mc{Z}_N(\x)$ is a GNE  of~\eqref{eq:GNEP} if 
\begin{align*}
\forall v\in \mc{V}: \, J_N^v(x^{*}, u^{v,*}, u^{-v,*} ) \leq J_N^v(x, u^{v}, u^{-v*}) 
\end{align*} holds for all $(x,u^v) \in \mc{Z}^v_N( \x, u^{-v,*})$. The corresponding solution set for fixed $N\in \bb{N}$ and $\x$ is denoted as
\[
(x^*,u^*) \in \mc{S}^{\text{\tiny GNE}}_N(\x) \subset \R^{(N+1)n_x + N n_u}.%
\]
\end{dfn}
The solution concept combines two notions: (i)~\textit{Nash}, strategic stability such that no agent can lower its cost by unilaterally changing its decision; and (ii)~\textit{generalized}, referring to the coupled constraint sets, meaning no agent wants to deviate given their feasible actions that depend on $u^{-v}$. It thus generalizes the classic Nash equilibrium concept.\\[2mm]

A special subclass of equilibria is \textit{variational GNEs} (v-GNEs), which correspond to solutions of the following generalized equation:
\begin{align}\label{eq:GeneralizedEquation}
\text{F}( u^*, \mbf{x}) + \mc{N}_{\tilde{\mathcal{Z}}(\x)}(u^*) \ni 0,
\end{align}
or equivalently to solutions of the following variational inequality (VI)
\begin{align}\label{eq:VI}
 u^*\in \tilde{\mc{Z}}(\x):\, (u-u^*)^\top \text{F}(u^*, \mbf{x}) \geq 0,\quad \forall u \in \tilde{\mc{Z}}(\x) \tag{VI}
\end{align}
where F$(u, \x) = \text{col}(\nabla_{u^v} J^v(u^v, u^{-v}, \mbf{x}))_{v\in \mathcal{V}}$ is the pseudogradient of~\eqref{eq:GNEPcompact} if we substitute the dynamics~\eqref{eq:Constr1} and initial condition~\eqref{eq:Constr4} into the stage cost~\eqref{eq:RunningCost}, and $\mc{N}_{\tilde{\mathcal{Z}}(\x)}$ is the normal cone~\cite[Def.~6.38]{bauschke2017convex} of the global action set~\eqref{eq:GlobalFeasibleGNESet} defined solely in the $u$ space and is required to be closed and convex. Variational GNEs are a desirable solution concept in dynamic resource allocation problems as they are deemed \textit{economically fair}: they impose homogeneous Lagrange multipliers across the agent population, so each agent incurs the same marginal loss due to the presence of coupling constraints~\cite{facchinei2009nash}. In addition, v-GNEs have desirable computational properties as VIs can be efficiently solved using operator splitting methods which is laid out in Section~\ref{sec:rec_hor_clNE}.

One may ask if GNEPs also lead to structured optimality systems. Indeed, under suitable assumptions
such as (i) differentiable problem data (for fixed $u^{-v}$, cost, constraint, and dynamic functions are continuously differentiable in $x$ and $u^v$), and (ii) convexity of the cost and a closed and convex constraint set for fixed $u^{-v}$, it can be shown that \eqref{eq:GNEP} gives rise to the following per-agent KKT system which needs to hold $\forall v\in \mc{V}$ and $k\in \bb Z_N$, cf.~\cite[Thm. 4.6]{facchinei2009generalized}
\begin{subequations} \label{eq:KKT_GNEP}
\begin{align} 
  x_{k+1} &= f(x_k,u_k^v, u_k^{-v}) \label{eq:KKT_GNEPx}\\
\lambda^v_{k} &= \ell_x^v +  (g_{x}^v)^\top \mu^v_{k} + f_x^\top \lambda_{k+1}^v, \quad \label{eq:KKT_GNEPlam}\\ 
0 &=   \ell_{u^v}^v  +  (g_{u^v}^v)^\top \mu^v_{k}  + f_{u^v}^\top \lambda_{k+1}^v +  h_{u^v}^\top \eta_k^v, \label{eq:KKT_GNEPu}
\end{align} 
as well as the split boundary conditions 
\begin{equation} \label{eq:KKT_GNEP_boundary}
   x_0 = \x  \quad\text{and}\quad\lambda_N^v = 0%
\end{equation}
\edit{and primal feasibility and complementarity slackness
\begin{align}\label{eq:KKT_GNEPslack1}
0 \leq \mu_k^v &\quad \perp \quad  -g^v(x_k,u_k^v,u_k^{-v})\geq 0,\\
0 \leq \eta_k^v&\quad\perp\quad -h^v(u^v_k) \geq 0.\label{eq:KKT_GNEPslack2}
\end{align}}
 \end{subequations} 

\subsection{Delineation of Games and OCPs} 
Notice that the set $\mc{Z}_N(\x)$ from \eqref{eq:GlobalFeasibleGNESet} corresponds to the set $\mc{Z}^{\text{\tiny OCP}}_N(\x)$ from \eqref{eq:FeasibleSetOCP} if the per-agent dynamics \eqref{eq:sysv} are considered as one holistic system \eqref{eq:sysu}. Yet, despite this link in terms of feasible sets for the entire agent population, the generalized Nash equilibrium concept from Definition~\ref{dfn:GNE} is far more general than the usual optimality notion from \eqref{eq:optimalityOCP}.
In order to clearly distinguish OCP solutions---which as per \eqref{eq:optimalityOCP} are optimal---from GNE solutions, we use the superscript $\cdot^\diamond$ for the former and  $\cdot^*$ for the latter.

\begin{ex}[Differences of OCP and GNEP solutions]\label{ex:diff_OCPGNEP} We illustrate how these solution concepts lead to different numerical results in the following example. We consider the linear-quadratic OCP from~\cite{gruene2013economic} 
 \begin{equation}\label{eq:OCP_example} 
\begin{array}{r l}
\displaystyle \min_{x,u}  &\displaystyle \sum_{k= 0}^{N-1}  \displaystyle (u_k)^2 
\\[1em]
\textrm{s.t.} &  x_{k+1} = 2x_k +  0.25u_k,  \hspace{4.6em}  k \in \bb{Z}_{N}\\
&  - 2\leq x_{k+1}\leq 2,   \x = x_0 =\pm1.9,  \hspace{0.3em} k \in \bb{Z}_{N}.
\end{array}
\end{equation}
Its game-theoretic sibling is obtained by  adding coupling in the cost function and the dynamics. For a set of two agents $v\in\mc{V}=\{1,2\}$ we consider
\begin{equation}\label{eq:GNEP_example}
\left\{
\begin{array}{r l}
\displaystyle \min_{x,u^v}  &\displaystyle \sum_{k= 0}^{N-1}  \displaystyle u_k^v \sum_{j\in \{1,2\}} R^{v,j} u_k^{j}
\\[1em]
\textrm{s.t.} &  x_{k+1} = 2x_k +  0.1 u_k^1 + 0.15u_k^2,  \hspace{.3em}  k \in \bb{Z}_{N}\\
&  - 2\leq x_{k+1}\leq 2,   x_0 =\pm 1.9, \hspace{1.4em} k \in \bb{Z}_{N}.
\end{array}
\right.
\end{equation}

We plot the open-loop trajectories for increasing horizon length $N=2,4,6,\dots,18$ and the positive and negative initial condition. We  observe that all trajectories approach a neighborhood of $(x_s, u_s) = (0,0)$ during the middle of the horizon before they depart towards the end of the horizon. As we will explore in Section~\ref{sec:turnpike} this is a classic turnpike behavior for both the OCP (Figure~\ref{fig:OpenLoopOCP}) as well as the GNEP (Figure~\ref{fig:OpenLoopGNE}). Note that in the game-theoretic example, the two agents carry a different ``burden" (input magnitude) of bringing the system close to the steady-state GNE $(x_s, u_s) = (0,0)$ as per Definition~\ref{dfn:SteadyStateGNEP}.

\begin{figure}
  \centering
  \begin{subfigure}{\columnwidth}
    \centering
    \includegraphics[width=\columnwidth]{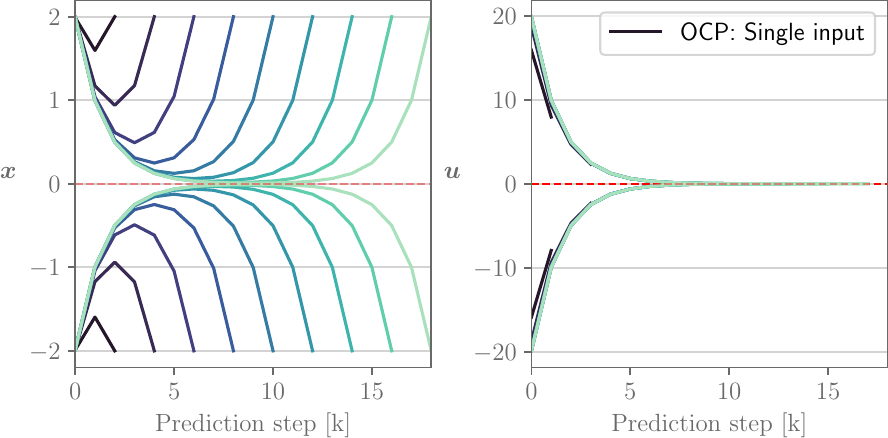}
    \caption{OCP~\eqref{eq:OCP_example} }
    \label{fig:OpenLoopOCP}
  \end{subfigure}
  \begin{subfigure}{\columnwidth}
    \centering
    \includegraphics[width=\columnwidth]{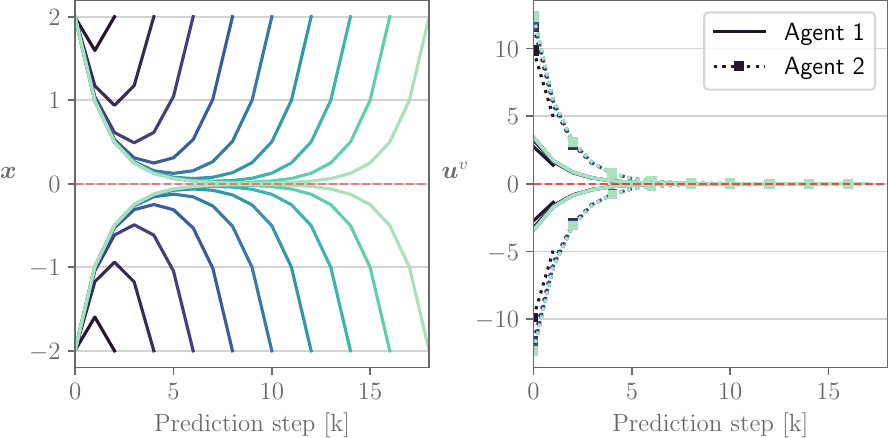}
    \caption{Dynamic GNEP~\eqref{eq:GNEP_example}}
    \label{fig:OpenLoopGNE}
  \end{subfigure}
  \caption{A comparison of open-loop trajectories shown for $N=2,4,6,\dots,18$. }
  \label{fig:OpenLoopComparison}
\end{figure}

\end{ex}

\paragraph{Objective gradients}
An important first-order object in optimization is the gradient of the objective functional $\nabla_u J_N(x, u)$, whereby one regards the state trajectory $x=x(u)$, i.e., as being generated by the input decisions. Interestingly, in games a parallel concept exists, yet instead of taking the gradient of a single objective with respect to the full input vector $u$, one stacks each agent's objectives $J_N^v(x, u^v, u^{-v})$ differentiated with respect to their own decisions $u^v$. The intuition is that each agent can only influence and actuate its own input $u^v$, so in the unconstrained case it tries to approach $\nabla_{u^v} J^v = 0$ (which is akin to first-order stationarity of its gradient). Stacking these per-agent gradients yields the \emph{pseudogradient}
\[
F(u, \x) = \mathrm{col}\big(\nabla_{u^v} J^v(x, u^v, u^{-v})\big)_{v\in \mc{V}},
\]
which plays a fundamental role  in game-theoretic problems -- its properties (e.g., monotonicity) influence existence and uniqueness of solutions, similar to a gradient in OCPs. Note that $F$ is vector-valued, so its derivative is the Jacobian $\nabla_u F$, which is in general asymmetric, meaning $F$ is not the gradient of any scalar potential. The fact that it cannot be integrated is exactly why, even in the unconstrained case NEs cannot in general be equivalently recast as a single optimization problem. An important class of games in which the symmetry of $\nabla_u F$ holds, and thus an OCP representation exists, are~\textit{potential games}, where the pseudogradient integrates to a joint potential for the agent population~\cite{monderer1996potential}.

\paragraph{Constraints} Comparing the KKT system of the OCP in~\eqref{eq:KKT_OCP} with that of the GNEP in~\eqref{eq:KKT_GNEP}, we directly observe why, also in the constrained case, a GNE is usually not directly a solution to an OCP. Specifically, in GNEs the dual variables associated with the coupling constraints, such as the dynamics~\eqref{eq:Constr1} and the inequality constraints~\eqref{eq:Constr2}, are heterogeneous across agents. For the multipliers related to the dynamics $\lambda^v$, this implies that their trajectories may follow entirely different paths for each agent $v\in \mc V$ governed by the evolution of~\eqref{eq:KKT_GNEPlam}. Intuitively, this means that agents contribute differently to ensuring that the dynamic constraints are satisfied. Similarly, for the multipliers $\mu^v$ associated with the coupled inequality constraints, the differing values lead to agents experiencing a different burden in ensuring they are satisfied.  In resource allocation problems, where $\mu^v$ is often interpreted as the shadow price of the resource, this is interpreted as resources being priced heterogeneously across the agent population. Thus, in terms of constraints, the solutions coincide only if we impose $\lambda^v = \lambda,\; \mu^v = \mu \;\, \forall v \in \mc{V}$, which corresponds to selecting a v-GNE as in~\eqref{eq:VI}~\cite{facchinei2009nash}.\vspace{2mm}

We close this section by pointing towards the following crucial relation between the optimal value function of the OCP~\eqref{eq:OCP} and the overall performance implied by GNEP~\eqref{eq:GNEP}.
\begin{pthm}[Performance of OCPs and GNEPs~{\edit{\cite[Rmk. 2]{hall2025system}}}] \label{pthm:performance}
Consider 
\begin{equation}\label{eq:aggregatedCost}
    \ell(x,u):= \sum_{v\in \mc V}\ell^v(x,u^{v}, u^{-v})
\end{equation} and let
 $\mc{Z}^{\text{\tiny OCP}}_N(\x) = \mc{Z}_N(\x) \neq \emptyset$ for any $N\in\bb N$, then 
\[
V_N^\diamond(\x) \leq\sum_{v\in \mc V} J^v_N(x^*,u^{v,*}, u^{-v,*}),\qquad \forall N\in\bb N.
\]
\end{pthm}
First notice that by construction the constraint sets $\mc{Z}^{\text{\tiny OCP}}_N(\x)$ and $\mc{Z}_N(\x)$ are equivalent. 
This insight is grounded in the different solution concepts for OCPs and GNEPs: the former require optimality \eqref{eq:optimalityOCP} (i.e. the smallest possible objective) while the latter asks for finding strategic equilibrium (Definition~\ref{dfn:GNE}).

\section{Receding Horizon Frameworks}\label{sec:recedinghorizon}

The dominating perspective of closed-loop control is an asymptotic one: we are interested in designing algorithms which influence a system automatically and over an infinite horizon, i.e., there is no pre-determined stopping time. This leaves us with three main options:
\begin{enumerate}
    \item[(i)] \textit{extend the horizon $N \to\infty$} in the open-loop setting,      
    \item[(ii)]\edit{ compute an explicit infinite-horizon feedback law off\-line via \textit{dynamic programming}}, or
    \item[(iii)]  solve a finite-horizon problem (OCP or GNEP) \edit{repeatedly online} (a.k.a. \textit{the receding horizon approach}). 
\end{enumerate}

For OCPs and GNEPs alike, approach (i) leads to fundamental issues in terms of numerical solutions. That is, except for very specific problem structures (mostly linear systems and quadratic objectives) infinite-horizon OCPs and GNEPs are not computationally tractable since they correspond to infinite-dimensional decision problems.

\edit{Note that both (ii) and (iii) result in closed-loop control. The difference is that (ii) yields an explicit feedback law, whereas in (iii) feedback is introduced implicitly by repeatedly solving an OCP/GNEP.} While (ii) is conceptually very elegant, except for very specific settings, the computation of (finite and infinite-horizon) value functions is also subject to the curse of dimensionality.

\begin{figure*}
  \centering
  \includegraphics[width=\textwidth]{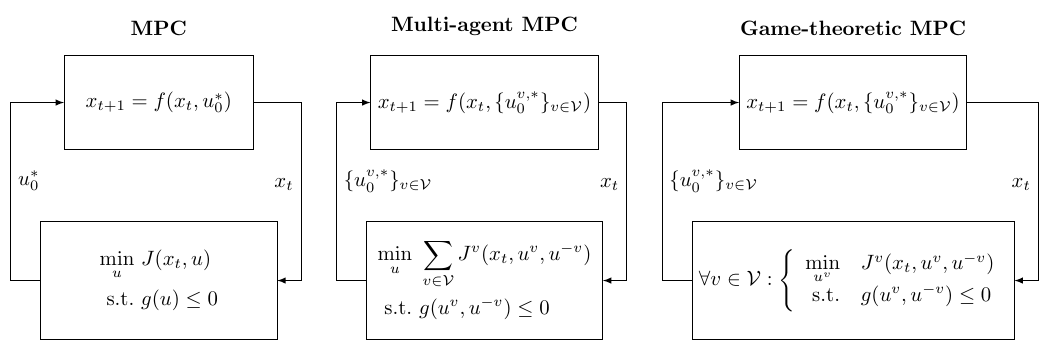}
  \caption{Overview of optimization-based approaches to receding horizon control}
  \label{fig:Comparison_schematic}
\end{figure*}

Approach (iii) exploits that the open-loop inputs $u^\diamond(\x)$ and $u^*(\x)$ depend implicitly on the initial condition of the dynamics, cf. \eqref{eq:OCP} and \eqref{eq:GNEP}. Hence for OCPs \eqref{eq:OCP} and GNEPs \eqref{eq:GNEP} the map
\[
\mu_N^\dagger: \x \mapsto u_0^\dagger, \qquad \dagger \in \{\diamond, *\}
\]
defines an implicit state feedback. 
This feedback is implicit as once $\x$ is known the feedback actions arise through the computation of an appropriate solution. Indeed, in case of GNEPs we will have that
\[ u^* = \begin{bmatrix} u^{v, *}_0 & u^{-v, *}_0\end{bmatrix}^\top =: \mu^*_N(\x). \]
Independently of whether OCPs \eqref{eq:OCP} or GNEPs \eqref{eq:GNEP} are considered, the closed-loop system becomes
\begin{equation}\label{eq:sys-cl}
    \x_{t+1} = f(\x_t, \mu_N^\dagger(\x_t)), \qquad  \dagger \in \{\diamond, *\}. 
\end{equation}
Hence, we have the situation that the trajectories of the closed-loop system \eqref{eq:sys-cl} have to be distinguished from the ones obtained during the solution of OCPs \eqref{eq:OCP} / GNEPs \eqref{eq:GNEP}. Borrowing notation which originally arose for filtering/estimation problems, we use the shorthands
\begin{align*}
     x^\dagger_{k|t} := x^\dagger_{k|t}(\x_t) \quad\text{and}\ \quad  u^\dagger_{k|t}:= u^\dagger_{k|t}(\x_t),\qquad \dagger \in \{\diamond, *\}
\end{align*}
whereby the subscript $\cdot_{k|t}$ indicates that a quantity has been computed using the state $\x_t$ and the index $k\in \bb Z_N$ picks a specific element from the $N$-step horizon.\footnote{Rudolf E. Kalman used the notation $x(k|t)$ to refer to an estimate for time step $k$ which is based on information up to time step $t$~\cite{Kalman60b}.} When we refer to the entire predicted sequence or trajectory we denote it as $\cdot_{\cdot|t}$. The left and right parts of Figure~\ref{fig:Comparison_schematic} illustrate the core idea of receding horizon approaches using OCPs and GNEPs.

\subsection{Model Predictive Control}
The first conceptualizations of the receding-horizon ideas appeared in early texts on optimal control. Stuart Dreyfus suggested the idea in \cite{Dreyfus60}\footnote{Indeed, Dreyfus later discredited the concept of receding horizon optimal control in \cite{Dreyfus77}.} and shortly after it was also mentioned in the early textbook of Lee and Markus \cite{Lee67}. The first technical paper on what nowadays is called model predictive control (MPC) is \cite{Propoi63}. 
However, before MPC became an intensively investigated control method,  it saw first applications in process industries in the 1970s. When in-depth exploration of closed-loop properties commenced in the late 1980s, e.g.~\cite{Keerthi88}, it became apparent that  Kalman's aphorism \textit{optimality does not imply stability}~\cite{Kalman60a} does not only apply to infinite-horizon optimal control but also to its receding-horizon variant. 
In 2001 Jan Maciejowski claimed that MPC has \textit{had a significant and widespread impact on industrial process control}~\cite{Maciejowski02a}. Hence it is not surprising that MPC theory for deterministic and finite-dimensional systems is by now available in a number of excellent textbooks~\cite{kouvaritakis2016mpc,Rawlings17,Gruene17a} and it is part of many engineering curricula~\cite{tudo:faulwasser24b}. 

Clearly it is beyond the scope of this tutorial to summarize more than three decades of research on closed-loop properties of MPC. However, for deterministic systems a compact gist can be derived. To this end, we focus on the setpoint stabilization problem, i.e., a target setpoint $(\xs, \us) \in \inte \mc Z^{\text{\tiny{OCP}}}_0(\x)$ is given and the control task is its asymptotic stabilization. Stability results for MPC  can actually be read as design guidelines for the underlying OCP.
\begin{pthm}[OCP design \& stability of MPC]~\label{pthm:stabMPC}\\
Consider the closed-loop system \eqref{eq:sys-cl} arising from OCP \eqref{eq:OCP} and the target setpoint $(\xs, \us) \in \inte \mc Z^\text{\tiny{OCP}}_0(\x)$. Asymptotic stability of \eqref{eq:sys-cl} at $\x = \xs$ can be fostered as follows:
\begin{itemize}
    \item[(i)] Design the  OCP \eqref{eq:OCP} such that for any 
    $(x^\diamond, u^\diamond) \in \mc{S}^{\text{\tiny OCP}}_\infty(\x)$ it holds that
    \[\lim_{k\to \infty} x_k^\diamond(\x) = \xs.\]
     \item[(ii)]  The optimal value function $V^\diamond_N$ from \eqref{eq:VN} is a natural candidate for a Lyapunov function of \eqref{eq:sys-cl}.
     \item[(iii)]  Design the  OCP \eqref{eq:OCP} such that $V^\diamond_N$ approximates $V^\diamond_\infty$ sufficiently well. 
\end{itemize}
\end{pthm}
Item (i) \edit{is based on} the observation that if infinite-horizon optimal solutions do not deliver the desired outcome ($ x_k^\diamond(\x) = \xs$ for $k\to\infty$) then the chosen OCP does not reflect the considered control task. 
Item (ii) is deeply rooted in Bellman's principle of optimality: \textit{end pieces of optimal solutions are optimal}. By this principle we have that
\begin{align*}
V_\infty(\x_{t+1}) - V_\infty(\x_t) &= \sum_{k = 0}^\infty  \ell(x^\diamond_{k|t+1}, u^\diamond_{k|t+1})  -  \ell(x^\diamond_{k|t}, u^\diamond_{k|t})   \\
&= -\ell(x^\diamond_{k|t}, u^\diamond_{k|t}) \\
&=-\ell(\x_t, \mu_\infty(\x_t)), 
\end{align*}
\edit{see, e.g., \cite{Keerthi88}, \cite[Thm. 4.8]{Gruene17a}. 
Using the optimal value function as Lyapunov function can be traced back to Kalman's  continuous-time stability analysis for the linear quadratic regulator~\cite{Kalman60a}.}
Hence a typical MPC design condition for setpoint stabilization is 
\begin{align}\label{eq:stagecostLB}
    \ell(x,u) \geq \alpha_\ell(\|x-\xs\|)
\end{align}
where $\alpha_\ell$ is a class $\mc K$ function lower bounding the stage cost~\cite[Sec. 1.2]{Mayne00a}, \cite{Rawlings17,Gruene17a}.\footnote{That is, $\alpha_\ell: \bb R_{\geq 0}\to \bb R_{\geq 0}$ is strictly monotonously increasing and satisfies $\alpha_\ell(0) = 0$, cf. \cite{Kellett14}.} As infinite-horizon optimal control is most often computationally intractable, item (iii) arises from the practical need to consider \edit{finite-horizon optimization problems}. A classic approach is to add so-called terminal ingredients to \eqref{eq:OCP} such that predicted trajectories end in a set $\bb X_{\mathrm f}$ and on this set the approximation $V_{\mathrm f}(x) \geq V_\infty(x), \forall x \in \bb X_{\mathrm f}$ and hence $\min J_N(x^\diamond(\x_t), u^\diamond(\x_t)) + V_{\mathrm f}(x^\diamond_{N|t}) \approx V_\infty(\x_t)$ holds. The seminal papers \cite{Mayne00a,Chen98} provide concise introductions to this approach. \vspace{2mm}

\edit{An alternative approach is rooted in relaxed dynamic programming~\cite{lincoln2006relaxing}. That is, instead of \eqref{eq:BE} one considers
\[
 V^\diamond_N(\x) \geq   \alpha \ell(\x, \mu_N(\x)) + V^\diamond_N(f(\x, \mu_N(\x))).
\]
for some $\alpha\in \bb R$. Supposing that $\alpha \in (0,1]$  gives
\[
 \alpha \ell(\x, \mu_N(\x)) \leq  V^\diamond_N(\x) - V^\diamond_N(f(\x, \mu_N(\x)))
\]
and summing from $t=0$ to $\infty$ combined with $  V^\diamond_N(\x) \geq 0$ (due to \eqref{eq:stagecostLB}) allows to show that
\[
V_\infty(\x_0) \leq \sum_{t=0}^{\infty} \ell(\x_t, \mu^\diamond_N(\x_t)) \leq \frac 1 \alpha V^\diamond_N(\x_0).
\]
Put differently, the performance realized by the receding horizon optimal control approach is a sufficiently good approximation of $ V^\diamond_\infty$ for closed-loop stability  if the suboptimality degree satisfies $\alpha \in (0,1]$, cf. \cite[Thm. 4.11]{Gruene17a}, \cite{Gruene09a,Gruene10a}.
Hence, this approach directly leads to closed-loop performance bounds for MPC~\cite{Gruene17a} and the prediction horizon $N$ becomes the crucial tuning knob to achieve $V^\diamond_N \approx V^\diamond_\infty$, cf. (iii) of Insight~\ref{pthm:stabMPC}. For further details and references we refer to the monograph~\cite{Gruene17a}.}

As we consider problems with constraints on states and inputs, any receding horizon approach also has to touch upon the issue of recursive feasibility, i.e., the question of whether existence of an optimal (and hence feasible) solution for the OCP solved for initial condition $\x_0$ implies its feasibility for all $t\in \bb N$. It turns out that both main avenues---terminal ingredients and sufficiently long horizons---can be used to enforce recursive feasibility. We refer, e.g., to \cite{Mayne00a} for the former and to \cite{Boccia14a} for the latter.

\subsection{Multi-Agent and Distributed MPC}
So far we have discussed the left hand side part of Figure~\ref{fig:Comparison_schematic}. Before we can turn to the right hand side part, which illustrates a game theoretic approach, we briefly discuss how one may handle multi-agent control problems using OCPs, cf. the middle part of Figure~\ref{fig:Comparison_schematic}. 

Indeed, considering a dynamic system subject to control actions of multiple agents \eqref{eq:sysv} we could stick to the OCP perspective and formulate the following problem 
\begin{subequations}
\label{eq:dOCP}
\begin{align}
\label{eq:dOCPRunningCost}
\displaystyle \min_{x,\,u^v, v\in \mc V}  &\;\sum_{v\in \mc V}\sum_{k= 0}^{N-1} \ell(x_k, u^v_k, u^{-v}_k) \\
\textrm{s.t.} \quad &  x_{k+1} =  f(x_k,u^1_k, \dots, u^M_k)  \hspace{1.25em}  k \in \bb{Z}_{N} \label{eq:dOCPConstr1}\\
&g(x_k,u^1_k, \dots, u^M_k) \leq 0, \hspace{2.5em} k \in \bb{Z}_{N} \label{eq:dOCPConstr2}\\
 & h(u^1_k, \dots, u^M_k)\leq 0,      \hspace{4.em} k \in \bb{Z}_{N} \label{eq:dOCPConstr3}\\ 
&\; x_0 = \x. \label{eq:dOCPConstr4}
\end{align}
\end{subequations}
In contrast to OCP \eqref{eq:OCP} here the multi-agent structure is made explicit, yet the problem remains \textit{one OCP} which asks to optimize over the summed cost of all agents in a cooperative manner. Thus, for ~\eqref{eq:dOCP} the usual optimality solution concept from \eqref{eq:optimalityOCP} applies. Moreover, if one solves \eqref{eq:dOCP} in cooperative fashion to optimality then our previous discussion on stability of MPC remains valid. 

There exist two main motivations for considering a multi-agent structure in an OCP setting: (i) when a large number of systems shall be coordinated with each other while the control actions for each system shall be computed locally, see, e.g., \cite{keviczky2006decentralized,camponogara2002distributed}, or (ii) when the problem size requires a distributed implementation to render it numerically tractable.

With respect to (i), it is clear that the dynamics \eqref{eq:sysv}/\eqref{eq:dOCPConstr1} and the inequality constraints will have additional structure which captures that individual dynamic agents interact with each other. Prime examples are energy grids or cooperating robots. In such settings, one often strives to approximate the centralized solution to \eqref{eq:dOCP} through tailored distributed numerical algorithms which exploit the interconnection structure. Put differently, in such situations a distributed optimization algorithm may be relied upon to ensure coordination. Examples are ADMM\edit{~\cite{Boyd2011}} and dual decomposition for linear dynamics and convex constraints, or ALADIN~\cite{Houska2016} and SQP-like methods~\cite{Stomberg2025b} for nonconvex problems. We refer to \cite{Scattolini09a,muller2017economic} for introductory overviews on \textit{distributed MPC}. Indeed, an interesting aspect of such problems is that distributed optimization algorithms often exhibit slower convergence than their centralized counterparts. Hence the interplay of in-exact distributed optimization and the multi-agent dynamics becomes relevant as pointed out in~\cite{Stomberg2025b}. Alternatively, one can partially alleviate the centralized computational burden by considering the influence of neighboring agents, e.g., through sensitivities or compatibility constraints \cite{dunbar2006distributed}. 

With respect to (ii), we remark that early works on this include \cite{Venkat08a} while recently \cite{tuhh:stomberg25b} provided a case study on numerical scalability ranging up to more than $10^6$ decision variables. A recent overview on tailored algorithms for convex linear-quadratic problems can be found in \cite{tudo:stomberg22a}.
We conclude by noting that distributed MPC approach do not allow to model settings where agents are self-interested and thus are not willing to cooperate. Hence  we next turn to the game-theoretic point of view.

\subsection{Receding Horizon Games}
Receding horizon games (RHGs) are a control approach for multi-agent systems that generates control actions by solving a dynamic game with coupling constraints in a receding-horizon fashion. They are therefore able to model the competitive nature of self-interested agents with shared resources while incorporating future predictions, dynamic models, and constraints into the decision-making process. Applications of RHGs include robotics~\cite{gu2008differential}, autonomous driving~\cite{liniger2020noncooperative, wang2021game, lecleach2022algames}, electric vehicle charging~\cite{mignoni2023distributed},  traffic routing~\cite{benenati2024probabilistic}, supply chains~\cite{hall2024receding}, as well as smart grids~\cite{paola2018distributed, stephens2015game, hall2022receding}. 

In technical terms, an RHG controller solves the dynamic, finite-horizon GNEP in~\eqref{eq:GNEP} at each step and applies it in receding-horizon fashion. This yields an implicit feedback policy, the \textit{receding-horizon game} feedback law
\begin{align}\label{eq:FeedbackLaw}
u^*_{0|t} = \mathrm{col}(u_{0|t}^{*,v})_{v \in \mc{V}}  
=:  \mu^*_N(\x_t) = \kappa(\mc{S}^{\text{\tiny GNE}}_N(\x_t)),
\end{align}
where the GNE solution map $\mc{S}^{\text{\tiny GNE}}_N(\x_t)$ returns the (possibly infinite) set of GNEs, and $\kappa$ is a deterministic selection mechanism that picks a unique one and extracts the first input of each agent, $\{u_{0|t}^{*,v}\}_{v \in \mc{V}}$. This selection mechanism is necessary if one does not impose conditions ensuring uniqueness of the GNE in~\eqref{eq:GNEP}.\footnote{We note that also in the MPC context, situations with multiple optimal solutions formally necessitate the use of such a selection mechanism or tie-breaker rule. \edit{The effect of selection mechanisms for GNEPs on their closed-loop trajectories is still an open research topic.}} Various selection mechanisms exist in the \edit{literature}~\cite{benenati2023optimal, hall2025limits, hall2026solving}.

Naturally, one may ask what equilibria such a receding-horizon implementation of GNEs converges to. Interestingly, the closed-loop attractors in GNEPs, in contrast to OCPs, are \edit{two-fold equilibria}: a point that is both (i) a steady-state of the dynamics~\eqref{eq:Constr1}, i.e., $\bar x = f(\bar x, \bar u^v, \bar u^{-v})$ (as in standard OCPs); and (ii) a strategic (decision) equilibrium of the 1-step GNEP in~\eqref{eq:GNEP} \edit{such that no agent wants to unilaterally deviate (the classic Nash equilibrium)}. Formally, \edit{we refer to this point as} a \textit{steady-state GNE} and is computed by solving the 1-step version of~\eqref{eq:GNEP} with an additional \edit{steady-state constraint} as stated next.

\begin{dfn}[Steady-state GNE]\label{dfn:SteadyStateGNEP} The pair $(x_s,u_s)$ is called a steady-state GNE if it solves 
\begin{align}\label{eq:SteadyStateGNEP}
v\in \mc{V}: \left\{
\begin{array}{r l}
\displaystyle \min_{\bar{x}, \bar{u}^v} & \; \ell^v(\bar{x}, \bar{u}^v, \bar{u}^{-v}) \\ 
 \subjectto  &  f(\bar{x},\bar{u}^v,\bar{u}^{-v})- \bar{x}=0\\ 
            &  g(\bar{x},\bar{u}^v, \bar{u}^{-v}) \leq 0,\\
            &h^v(\bar u^v)\leq 0,\\
\end{array} 
\right.
\end{align}
with the corresponding solution set $\mc{S}^{\text{\tiny GNE}}_s \subset \R^{n_x + n_u}.$ Problem \eqref{eq:SteadyStateGNEP} is called a steady-state GNEP.
\end{dfn}

We will showcase closed-loop RHG trajectories and their convergence to the steady-state GNE in the following nonlinear example.  

 \begin{ex}[A nonlinear receding horizon game]\label{ex:NonlinearRHG}
We adapt the nonlinear example from \cite{gruene2013economic, gruene2014asymptotic} to a game-theoretic setting with two agents and local dynamics: 

\begin{equation*}
\text{\small $\forall v \in \mathcal{V}$}  
\left\{
\begin{array}{r l}
\displaystyle \min_{x,u^v}  &\displaystyle \sum_{k= 0}^{N-1}  \displaystyle 
-\ln\Big(q^v(x_k^v)^{\alpha^v} -r^v  u_k^v \sum_{j\in \mc{V}} u_k^j\Big) \\
\textrm{s.t.} & \displaystyle  x_{k+1}^v =u_k^v  \hspace{4.4em}  k \in \bb{Z}_{N}\\
& 0.1\leq \displaystyle \sum_{j\in\mc{V}} u_k^j\leq 5, \hspace{1.3em} k \in \bb{Z}_{N} \\ 
& 0\leq \displaystyle x_k^v\leq 10, \hspace{0.3em}
\hspace{2.9em} k \in \bb{Z}_{N+1} \\ 
 & x_0^v =1. 
\end{array}
\right.
\end{equation*}
with parameter values $q^1 = 5$, $q^2= 4$, $r^1 = 1$, $r^2= 1.5$, $\alpha^1 = 0.3, \alpha^2=  0.2$, and a horizon length of $N=12$. We solve the GNEP using the~\textit{NashOpt} library~\cite{bemporad2025nashopt}. The closed-loop RHG trajectories are plotted in Figure~\ref{fig:ClosedLoopEconGrowth} \edit{as well as the OCP counterpart} and we want to point out two things: (i) we see the convergence of the closed-loop to their respective steady states $x_s^{*}$ and $x_s^{\diamond}$, and (ii) the GNE trajectories clearly exhibit the turnpike property in the open-loop predictions (red dashed lines).

\edit{
For this example, we can also give the explicit solutions for steady states. 
To this end, we consider the special case of $M$ agents with $\alpha^v = \alpha \in (0,1)$ and $q^v/r^v = c$, and
assume the constraints are inactive. Then the steady states
of~\eqref{eq:SteadyStateGNEP} and of its OCP counterpart can be computed explicitly as follows
\[
    u_s^{v,*} = \Big(\frac{\alpha \,c}{M+1}\Big)^{\frac{1}{2-\alpha}},
    \quad
    u_s^{v,\diamond} = \Big(\frac{\alpha \, c}{2M}\Big)^{\frac{1}{2-\alpha}}.
\]
The difference between GNE and OCP vanishes for $M = 1$, as expected. It is caused by the cost
coupling $-r^v u^v_k \sum_j u^j_k$, since agent $v$ accounts only for the effect on
itself, $-r^v(u^v_k)^2$, and disregards the effect on the others.
}

\begin{figure}
\centering
\includegraphics[width=\columnwidth]{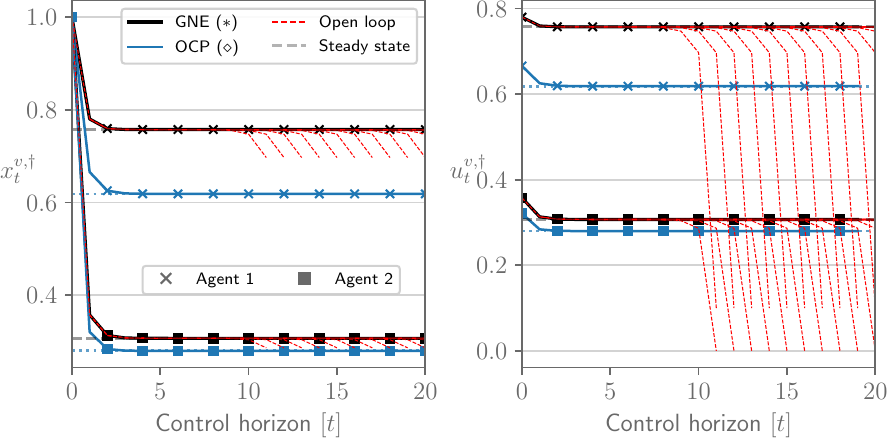}
\caption{\edit{Closed-loop GNE and OCP trajectories (solid) and open-loop GNE predictions (dashed) for Example~\ref{ex:NonlinearRHG}}.} \label{fig:ClosedLoopEconGrowth}
\end{figure}
 \end{ex}%

\section{Dissipativity and Turnpikes} \label{sec:diss}
After our introduction to main concepts for OCP and GNEPs and to the receding horizon perspective on both, we now turn towards dissipativity as it provides a powerful approach for the analysis of OCPs, GNEPs and their receding horizon variants.

\subsection{The Perspective of Willems}
In his own words, Willems' motivation for exploring dissipativity was \textit{a
generalization of Lyapunov functions to open systems, to systems with inputs and outputs} \cite{Willems07a}.
Hence, he considered non-negative functions $\Lambda:\bb R^{n_x}\to\bb R_{\geq 0}$\footnote{\edit{Willems considered non-negative storage functions, while for the results presented here only boundedness from below is required.}} and systems with inputs and outputs. Transferred to the discrete-time setting of this paper,  he considered
\[ x_{t+1} = f(x_t, u_t), \quad  y_t = h(x_t)\]
and conceptualized the crucial inequality
\[
\Lambda(f(x,u)) - \Lambda(x)  \leq s(y, u),
\]
where $\Lambda$ is called \textit{storage function} and $s:\bb R^{n_y}\times \bb R^{n_u} \to \bb R$ is called \textit{supply}. It expresses  that 
\[\text{change of stored energy} \leq \text{supplied energy}\]
holds in open physical systems, but translated to a more abstract control-specific language.
While 1972 Willems' foundational paper considers the continuous-time setting~\cite{Willems72a}, the first approach towards the discrete-time extension is due to \cite{Byrnes94}.
The use cases of dissipativity concepts in systems and control are ubiquitous, we refer to the two-part special issue in IEEE Control Systems~\cite{IEEECSM02a,IEEECSM02b} for an overview.

One fundamental link between optimal control and dissipativity has been identified by Willems early on.
Consider  $s(x, u) := s(h(x),u)$ and the free end-time OCP
\begin{equation}\label{eq:availStor}
\begin{aligned}
    \Lambda^a(\x) &:= \sup_{x, u, N\in \bb N} -\sum_{k=0}^N s(x_k,u_k)\\
    \text{ s.t. } &(x,u) \text{ satisfy } \eqref{eq:sysu}\quad\text{and}\quad
    x_0 = \x.%
\end{aligned}
\end{equation}
This problem expresses the idea to compute the maximum storage which can be extracted by dynamic operation starting from the initial condition $\x$. Hence the value function $\Lambda^a:\bb R^{n_x} \to \bb R_{\geq 0} \cup \{\infty\}$ is called \textit{available storage}. 

\begin{pthm}[Available storage and dissipativity\edit{~\cite[Thm. 1]{Willems72a}}]\label{pthm:availStor}
Consider the dynamic system \eqref{eq:sysu} and a supply rate $s:\bb R^{n_x}\times \bb R^{n_u} \to \bb R$. 
\begin{itemize}
   \item[(i)]  The available storage is bounded from below by 0. 
    \item[(ii)]  The system is dissipative on $\bb R^{n_x}\times \bb R^{n_u}$ if and only if $\Lambda^a(\x) < \infty$ for all $\x \in \bb R^{n_x}$.   
\end{itemize}
\end{pthm}
Item (i) follows from the fact that for $N=0$, the objective of \eqref{eq:availStor} equates to $0$. Item (ii) can be shown as follows:

\textit{$\Lambda^{a} < \infty \Rightarrow$ dissipativity:} \qquad Consider some trajectory $(x, u)$ which connects $x_0 = \x$ with $x_m$ in $m \geq 1$ steps. 
One suboptimal possibility to extract the storage at $x_m$ is to first use $(x, u)$ to traverse from $\x$ to $x_m$ and then apply the inputs suggested by \eqref{eq:availStor}. 
This extracts the storage equivalent to 
$
\sum_{k=0}^m  -s(x_k,u_k) + \Lambda^a(x_m)$.
The definition of the available storage \eqref{eq:availStor}, however, requires to find the supremum of
$\sum_{k=0}^N  - s(x_k, u_k)$ and hence
$\Lambda^a(\x) \geq \sum_{k=0}^m    -s(x_k,u_k) + \Lambda^a(x_m)$.

\textit{Dissipativity $\Rightarrow\Lambda^{a} < \infty$:} \quad 
For the sake of contradiction, suppose that at some $\x$ a dissipation inequality holds with a storage function $\Lambda$ bounded from below while $\Lambda^a(\x) = \infty$.
At $x_0 = \x$ and for any storage function $\Lambda$ we have
$\Lambda(\x)+ \sum_{k=0}^m s(x_k,u_k) \geq \Lambda(x_m) \geq -c >-\infty$.
The right hand side inequalities follow from the lower boundedness of  storage functions $\Lambda$. 
Hence, we set $m=N$ and rearrange the inequalities  such that
$\Lambda(\x) \geq -c + \sup_{x, u, N\in \bb N} - \sum_{k=0}^N s(x_k,u_k)$.
Here we have used that $-c$ is not affected by the optimization over the free end time $N$.
With $x_0 = \x$, the previous inequality is equivalent to
$
\Lambda(\x) \geq -c + \Lambda^a(\x)$.
Since $c \in \mathbb R$, $\Lambda^a(\x) =\infty$ would contradict $\Lambda(\x)\in \mathbb R$.\vspace{2mm}

Notice that the structure of this argument is quite general. That is, one could restrict the class of considered trajectories in \eqref{eq:availStor} (and thus also the set of trajectories for which a dissipation inequality holds). This arises, e.g., when state and input constraints are considered, see \cite{Gruene16a} for a detailed proof in the optimal control context. 
Furthermore, we remark without elaboration that \cite{Willems72a} also introduced the  \textit{required supply} as a value function of an OCP with prescribed terminal constraint and as an alternative dissipativity characterization using reachability properties.

\subsection{Dissipativity with Stage Costs}
In his 1971 paper on least-squares infinite-horizon optimal control~\cite{Willems71a}, Willems did already use dissipation inequalities to analyze optimal control problems. In particular he explores the link between dissipation inequalities and the existence of stabilizing solutions to the Riccati equation underlying the OCP. Yet, he swapped the sign of the inequality~\cite{tudo:faulwasser22b}.
Nevertheless, the key idea is to consider the stage cost $\ell$ to define the supply $s$.

Consider
\begin{equation}\label{eq:SOP}
\begin{aligned}
(\xs^\diamond, \us^\diamond) &\in \argmin_{x,u} \ell(x,u) \\
\text{ s.t. } x &= f(x,u),~ \edit{\eqref{eq:OCPConstr2}, \eqref{eq:OCPConstr3}},
\end{aligned}
\end{equation}
that is, $(\xs^\diamond, \us^\diamond)$ is the steady state minimizing the stage cost $\ell$. 
The crucial dissipation inequality for optimal control problems then reads
    \begin{equation}\label{eq:DI}
    \Lambda(f(x,u)) - \Lambda(x)  \leq \ell(x,u) - \ell(\xs^\diamond, \us^\diamond). \tag{DI}
\end{equation}
\begin{pthm}[Dissipativity, $V^\diamond_N$, and \eqref{eq:BE}]~\label{pthm:LBonVN}\\
Consider the OCP \eqref{eq:OCP} and suppose that \eqref{eq:DI} holds for any optimal pair $(x^\diamond, u^\diamond) \in \mc{S}^{\text{\tiny OCP}}_N(\x), N\in \bb N$.
Then, any storage function $\Lambda$ satisfies:
\begin{itemize}
   \item[(i)] $ \Lambda(f(x^\diamond_{N-1},u^\diamond_{N-1})) - \Lambda(\x) \leq V_N^\diamond(\x) -N \ell(\xs^\diamond, \us^\diamond)$ and
\item[(ii)]  $\displaystyle -\Lambda(\x) \leq \min_u\, \ell(\x, u) - \ell(\xs^\diamond, \us^\diamond) - \Lambda(f(\x, u))$.
\end{itemize}
\end{pthm}
Item (i): If one considers some optimal pair $(x^\diamond, u^\diamond) \in \mc{S}^{\text{\tiny OCP}}_N(\x)$ a telescopic sum argument shows that
\[
\Lambda(f(x^\diamond_{N-1},u^\diamond_{N-1})) - \Lambda(\x)  \leq \sum_{k=0}^{N-1}\ell(x^\diamond_k,u^\diamond_k) - \ell(\xs^\diamond, \us^\diamond).
\]
That is, any storage function which corresponds to the stage cost supply $\ell$ provides a lower bound on the optimal value function. Item (ii): 
Rearranging the last inequality and \edit{considering $N=1$ gives the Bellman inequality of (ii). Indeed, setting $\ell(\xs^\diamond, \us^\diamond) = 0$ highlights the structural link to \eqref{eq:BE}.} Hence another explanation for (i) is that negative storage functions ($-\Lambda$) constitute \textit{subsolutions} to Bellman equations \eqref{eq:BE}. For OCPs and MPC the link between \eqref{eq:BE} and \eqref{eq:DI}/\eqref{eq:sDI} is discussed in \cite{zanon2026rethinking,tudo:faulwasser21a}.

In its strict form the crucial inequality reads
 \begin{multline}\label{eq:sDI}
    \Lambda(f(x,u))- \Lambda(x)  \\ \leq -\alpha(\left\|\begin{smallmatrix}
    x-\xs \\ u-\us
\end{smallmatrix}\right\|)+\ell(x,u) - \ell(\xs, \us). \tag{sDI}
\end{multline}
Clearly, any storage function satisfying \eqref{eq:sDI} also satisfies \eqref{eq:DI}. Moreover, the available storage characterization of dissipativity carries over to \eqref{eq:sDI} if one sets $s(x,u):=  -\alpha(\left\|\begin{smallmatrix}
    x-\xs \\ u-\us
\end{smallmatrix}\right\|)+\ell(x,u) - \ell(\xs, \us)$ in Insight~\ref{pthm:availStor}. 
\begin{rmk}[Why is the inequality \eqref{eq:sDI} called strict?]~\\
    Given a closed inequality $a\leq b$, the limiting case is the equality $a=b$. However, the name strict dissipativity inequality for \eqref{eq:sDI} does not refer to this limiting case. Rather the fact that $\alpha:\bb R_{\geq 0} \to \bb R_{\geq 0}$ is strictly monotonous leads to the situation that $\eqref{eq:sDI}$ only holds with equality at points where $\alpha(\left\|\begin{smallmatrix}
    x-\xs \\ u-\us
\end{smallmatrix}\right\|)=0$. For \eqref{eq:sDI} as given above this is equivalent to the singleton set $(\xs,\us) = (x,u)$. 

Notice that one could also work with $\alpha(\left\|x-\xs\right\|)$ thus require strictness in $x$ only. Here we simplify the analysis and ask for strictness in $x$ and $u$.
\end{rmk}
\begin{pthm}[Strict dissipativity and \eqref{eq:SOP}]~\label{pthm:sDISOP}\\
Let \eqref{eq:sDI} hold on the feasible set of \eqref{eq:SOP}.
Then $(\xs,\us)$ is the unique global minimizer of \eqref{eq:SOP}.
\end{pthm}
This insight is obtained by rearranging \eqref{eq:sDI} into \[\alpha(\left\|\begin{smallmatrix}
    x-\xs \\ u-\us
\end{smallmatrix}\right\|) \leq\ell(x,u) - \ell(\xs, \us)\] and exploiting that for steady states $\Lambda(x) = \Lambda(f(x,u))$. \edit{Early observations of this are made in \cite{Diehl11a}, formal results can be found in \cite{Angeli12a,epfl:faulwasser15h,Stieler14a}}. That is, strict dissipativity may serve as an abstract generalization of strict convexity properties which would otherwise be the go-to choice for establishing uniqueness of global minimizers. \edit{It is worth noting that the inequality \eqref{eq:sDI} relaxes the classic stage cost lower bound \eqref{eq:stagecostLB} to the set of constrained steady-states of the system. }

\begin{rmk}[System property or design requirement?]\label{rmk:sysProporDesign}~\\
At this point, it is fair to ask which object carries the dissipativity property: Willems defined dissipativity  as a system property. However, in Insight~\ref{pthm:LBonVN} we only required it for trajectory pairs from $\mc{S}^{\text{\tiny OCP}}_N(\x)$, i.e., \eqref{eq:DI} was needed along optimal pairs.  If (strict) dissipativity is imposed for all $(x,u) \in  \mc Z^{\text{\tiny OCP}}_0$, then it turns out to be the property of a constrained system. If it is imposed on $\mc{S}^{\text{\tiny OCP}}_N(\x) \subseteq \mc Z^{\text{\tiny OCP}}_N \subseteq \mc Z^{\text{\tiny OCP}}_0$, then it becomes a property of the considered OCP \cite{epfl:faulwasser14e}. Following Willems' original motivation (extending Lyapunov concepts to open systems), the former is a very natural approach. However, from the receding-horizon perspective it is also restrictive. In a nominal receding horizon setting, the inputs applied in closed loop \eqref{eq:sys-cl} are obtained by solving a finite-horizon open-loop problem. Thus, considering the OCP to be the carrier of the (strict) dissipativity property is considerably less restrictive. Moreover, as the formulation of OCPs involves a number of design choices (constraints, objective, horizon), dissipativity of OCPs can be enforced through these choices. In other words, dissipativity as a system property can only be influenced through the stage cost, while dissipativity as an OCP property can be affected through the \textnormal{stage cost}, the \textnormal{terminal costs}, and the \textnormal{constraints}.
\end{rmk}

So far, our discussion of dissipativity has focused on OCPs. Given the theme of this paper, it is natural to ask if and how one can extend it to GNEPs. 
\begin{pthm}[The strict dissipation obstacle for GNEPs \edit{\cite[Lem. 1]{hall2025system}}]\label{pthm:DISOb}
Consider the aggregated cost  \eqref{eq:aggregatedCost} and suppose that $(\xs^\diamond, \us^\diamond)$  $\neq (\xs^*, \us^*)$ solve \eqref{eq:SOP}, respectively, \eqref{eq:SteadyStateGNEP}.

Then the inequality
\begin{multline*}
     \Lambda(f(x,u))- \Lambda(x) \leq -\alpha\left(\left\|\begin{smallmatrix}
    x-\xs^* \\ u-\us^*\end{smallmatrix}\right\|\right)\\  +\ell(x,u) - \ell(\xs^*, \us^*)
\end{multline*}
cannot hold on $\mc Z^{\text{\tiny OCP}}_N = \mc Z^{\text{\tiny GNEP}}_N$, $N\in \bb N$.
\end{pthm}
As $(\xs^\diamond, \us^\diamond)$ satisfies an optimality condition, we have that
$\ell(\xs^*, \us^*) \geq \ell(\xs^\diamond, \us^\diamond)$ and thus
\[
0 \leq \alpha\left(\left\|\begin{smallmatrix} x-\xs^* \\ u-\us^*\end{smallmatrix}\right\|\right) \leq \ell(\xs^\diamond, \us^\diamond) - \ell(\xs^*, \us^*).
\]
The inequality proposed would need to hold on the feasible set $\mc Z^{\text{\tiny OCP}}_N$ and thus for all steady states. This would imply 
\[
\alpha\left(\left\|\begin{smallmatrix} x-\xs^* \\ u-\us^*\end{smallmatrix}\right\|\right)\leq \ell(x,u) - \ell(\xs^*, \us^*).
\]
which contradicts $\ell(\xs^*, \us^*) \geq \ell(\xs^\diamond, \us^\diamond)$. \vspace{2mm}

The previous considerations show that in case of GNEPs, conceptualizing strict dissipativity as a system property is extremely restrictive. It will only make sense if $(\xs^\diamond, \us^\diamond)\equiv (\xs^*, \us^*)$. However our previous considerations of dissipativity as a property of an OCP show the route to move forward---we need to consider a strict dissipation inequality (i.e. \eqref{eq:sDI} with $(\xs^\diamond, \us^\diamond)$ swapped for $(\xs^*, \us^*)$ in $\ell$) and $\alpha_\ell$ holds along the trajectories generated by the GNEP, i.e., on $\mc{S}^{\text{\tiny GNE}}_N(\x)$. 

We remark that the available storage characterization from Insight~\ref{pthm:availStor} can be extended to strict dissipativity of GNEPs \cite{hall2025system}. Likewise, a counterpart to Insight~\ref{pthm:LBonVN} can be derived if the aggregated cost from \eqref{eq:aggregatedCost} is evaluated along $(x,u) \in \mc{S}^{\text{\tiny GNE}}_N(\x)$.

\subsection{The Turnpike Phenomenon}\label{sec:turnpike}
\begin{figure}
  \centering
  \includegraphics[width=\columnwidth]{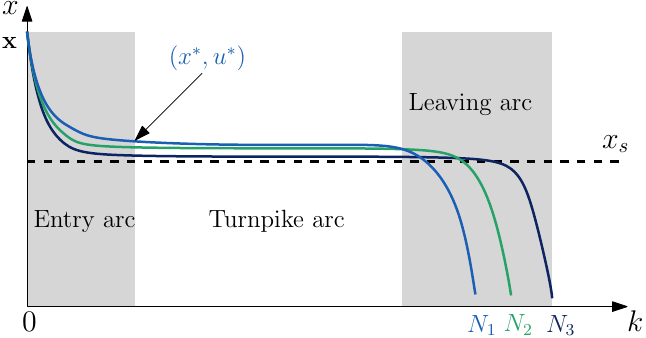}
  \caption{Turnpike phenomenon}
  \label{fig:Turnpike_schematic}
\end{figure}

In the numerical results of Example~\ref{ex:diff_OCPGNEP} as well as in Example~\ref{ex:NonlinearRHG} we observe the typical turnpike behavior of optimal or GNE solutions: for different initial conditions and different horizon lengths the solutions approach a neighborhood of one common steady state while the amount of time spent close to this equilibrium grows with increasing horizon length. An illustrative sketch is also provided in Figure~\ref{fig:Turnpike_schematic}. Hence we now discuss the relation of turnpike and dissipativity properties in OCPs and GNEPs. \vspace{2mm}

\subsubsection*{Turnpikes in OCPs}
In the context of economic MPC it had been observed that strict dissipativity allows to explain the turnpike phenomenon.  
\begin{pthm}[\eqref{eq:sDI} $\Rightarrow$ turnpike in OCPs]~\\
\label{pthm:DissiTurnpikeDeterministic}
Suppose that 
\begin{itemize}
    \item all $(x^\diamond, u^\diamond) \in \mc{S}^{\text{\tiny OCP}}_N(\x)$ satisfy \eqref{eq:sDI},  
    \item a suitable reachability condition holds, and the storage is bounded. 
\end{itemize}
    Then optimal solutions $(x^\diamond, u^\diamond) \in \mc{S}^{\text{\tiny OCP}}_N(\x)$
    satisfy
    \[
    |\text{ time } (x^\diamond, u^\diamond) \text{ spent outside } \mc B_\epsilon(\bar x)~ | \leq \dfrac{C(\x)}{\alpha(\epsilon)}
    \]
    where $C(\x)$ does not depend on $N$.
\end{pthm}
Formal proofs of this result can be traced back to \cite[Thm. 5.3]{Gruene13a}, while a recent literature overview is given in \cite{tudo:faulwasser22a}. We note that an early continuous-time precursor of this result (which does not explicitly use dissipativity  but conditions which imply it) can also be found in the book of Carlson, Haurie and Leizarowitz~\cite[Chap. 4.4]{Carlson91}. A continuous-time proof can be found in \cite{epfl:faulwasser15h}. Common to these proofs is that the constant $C(\x)$ is the sum of a bound on $ \Lambda(x) - \Lambda(f(x,u))$ and a performance bound deduced from the assumed reachability property.
Finally, notice that in view of Remark~\ref{rmk:sysProporDesign} the above implication also holds if the strict dissipation inequality is required for all feasible $(x,u)$ pairs instead of only optimal ones. This setting is, e.g., considered by \cite{Gruene13a}.

It should also be noted that the amount of time $(x^\diamond, u^\diamond)$ spent outside of the closed ball $ \mc B_\epsilon((\xs, \us))$ can be captured in different ways:
\begin{itemize}
    \item One may explicitly measure this amount of time. This leads to so-called cardinality turnpikes for discrete-time systems or to measure turnpikes for continuous-time ones. Comparison of Insight~\ref{pthm:DissiTurnpikeDeterministic} and Definition~\ref{dfn:GameTurnpike} shows that one may equivalently measure the time close to the turnpike or far away from it. 
    \item One bounds the distance of solutions pairs $(x^\diamond, u^\diamond)$ to  $(\xs, \us)$ from above by exponential bounds of the form $C(e^{-c k}- e^{-c(N-k)})$, $C,c>0$ \cite{Trelat15a}.
\end{itemize}
An overview of approaches to formalize the turnpike property for OCPs can be found in  \cite{tudo:faulwasser22a}.\vspace{2mm}

\subsubsection*{Turnpikes in GNEPs}
Turnpikes have been observed and discussed in the context of games since the 1980s~\cite{fershtman1986turnpike}. Early work studied infinite-horizon open-loop games in continuous~\cite{carlson1995turnpike} and discrete time~\cite{carlson1996turnpike}, addressing existence and uniqueness of a turnpike as well as convergence properties~\cite{carlson2000infinite}. Turnpike properties in games have recently received renewed attention. Stochastic differential LQ games with continuous-time dynamics are studied in the zero-sum~\cite{li2025turnpike} and non-zero-sum~\cite{cohen2025turnpike} settings. Several works derive turnpike properties for mean field games under a large-population assumption~\cite{cirant2021long, carmona2024leveraging, ersland2025long, fedorov2025studying}. For dynamic GNEs such as~\eqref{eq:GNEP} specifically, turnpikes have been observed in competitive dynamic supply chains~\cite{hall2024receding}.

Despite the long history of work on turnpikes in games, a formalization in line with the system-theoretic characterization in optimal control was, until recently, missing.
 \cite{hall2025system} is the first work establishing an overarching theory of turnpikes in GNEPs for nonlinear costs, nonlinear constraints, and discrete-time dynamics.
Formally, we define the property as follows:
\begin{dfn}[Turnpike in GNEPs]\label{dfn:GameTurnpike}~\\
The GNEP~\eqref{eq:GNEP} exhibits the turnpike property at $(x_s,u_s)$ if for each $\varepsilon>0$ there exists $C>0$ such that $\forall N\in \bb{N}, \forall \x\in  \bb{X}_0$, and for all game pairs $(x,u)\in \mc{S}^{\text{\tiny GNE}}_N(\x)$ it holds that
\begin{align}\label{eq:TurnpikeInequality}
Q_{\varepsilon} := \#\left\{k \in \bb{Z}_N |\,\left\| \begin{smallmatrix}x_k- x_s\\ u_k - u_s \end{smallmatrix} \right\|\leq \varepsilon\right\} \geq N - \frac{C}{\alpha(\varepsilon)}
\end{align}
for some $\alpha \in \mc K,$ and where $\#$ refers to the cardinality.
\end{dfn}
In order to shed some light on the generating mechanism for turnpikes in GNEPs, we introduce
\begin{equation}\label{eq:PoA}
    PoA(\x):=\frac{ \displaystyle\sup_{(x^*, u^*) \in \mc{S}^{\text{\tiny GNE}}_N(\x)} \sum_{v\in \mc V} J^v_N(x^*,u^{v,*}, u^{-v,*})}{V^\diamond_N(\mathbf{x})} 
\end{equation}
which is also known as the \textit{price of anarchy}~\cite{koutsoupias1999worst,papadimitriou2001algorithms}. It captures the decrease in performance (increase of the aggregated objective functional) of the ``worst" GNEP solution with respect to the OCP solution. Clearly, for this comparison to be valid we consider $\ell$ from \eqref{eq:aggregatedCost} and it must hold that $\infty > V^\diamond_N(\mathbf{x}) > 0$.

\begin{pthm}[\eqref{eq:sDI} $\Rightarrow$ turnpike in GNEPs \edit{\cite[Thm. 3]{hall2025system}}]~\\
\label{pthm:DissiTurnpikeDeterministicGNEP}
Suppose that 
\begin{itemize}
    \item a suitable reachability condition holds, 
    \item the storage is bounded, 
    \item all $(x^*(\x), u^*(\x)) \in \mc{S}^{\text{\tiny GNE}}_N(\x)$ satisfy \eqref{eq:sDI},  and  
    \item the price of anarchy is bounded. 
\end{itemize}
    Then all GNEP solutions $(x^*(\x), u^*(\x)) \in \mc{S}^{\text{\tiny GNE}}_N(\x)$
    satisfy the turnpike property from Definition~\ref{dfn:GameTurnpike} at $(\xs^*, \us^*)$.
\end{pthm}
It turns out that, besides restricting strict dissipativity to all $(x^*(\x), u^*(\x)) \in \mc{S}^{\text{\tiny GNE}}_N(\x)$, the ingredient used to go from a turnpike result for OCPs to one for GNEPs is the assumption on the bounded price of anarchy. Indeed, in \cite{hall2025system} it has been shown that \eqref{eq:PoA} implies that $\sum_{v\in \mc V} J^v_N(x^*,u^{v,*}, u^{-v,*}) \leq V^\diamond_N(\mathbf{x}) + P$ which combined with Insight~\ref{pthm:performance}
gives
\[V^\diamond_N(\mathbf{x})\leq \sum_{v\in \mc V} J^v_N(x^*,u^{v,*}, u^{-v,*}) \leq V^\diamond_N(\mathbf{x}) + P. \]

\begin{rmk}[Converse turnpike results?]
Insights~\ref{pthm:DissiTurnpikeDeterministic} \& \ref{pthm:DissiTurnpikeDeterministicGNEP} illustrate that strict dissipativity allows to deduce the turnpike property. Naturally, one may ask if there could be also other ways to prove the property. Indeed, in the OCP literature there exist results to verify the property exploiting the saddle-point nature of the optimality system~\cite{Trelat15a}.
Moreover, there also have been efforts to establish the converse implication, i.e., turnpike $\Rightarrow$ strict dissipativity: \cite{Gruene16a} explores this for discrete-time  OCPs, while \cite{epfl:faulwasser15h} considers continuous-time problems using a local cost bound which is related to second-order sufficient conditions for \eqref{eq:SOP}. Moreover, the proof technique from the latter paper has been transferred to the GNEP setting by \cite{hall2025system}. 
\end{rmk}

\section{Closed-Loop  Analysis --  Dissipativity and Beyond}
\subsection{Model Predictive Control}
As mentioned earlier in the context of economic MPC, dissipativity-based concepts have led to a substantial generalization of analysis tools and concepts. The early works \cite{Diehl11a,Angeli12a} use terminal constraints and different variations of strict dissipativity assumptions, \cite{Gruene13a} explores settings without terminal ingredients and provides performance bounds. While an overview of the available closed-loop results can be found in \cite{kit:faulwasser18c} we now provide the dissipativity-based counterpart to the gist summarized in Insight~\ref{pthm:stabMPC}. 

\begin{pthm}[Dissipativity-based MPC analysis]~\label{pthm:sDIMPC}\\
Consider the closed-loop system \eqref{eq:sys-cl} arising from OCP \eqref{eq:OCP} and suppose that  $(x^\diamond, u^\diamond) \in \mc{S}^{\text{\tiny OCP}}_N(\x)$ satisfy \eqref{eq:sDI}. 
\begin{itemize}
    \item[(i)]   If, for any $N\in \bb N\cup \{\infty\}$ all $(x^\diamond, u^\diamond) \in \mathcal{S}^{\text{\tiny OCP}}_N(\x)$ satisfy \eqref{eq:sDI} 
    then it holds that
    \[\lim_{k\to \infty} x_k^\diamond(\x) = \xs^\diamond.\]
     \item[(ii)]  The optimal value function $V^\diamond_N$ from \eqref{eq:VN} and any storage $\Lambda$ imply  a natural candidate for a Lyapunov function for (practical) asymptotic stability of \eqref{eq:sys-cl}: $W(\x):=V^\diamond_N(\x) + \Lambda(\x)$.
     \item[(iii)]  Design the  OCP \eqref{eq:OCP} such that $V^\diamond_N$ approximates $V^\diamond_\infty$ sufficiently well. 
\end{itemize}
\end{pthm}
Observe that in the dissipativity-based setting, item (i) states that infinite-horizon optimal trajectories converge to the turnpike steady state $x^\diamond_s$. Formal results are \cite[Thm. 3.1]{grunekellett2017} and~\cite[Thm. 2]{tudo:faulwasser21a}. However, the corresponding item (i) of Insight~\ref{pthm:stabMPC} is phrased as a design guideline. This is due to the close relation between the turnpike property and strict dissipativity. Item (ii) stems from the crucial observation that the gradients of value and storage functions are both closely related to the Lagrange multipliers of the dynamics in \eqref{eq:SOP}, see, e.g.,  \cite{kit:faulwasser18e_2,kit:zanon18a}. Indeed, one can show that, for sufficiently large $N$, $\nabla V^\diamond_N(x_s^\diamond) \approx-\nabla \Lambda(x_s^\diamond)$.  This is also related to the concept of rotated stage costs \cite{Diehl11a,Angeli12a}, \cite[Chap. 8]{Gruene17a}. In particular,  for OCPs with suitable terminal constraints, \cite[Thm. 1]{Diehl11a} appears to be the first construction of a Lyapunov function of the form $V^\diamond_N(\x) + \Lambda(\x) + c$ with linear storage $\Lambda$ and a constant offset $c$ ensuring $W(x^\diamond_s) = 0$. Conditions establishing linear storage functions for specific convex OCPs can be found in \cite{Stieler14a}. Finally, observe that in phrasing the last insight, we did not mention asymptotic stability. The reason is that whenever the open-loop optimal solutions show a turnpike leaving arc, then one usually has that $\mu^\diamond_N(\xs^\diamond) \neq \us^\diamond$ and hence one will not obtain asymptotic stability of the closed loop at $\xs^\diamond$ but rather stability of a neighborhood whose size depends on the considered horizon length. We note that a number of works have analyzed and tackled this problem, e.g., \cite{Gruene17a,kit:faulwasser18e_2}. The mechanisms generating turnpike leaving arcs are discussed in \cite{tudo:faulwasser22a}.

\subsection{Receding Horizon Games}

The closed-loop behavior of receding-horizon controllers is counter-intuitive even in the optimization-based case. Thus, clearly when adding competition between agents to the mix, having theoretical guarantees is essential to guarantee safe deployment of RHG. However, while closed-loop properties of optimization-based MPC have been studied for 30+ years across linear, nonlinear, economic, adaptive, stochastic, and robust settings, the game-theoretic case is much less developed:

First results were derived under a potential game assumption in~\cite{hall2022receding}. In potential games, agents simultaneously and unknowingly minimize a higher-level objective: the \textit{potential function}. This structure allows to reformulate~\eqref{eq:GNEP} as an equivalent centralized optimal control problem, connecting it directly to economic MPC theory and its stability results.

The first closed-loop stability results for non-potential RHG have been presented in~\cite{hall2026stability}\footnote{There exists an extensive line of research analyzing the stability of open-loop dynamic Nash equilibria, however, here we focus specifically on the receding-horizon implementation of GNEPs.}. They rely on dissipativity techniques but define it for the entire pseudogradient operator which is much more restrictive than requiring dissipativity of the system with respect to the stage cost. Following this, \cite{benenati2026linear} proved closed-loop stability for LQ games for both open-loop and feedback Nash equilibria.

Dissipativity and turnpike theory allowed to establish the first recursive feasibility and closed-loop stability results for nonlinear GNEPs without relying on terminal constraints~\cite{hall2026towards} which we summarize in the following insights.

\begin{pthm}[Recursive feasibility of GNEPs \edit{{\cite[Lem. 2]{hall2026towards}}}]\label{pthm:recursiveFeasibility}
Suppose the following holds
\begin{itemize}
        \item the functions defining~\eqref{eq:GNEP} are continuous and $\mathcal{Z}_N(\x)$ is compact,
        \item suitable controllability assumptions hold,
        \item the price of anarchy is bounded,
        \item all $(x^*, u^*) \in \mathcal{S}^{\text{\tiny GNE}}_N(\x)$ satisfy  \eqref{eq:sDI} with respect to  $(x_s,u_s)$, 
        \item the storage $\Lambda(\x)$ is bounded along closed-loop RHG trajectories.
\end{itemize}
      Then, there exists a finite horizon $\hat N$, such that for all $N\geq \hat N$, the GNEP is recursively feasible, i.e., if \eqref{eq:GNEP} is feasible for $\x_t$ then it is also feasible for $\x_{t+1}$.
\end{pthm}

We introduce the following candidate Lyapunov function to state the convergence results
\begin{equation*}\label{eq:LyapCand}
  W(\x) := \sum_{k=0}^{N-1}  \sum_{v\in \mc V}\ell^v(x^*_k, u_k^{v*}, u_k^{-v*}) + \Lambda(\x).
\end{equation*}

\begin{pthm}[Practical asymptotic convergence \edit{{\cite[Thm. 2]{hall2026towards}}}]\label{pthm:practicalConv}
Suppose the conditions of Insight~\ref{pthm:recursiveFeasibility} hold and $W(\x)$ is bounded \edit{on a neighborhood $\mathcal B_{\tilde \rho}(x_s)$}.

Then, there exists $\bar{N}\in \bb{N}$ such that for all $N\geq \bar N$ the closed RHG loop satisfies
\[
\lim_{t\to\infty} \dist(\x_t, \mathcal{B}_{\hat\rho}(x_s)) = 0
\]
for some $\hat\rho \geq \tilde \rho$.
\end{pthm}

Going from the convergence results to practical asymptotic stability follows directly if \edit{one} assumes the existence of an upper bound on $W$ in the RHG setting, as done in~\cite[Prop. 1]{hall2026towards}. Yet, verifying \edit{its} existence is still an open research topic.

\section{Uncertainty in OCPs} \label{sec:uncertainty}
In many real-world applications, model predictive control must operate in the presence of uncertainty, for instance due to process noise, external disturbances, or modeling errors. 
In such situations, a deterministic prediction model is no longer sufficient, since future system behavior cannot be described by a single trajectory. 

Hence, stochastic MPC has received considerable attention in recent years, cf.~\cite{mesbah2016stochastic} and the references therein, where the deterministic open-loop optimal control problem is replaced by a stochastic optimal control problem (SOCP). 
In this setting, predictions and control actions must account for random perturbations of the system evolution, which leads to dynamics of the form
\begin{equation} \label{eq:stochSysSample}
    x_{k+1} = f(x_k, u_k, w_k),
\end{equation}
where $w_k \in \mathbb{R}^{n_w}$ denotes a realization of a stochastic disturbance. The pair $(x_k, u_k) \in \mathbb{R}^{n_x} \times \mathbb{R}^{n_u}$ represents the corresponding state-control realization at time $k$.

When switching to the stochastic setting, several nontrivial issues arise. First, the underlying optimal control problem must be reformulated appropriately, since one is no longer minimizing a single trajectory cost but rather a functional defined over random outcomes. This typically requires the introduction of operators such as expectations (or more general risk measures) in order to aggregate costs across different realizations of the disturbance.

Second, the treatment of constraints becomes more delicate. One may either enforce constraints almost surely, which leads to robust formulations, or relax them for instance in a probabilistic sense via chance constraints. \edit{Almost-sure formulations, though often more tractable, become overly conservative or even infeasible under unbounded disturbances.} 

In order to derive a formulation of the SOCP that is suitable for turnpike and dissipativity analysis in the stochastic setting, we lift the dynamics to the space of random variables, leading to the stochastic system
\begin{equation} \label{eq:stochSys}
    X_{k+1} = f(X_k, U_k, W_k), \quad X_0 = \mathbf{X}.
\end{equation}
Here, the states $X_k$ are random variables taking values in $\mathbb{R}^{n_x}$, and we denote by $\mathcal{R}(\mathbb{R}^{n_x})$ the corresponding space of such random variables. Similarly, the controls $U_k$ are considered as elements of $\mathcal{R}(\mathbb{R}^{n_u})$. The sequence $(W_k)_{k \ge 0}$ is assumed to be i.i.d.\ random variables with known distribution, modeling the stochastic disturbances. Moreover, $W_k$ is assumed to be independent of $(X_k, U_k)$ for all $k \in \mathbb{N}_0$.
\edit{
Furthermore, in \eqref{eq:stochSys}, the initial state $\mathbf{X} \in \mathcal{R}(\mathbb{R}^{n_x})$ can in general be a random variable, since the turnpike and dissipativity analysis below is carried out on the layer of random variables. Hence, we also allow for random initial conditions. At the same time, deterministic initial conditions at time $k=0$ are included by setting $\mathbf{X} = \mathbf{x} \in \mathbb{R}^{n_x}$ almost surely.
}

Although working with \eqref{eq:stochSys} instead of \eqref{eq:stochSysSample} requires the analysis of stochastic systems on the infinite-dimensional space of random variables, it offers a significant advantage. 
Indeed, for a fixed i.i.d. disturbance sequence $(W_k)_{k\ge 0}$, the dynamics \eqref{eq:stochSys} can be interpreted as a time-varying system evolving on the space of random variables. 
The resulting time dependence is induced by the disturbance process, since the system map at time $k$ is given by $(X,U)\mapsto f(X,U,W_k)$.

To emphasize this time dependence, we adopt the notation $X_{k|t}$ and $U_{k|t}$ introduced in Section~\ref{sec:recedinghorizon}, where $X_{k|t}$ denotes the state at prediction step $k$ computed at initial time $t$, starting from the initial condition $\mathbf{X}_t$ and driven by the disturbance realization $(W_{t}, W_{t+1}, \dots)$. 
This perspective enables the transfer of concepts and proof techniques from deterministic time-varying optimal and predictive control to the stochastic setting.

The stochastic counterpart of the deterministic OCP~\eqref{eq:OCP}, posed on the space of random variables and initialized at $(t,\mathbf{X}_t) \in \N_0 \times \mc{R}(\R^{n_x})$, can be formulated as
\begin{subequations}
\label{eq:sOCP}
\begin{align}
\displaystyle \min_{X,\, U}  &\; J_N(X,U) := \sum_{k=0}^{N-1} \bb{L}[\ell(X_{k|t}, U_{k|t})] \label{eq:sOCPRunningCost} \\
\textrm{s.t.} \quad &  X_{k+1|t} =  f(X_{k|t},U_{k|t},W_{k+t})  \hspace{1.25em}  k \in \bb{Z}_{N} \label{eq:sOCPConstr1}\\
&\bb{T}[g(X_{k|t},U_{k|t})] \leq 0 \hspace{5.25em} k \in \bb{Z}_{N} \label{eq:sOCPConstr2} \\
&\bb{T}[h(U_{k|t})] \leq 0 \hspace{7.5em} k \in \bb{Z}_{N} \label{eq:sOCPConstr3}\\
&\; X_{0|t} = \mathbf{X}_t \label{eq:sOCPConstr4} \\
&\sigma(U_{k|t}) \subseteq \sigma(X_{k|t}), \hspace{5.5em} k \in \bb{Z}_{N}. \label{eq:sOCPCFiltration}
\end{align}
\end{subequations}
Equation~\eqref{eq:sOCPRunningCost} is the stochastic version of the stage costs~\eqref{eq:OCPRunningCost}. 
Since $X_{k|t}$ and $U_{k|t}$ are random variables, the quantity $\ell(X_k,U_k)$ is itself a random variable and can therefore not be minimized directly. 
Instead, a law-invariant mapping $\mathbb{L}$, i.e., a mapping that depends only on the joint distribution of $(X_{k|t},U_{k|t})$, is applied in order to obtain a real-valued objective.
The most common choice is the expectation operator, leading to a risk-neutral formulation. 
However, more general risk-averse formulations can also be accommodated, for instance by choosing $\mathbb{L}$ as a conditional value-at-risk (CVaR) or another coherent risk measure.

In contrast to~\eqref{eq:OCPConstr1}, the state is propagated according to the stochastic dynamics~\eqref{eq:sOCPConstr1} on the space of random variables. 
As discussed above, this introduces an explicit time dependence into the optimal control problem. 
Indeed, the system dynamics at prediction step $k$ depend on the particular disturbance random variable $W_{k+t}$ entering the dynamics.
Consequently, even though the disturbance sequence is i.i.d., the lifted dynamics are represented by a time-varying system on the space of random variables. As a result, the predicted state trajectory depends not only on the initial condition $\mathbf{X}_t$ but also explicitly on the initial time $t$.
\edit{
Additionally, it should also be noted that, in a closed-loop simulation on the layer of random variables, even a deterministic initial state at time $t=0$ generally leads to a random state $\mathbf{X}_t$ for $t>0$ due to the stochasticity of the system.
}

The constraints \eqref{eq:sOCPConstr2} and \eqref{eq:sOCPConstr3} are the stochastic counterparts of the deterministic constraints \eqref{eq:OCPConstr2} and \eqref{eq:OCPConstr3}. 
Similar to the cost functional, we introduce a law-invariant mapping $\mathbb{T}$ to define the inequality constraints. 
This mapping can be interpreted as a design choice that determines whether constraints are enforced almost surely or in a probabilistic manner, for instance via chance constraints, or more generally through other risk-averse formulations.

Equation~\eqref{eq:sOCPConstr4} specifies the initial condition. In contrast to~\eqref{eq:OCPConstr4}, this initial condition is a random variable and hence need not be deterministic.

The additional condition in \eqref{eq:sOCPCFiltration} is a so-called non-anticipativity constraint. Mathematically, it requires that the control filtration $\sigma(U_{k|t})$ is contained in the state filtration $\sigma(X_{k|t})$, which can be interpreted as a causality requirement. 
It enforces that the current control action may only depend on information that is observable through the current state and must not depend on future disturbances.
This condition is commonly referred to as adaptedness of the control process to the state filtration. 
Furthermore, this is equivalent to requiring that $U_{k|t}$ can be represented by a deterministic feedback law (or policy), i.e., $U_{k|t} = \pi(X_{k|t})$ for some measurable mapping $\pi:\mathbb{R}^{n_x}\to\mathbb{R}^{n_u}$.
For more details on stochastic filtrations we refer to \cite{fristedt1997modern,Protter2005stochastic}.

Analogously to the deterministic case, we denote the set of optimal solutions of \eqref{eq:sOCP} with initial condition $(t,\mathbf{X}_t)$ by $\mathcal{S}^{\text{\tiny SOCP}}_N(t,\mathbf{X}_t)$ and the set of feasible state-control trajectories is defined as
\begin{equation*} 
\begin{split} 
\mc{Z}^{\text{\tiny SOCP}}_N(t,\mathbf{X}_t)  &= \{(X, U) \in \mc{R}(\R^{n_x})^{N+1} \\&\qquad \times \mc{R}(\R^{n_u})^N~|~ \eqref{eq:sOCPConstr1} - \eqref{eq:sOCPCFiltration}\}. 
\end{split} 
\end{equation*}

\subsection{The Stochastic Turnpike Property}

While turnpike properties are by now well understood for deterministic optimal control problems, considerably fewer results are available in the stochastic setting.

Most existing works analyze stochastic turnpike phenomena on the level of probability distributions or moments, cf.\ \cite{Kolokoltsov2012, Marimon1989, Sun2022}. This viewpoint is particularly appealing since the SOCP can then be reformulated as a time-invariant Markov decision process (MDP). In this setting, invariant distributions naturally take the role of the deterministic equilibrium, allowing turnpike properties to be formulated in terms of distances between probability laws or their moments.

More recently, however, it has been observed that SOCPs may also exhibit turnpike phenomena on the level of sample paths \cite{schiessl2023pathwise,ou2021simulation,sun2023stochturnpikepaths}. 
These findings suggest that turnpike behavior is not solely a distributional phenomenon and motivate a closer investigation of SOCPs on the space of random variables.
To illustrate the different layers of stochastic turnpike properties we use the following example which is a stochastic version of the OCP~\eqref{eq:OCP_example} from Example~\ref{ex:diff_OCPGNEP}.
\edit{Note that the disturbance in this example follows a uniform distribution, highlighting that the dissipativity approach is not restricted to Gaussian settings.}

\begin{ex}[Turnpike properties in stochastic OCPs] 
\label{ex:SOCP_turnpike}~\\
We consider the linear-quadratic SOCP 
 \begin{equation*}\label{eq:SOCP_example} 
\begin{array}{r l}
\displaystyle \min_{X,U}  &\displaystyle \sum_{k= 0}^{N-1}  \displaystyle \bb{E}[(U_{k|t})^2] 
\\[1em]
\textrm{s.t.} &  X_{k+1|t} = 2X_{k|t} +  0.25U_{k|t} + W_{k+t},  \hspace{.5em}  k \in \bb{Z}_{N}\\
&  \bb{P}(- 2 \leq X_{k|t} \leq 2) \geq 0.8, \hspace{5.em} k \in \bb{Z}_{N+1} \\
& X_{0|t} = \mathbf{X}_t, \\
& \sigma(U_{k|t}) \subseteq \sigma(X_{k|t}), ~  \hspace{7.4em} k \in \bb{Z}_{N}.
\end{array}
\end{equation*}
with Gaussian-distributed initial state $\mathbf{X}_t \sim \mc{N}(1.9,0.1)$ and uniformly distributed disturbance $W_k \sim \mc{U}(-0.5,0.5)$.
To solve this problem numerically, we use \emph{PolyOCP.jl} \cite{ou25polyocp} and \emph{PolyChaos.jl} \cite{muehlpfordt20polychaos}. 

The evolution of the probability density functions (PDFs) representing the distribution of the optimal state trajectory over a horizon of $N=50$ is shown in Figure~\ref{fig:SLQP_PDFs}. 
We observe that, after an initial transient (entry arc), the PDFs remain approximately constant over the middle of the horizon, followed by a terminal transient (leaving arc). 
This behavior is consistent with the schematic structure in Figure~\ref{fig:Turnpike_schematic} and thus indicates a turnpike property on the level of probability distributions.

In addition, Figure~\ref{fig:SLQP_paths} shows one specific realization of the optimal state trajectory for different prediction horizons $N$, computed under the same disturbance realization. 
As can be seen, the resulting realizations remain close to each other for most of the time horizon, indicating a pathwise turnpike property.

\begin{figure}[t]
    \centering
    \includegraphics[width=0.85\linewidth,trim={15.5cm 6.5cm 15.5cm 10.5cm},clip]{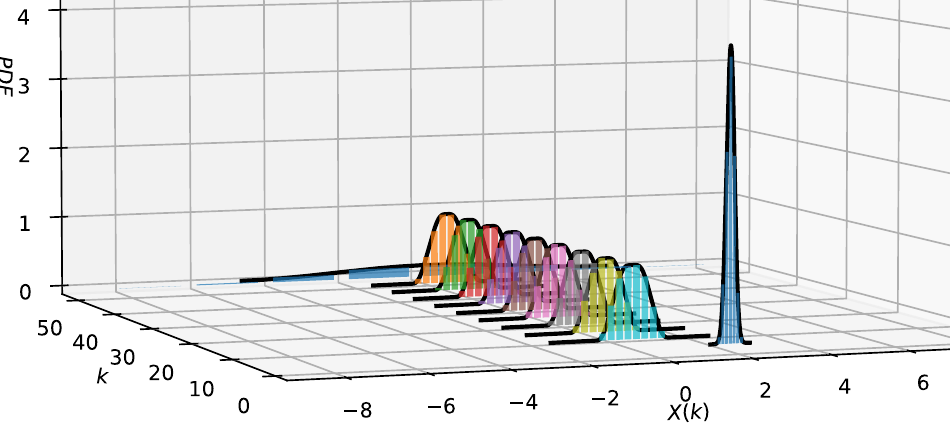}
    \caption{Evolution of the PDF of the optimal state on horizon $N=50$.}
    \label{fig:SLQP_PDFs}
\end{figure}

\begin{figure}[t]
    \centering
    \includegraphics[width=0.85\linewidth]{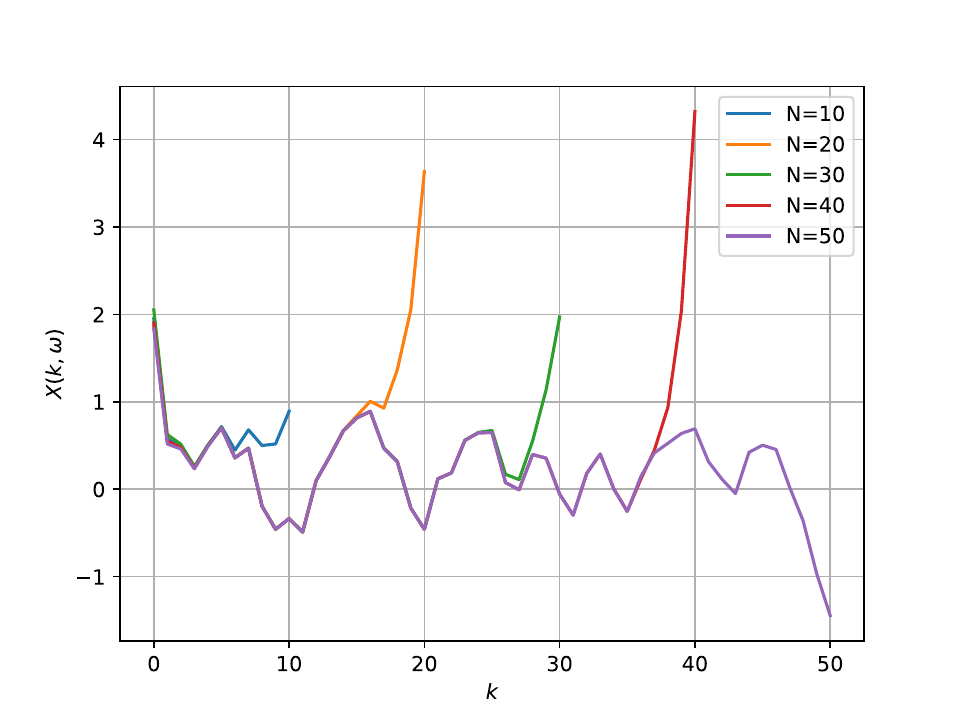}
    \caption{Evolution of one specific realization of the optimal state on different horizons $N$.}
    \label{fig:SLQP_paths}
\end{figure}

\end{ex}

While on the layer of distributions the turnpike is characterized by a stationary distribution, the pathwise observations from Example~\ref{ex:SOCP_turnpike} indicate that on the layer of random variables the turnpike has to be time-varying.

Hence, to derive a unified framework for the different turnpike properties illustrated in Example~\ref{ex:SOCP_turnpike}, the turnpike object should be modeled as a time-varying stochastic process with a time-invariant distribution. 
This motivates the following definition of a stationary process \cite{Doob1953,schiessl2024relationship}.

\begin{dfn}[Stationary stochastic processes] \label{dfn:stationaryProcess}
    A pair of stochastic processes $(X_s,U_s)$ with
    \begin{align*} 
        X_{s,k+1} &= f(X_{s,k}, U_{s,k}, W_k), \quad (X_s,U_s) \in \mc{Z}^{\text{\tiny SOCP}}_{\infty}(0,X^s_0)
    \end{align*}
    is called stationary for system \eqref{eq:stochSys} if there exist probability distributions $P_{X_s}$, $P_{U_s}$, and $P_{X_s,U_s}$ with
    \begin{equation*}
    \begin{split}
        X_{s,k} \sim P_{X_s}, \quad U_{s,k} \sim P_{U_s}, \quad (X_{s,k},U_{s,k}) \sim P_{X_s,U_s}, 
    \end{split}
    \end{equation*}
    for all $k \in \N_0$. 
\end{dfn}

The main difference to the stationary distributions commonly considered in the MDP literature is that Definition~\ref{dfn:stationaryProcess} does not merely characterize an invariant distribution. 
Instead, it requires the existence of a stochastic process realizing this law while simultaneously satisfying the dynamics~\eqref{eq:stochSys}. 
This additional structure enables the comparison of individual realization paths driven by the same disturbance sequence and is therefore crucial for the formulation of pathwise turnpike properties.

Turnpike properties on the different layers can then be characterized by the notion of distance to this stationary process. Depending on whether this distance is measured in a distributional or in a pathwise sense, different notions of stochastic turnpike arise. This is formalized in the following definition using the concept of a pseudometric on the space of random variables.

\begin{dfn}[Turnpike in SOCPs] \label{defn:SOCP_Turnpike}
    The SOCP~\eqref{eq:sOCP} exhibits the stochastic turnpike property at $(X_s,U_s)$ with respect to the pseudometric $d$ if for all $r> 0$ there exist $C > 0$ and $\alpha \in \mc{K}_{\infty}$ such that for each initial condition $(t,\mathbf{X}_t) \in \N_0 \times \mc{R}(\R^n_x)$ with $d(\mathbf{X}_t,X_{s,t}) \leq r$, each $\varepsilon > 0$, $N \in \N$, and any corresponding optimal solution $(X^{\diamond},U^{\diamond}) \in \mathcal{S}^{\text{\tiny SOCP}}_N(t,\mathbf{X}_t)$ the value
    \begin{equation*}
        Q_{\varepsilon} := \#\left\{k \in \bb{Z}_N |\, d \left( X^{\diamond}_k - X_{s,k+t} \right) \leq \varepsilon \right\} 
    \end{equation*}
    satisfies the inequality 
    $Q_{\varepsilon} \geq N - \frac{C}{\alpha(\varepsilon)}$.
\end{dfn}

As already indicated, different choices of the pseudometric in Definition~\ref{defn:SOCP_Turnpike} lead to different notions of turnpike properties with varying strength and interpretation.

For instance, choosing
\[
d(X,Y) = \mathbb{E}\big[\|X-Y\|^2\big]^{1/2}
\]
yields a mean-square turnpike property. 
With
\[
d(X,Y) = \inf \left\{ \mathbb{E}\big[\|\bar X - \bar Y\|^2\big]^{1/2} \;\middle|\; X \sim \bar X,\; Y \sim \bar Y \right\},
\]
one obtains a Wasserstein turnpike property on the level of distributions.
Finally, considering
\[
d(X,Y) = \left|\mathbb{E}[\|X\|] - \mathbb{E}[\|Y\|]\right|
\]
leads to a turnpike property formulated in terms of first-order moments.
For further details on this and on the hierarchy induced by different pseudometrics, we refer to \cite{schiessl2024relationship}.

\subsection{Dissipativity for Stochastic OCPs}

Although the deep connection between turnpike properties and dissipativity is well established in the deterministic case, stochastic dissipativity concepts tailored to optimal control problems remain comparatively underdeveloped.

As in the case of stochastic turnpike properties, initial approaches to stochastic dissipativity were formulated on the level of probability distributions and measures, see \cite{gros2022stochDissi}. 
More recently, however, dissipativity concepts have been extended to the space of random variables, in line with the pathwise perspective on stochastic turnpike phenomena, see \cite{schiessl2024relationship,schiessl2025turnpikeLQP}.

Based on this viewpoint, dissipativity on the space of random variables is defined as follows, thereby providing a unified framework consistent with the pathwise formulation of stochastic turnpike properties.

\begin{dfn}[Stochastic dissipativity] \label{defn:stochDissi}
    The SOCP \eqref{eq:sOCP} is called strictly stochastically dissipative at $(X_s,U_s)$ with respect to the pseudometric $d$, if there exists a storage function $\Lambda: \N_0 \times \mc{R}(\R^{n_x}) \rightarrow \R$ uniformly bounded from below and a function $\alpha \in \mc{K}_{\infty}$ 
    such that 
    \begin{multline} \label{eq:sSDI}
         \Lambda(k+1,f(X,U,W_k)) - \Lambda(k,X) \\\leq  - \alpha( d(X,X_{s,k}) )+ \bb{L}[\ell(X,U)] - \bb{L}[\ell(X_{s,k},U_{s,k})] 
    \end{multline}
    holds for all $k \in \N_0$ and all $(X,U) \in \mc{Z}^{\text{\tiny SOCP}}_{0}(k,X)$.
    
    The system is called dissipative if inequality \eqref{eq:sSDI} holds with $\alpha \equiv 0$.
\end{dfn}

Similar to stochastic turnpike properties, both the dissipation inequality and the storage function become time-dependent due to the time-varying nature of the SOCP~\eqref{eq:sOCP} on the space of random variables. The choice of pseudometric $d$ determines which aspects of the deviation from the stationary process are captured.

The following result, which is the stochastic counterpart of Insight~\ref{pthm:DissiTurnpikeDeterministic}, shows that strict dissipativity implies the turnpike property, as in the deterministic setting \cite{schiessl2024relationship}.
A key observation is that, in the stochastic setting, dissipativity and turnpike properties can typically only be related when formulated in a compatible way, i.e., with respect to the same pseudometric.

\begin{pthm}[Strict dissipativity $\Rightarrow$ turnpike for SOCPs]
    Suppose that 
    \begin{itemize}
        \item for all $k \in \N_0$ and $(X,U) \in \mc Z_0^{\text{\tiny SOCP}}(k,X)$ the strict dissipation inequality \eqref{eq:sSDI} is satisfied with respect to some pseudometric $d$ and 
        \item a suitable stochastic reachability condition holds. 
    \end{itemize}
        Then optimal solutions $(X^\diamond, U^\diamond) \in \mc{S}^{\text{\tiny SOCP}}_N(t,\mathbf{X}_t)$ have the stochastic turnpike property with respect to the same pseudometric $d$.
\end{pthm}

\edit{Detailed proofs can be found in \cite[Thm.~4.9]{schiessl26stability} under a cheap reachability assumption in a general setting and in \cite[Section III]{schiessl2024relationship} for specific pseudometrics.}

\subsection{Closed-Loop Results}

Having introduced stochastic turnpike and dissipativity concepts, a natural question is whether these properties can be exploited to derive closed-loop guarantees for stochastic MPC schemes. 
While the deterministic theory provides a rich collection of results linking dissipativity and turnpike properties to stability and performance guarantees, the corresponding stochastic theory is significantly less developed.

Existing approaches to stochastic MPC stability analysis often rely on concepts originating from robust control or require suitable terminal ingredients. Representative examples include practical stability results for multistage MPC schemes \cite{lucia2020stability}, expectation-based stability estimates \cite{mcallister2022stochMPC}, stability guarantees for linear systems with terminal constraints \cite{kouvaritakis2016mpc}, and drift-based approaches \cite{chatterjee2014stability}. 
More recently, probabilistic reachable sets and constraint-tightening techniques have enabled the treatment of chance-constrained stochastic MPC formulations \cite{hewing2018stochMPC,hewing2020stochMPC,koehler2025stochMPC}.

In contrast, dissipativity- and turnpike-based approaches seek to derive closed-loop properties directly from structural properties of the underlying SOCP. 
Although only a limited number of results are currently available, recent developments indicate that the deterministic interplay between dissipativity, turnpike properties, and MPC performance can be partially transferred to the stochastic setting \cite{kordabad2022MDP,schiessl26stability}. 
The following insight provides an overview of some dissipativity and turnpike based closed-loop guarantees for stochastic MPC.

\begin{pthm}[Stability and performance of stochastic MPC]%
\label{pthm:StabilityPerformance_SMPC}
    Suppose that
    \begin{itemize}
        \item the SOCP~\eqref{eq:sOCP} is strictly stochastically dissipative and has the stochastic turnpike property with respect to a pseudometric $d$ and
        \item the optimal value functions for certain auxiliary problems and the storage function $\Lambda$ satisfy suitable continuity and regularity assumptions.
    \end{itemize}
    Then the following properties hold:
    \begin{enumerate}[(i)]
        \item The stationary process $X_s$ is semi-globally practically asymptotically stable with respect to the optimization horizon $N$ and the pseudometric $d$ for the closed-loop system.
        \item The stochastic MPC closed-loop is approximately overtaking optimal and averaged performance near-optimal.
    \end{enumerate}
\end{pthm}

The stability property (i) in Insight~\ref{pthm:StabilityPerformance_SMPC} \edit{is proved in \cite[Cor. 5.6]{schiessl26stability}.
It} guarantees that, for every feasible initial condition, there exists a sufficiently large prediction horizon $N$ such that the resulting stochastic MPC closed-loop trajectory converges to a neighborhood of the stationary process $X_s$ with respect to the pseudometric $d$. Furthermore, the size of this neighborhood decreases as the prediction horizon increases and vanishes in the limit as $N\to\infty$. Thus, the result recovers asymptotic convergence to the stationary process in the infinite-horizon limit. 

The performance guarantees in Insight~\ref{pthm:StabilityPerformance_SMPC}(ii) can be interpreted in a similar spirit. 
Overtaking optimality, introduced in \cite{gale1967optimal}, provides a notion of optimality for trajectories whose infinite-horizon costs may not be finite. 
Roughly speaking, a trajectory is overtaking optimal if, over sufficiently long horizons, its accumulated cost is eventually no larger than that of any competing feasible trajectory. 
Consequently, whenever an infinite-horizon optimal solution exists, it is overtaking optimal.
\edit{The detailed proof of this performance result can be found in \cite[Cor. 6.7]{schiessl26stability}}

Since MPC computes only a finite-horizon approximation of the underlying infinite-horizon problem, one cannot expect exact overtaking optimality in general. 
Instead, the stochastic MPC closed loop is approximately overtaking optimal, where the suboptimality is quantified by an error term depending on the prediction horizon $N$. 
This error decreases as $N$ increases and vanishes in the limit as $N\to\infty$.

A similar interpretation applies to the averaged performance estimate. 
The average closed-loop performance achieved by stochastic MPC differs from the optimal infinite-horizon performance only by a horizon-dependent error term that decreases monotonically with increasing prediction horizon length. 
Hence, both overtaking optimality and average performance optimality are recovered in the infinite-horizon limit.
For further technical details on these results, we refer to \cite{schiessl26stability,schiessl2024nearOptimal}.

It should be emphasized that the above results are directly valid only in an abstract stochastic MPC setting in which, at each iteration, the full knowledge of the random variable representing the closed-loop state (or at least its distribution) is available. 
This assumption is a direct consequence of the fact that the turnpike and dissipativity analysis is carried out on the level of random variables or probability distributions.

While such a setting can be motivated, for instance, in mean-field-type control problems, in most practical applications only a realization of the current state is observable, rather than its full distribution. 
This raises the question of implementability of the proposed framework.

For simplified stochastic MPC formulations, in which only almost-sure constraints are considered and the expectation is used as optimization criterion, a pathwise dynamic programming principle can be established. 
In this case, the optimal control for a random initial state can be obtained by minimizing separately for each realization of the initial condition, see \cite{bertsekas1996stochasticDPP}. 
This observation can further be used to show that, in this setting, the closed-loop behavior and performance of an MPC algorithm based on full random-variable information coincide almost surely with those obtained using only pathwise measurements, see \cite[Corollary~3.1]{schiessl26stability}. 
Hence, in such cases the proposed results are not restricted to an abstract formulation but are also valid for an implementable algorithm.

However, this equivalence breaks down in more general settings. 
In particular, when risk measures are used as optimization criteria, the pathwise dynamic programming principle no longer holds, and additional error terms arise when applying implementable MPC schemes, cf.~\cite{schiessl2025riskcost}. 
Similarly, in the presence of probabilistic constraints, full distributional information is typically required to verify constraint satisfaction. 
Although stochastic MPC schemes based on partial distributional information exist and can guarantee closed-loop constraint satisfaction \cite{hewing2018stochMPC,hewing2020stochMPC,koehler2025stochMPC,schluter2022stochastic,schluter2023stochastic}, they are often conservative.
While dissipativity-based arguments can still ensure near-optimal averaged performance for such algorithms, see \cite{schiessl2026riskconstraints}, the transient performance of implementable algorithms has to be suboptimal compared to the abstract formulation. 
This discrepancy between analytical tractability and closed-loop measurability remains an important open research direction.\vspace*{2mm}

The following example illustrates the closed-loop results presented in this section. 
\edit{We emphasize that the above analysis applies to general unbounded, non-Gaussian uncertainties. For the numerical example below, however, we consider a two-point distribution, which allows the resulting stochastic problems to be solved exactly up to numerical accuracy.}

\begin{ex}[Stochastic MPC]
We consider the nonlinear SOCP 
 \begin{equation*}\label{eq:SMPC_example} 
\begin{array}{r l}
\displaystyle \min_{X,U}  &\displaystyle \sum_{k= 0}^{N-1}  \displaystyle \bb{E}[(X_{k|t})^2+ 25(U_{k|t})^2] 
\\[1em]
\textrm{s.t.} &  X_{k+1|t} = (U_{k|t} - X_{k|t})^2 + W_{k+t},  \hspace{1.em}  k \in \bb{Z}_{N}\\
& X_{0|t} = \mathbf{X}_t , \\
& \sigma(U_{k|t}) \subseteq \sigma(X_{k|t}) \hspace{7.5em} k \in \bb{Z}_{N}.
\end{array}
\end{equation*}
with disturbance 
\begin{equation*}
    W_k = \begin{cases}
        1 & \text{with probability } 0.7 \\
        0.25 & \text{with probability } 0.3.
    \end{cases}
\end{equation*}
\begin{figure}[t]
    \centering
    \includegraphics[width=0.5\linewidth]{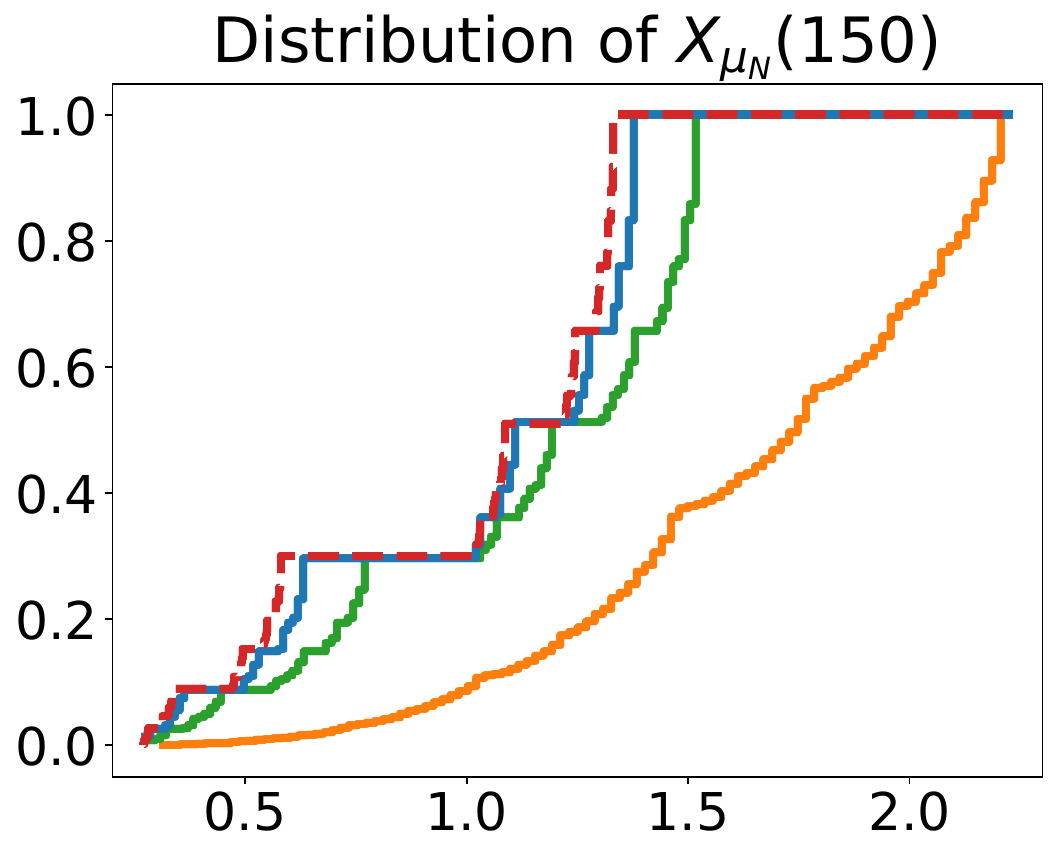}
    \caption{CDF of the closed-loop state at time $150$ for different horizons $N=3$ (orange), $N=4$ (green), and $N=5$ (blue) together with the stationary distribution (red).}
    \label{fig:cl_distribution}
\end{figure}
\begin{figure}[t]
\centering
   \begin{minipage}{0.49\columnwidth}
       \centering
       \includegraphics[width=0.95\textwidth]{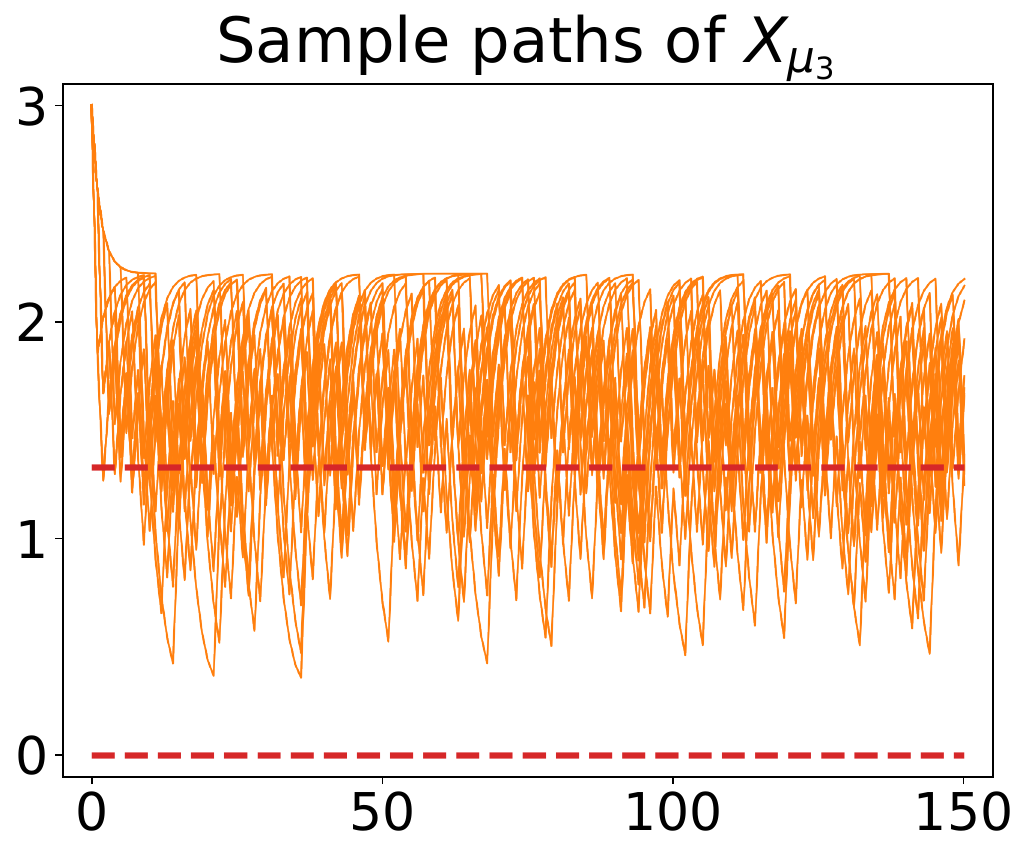}

    \end{minipage}
    \begin{minipage}{0.49\columnwidth}
       \centering
       \includegraphics[width=0.95\textwidth]{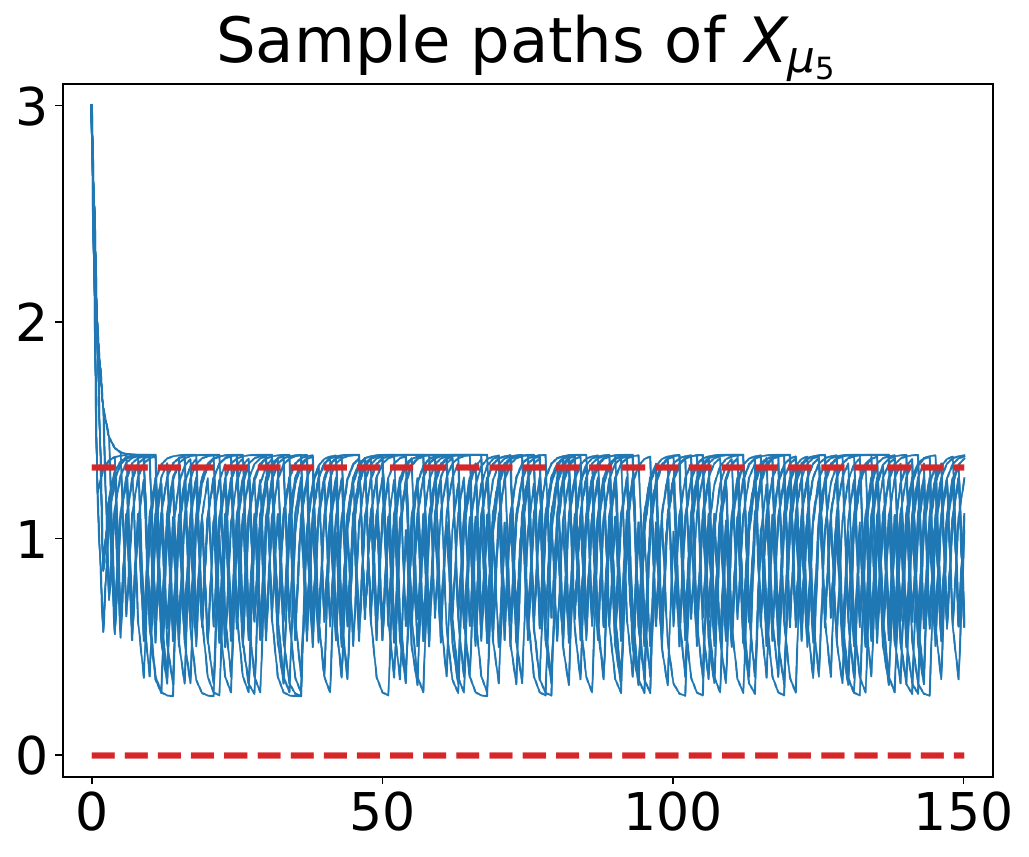}
   \end{minipage}
    \caption{Evolution of $15$ sample paths of the MPC closed loop for $N=3$ (orange) and $N=5$ (blue) together with the upper and lower bound of the support of the stationary process (red).}
    \label{fig:cl_paths}
\end{figure}
\begin{figure}[t]
\centering
   \begin{minipage}{0.49\columnwidth}
       \centering
       \includegraphics[width=0.95\textwidth]{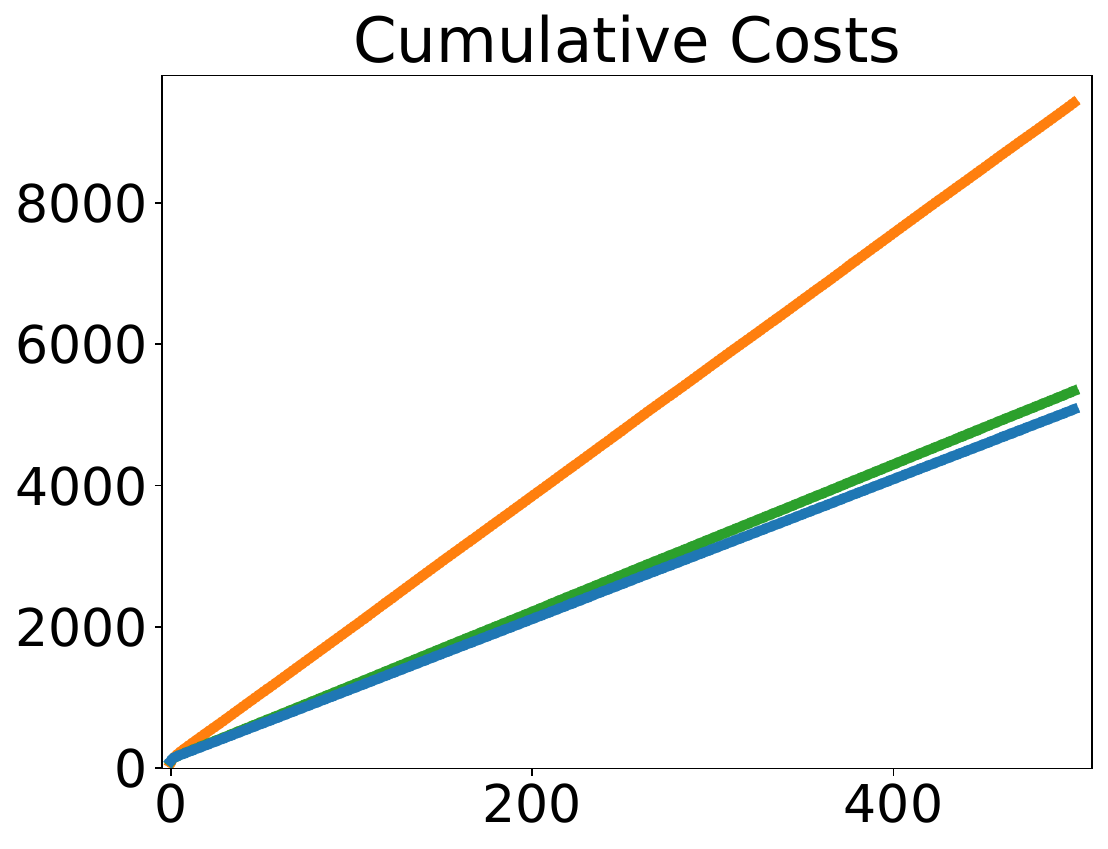}
   \end{minipage}
   \begin{minipage}{0.49\columnwidth}
       \centering
       \includegraphics[width=0.95\textwidth]{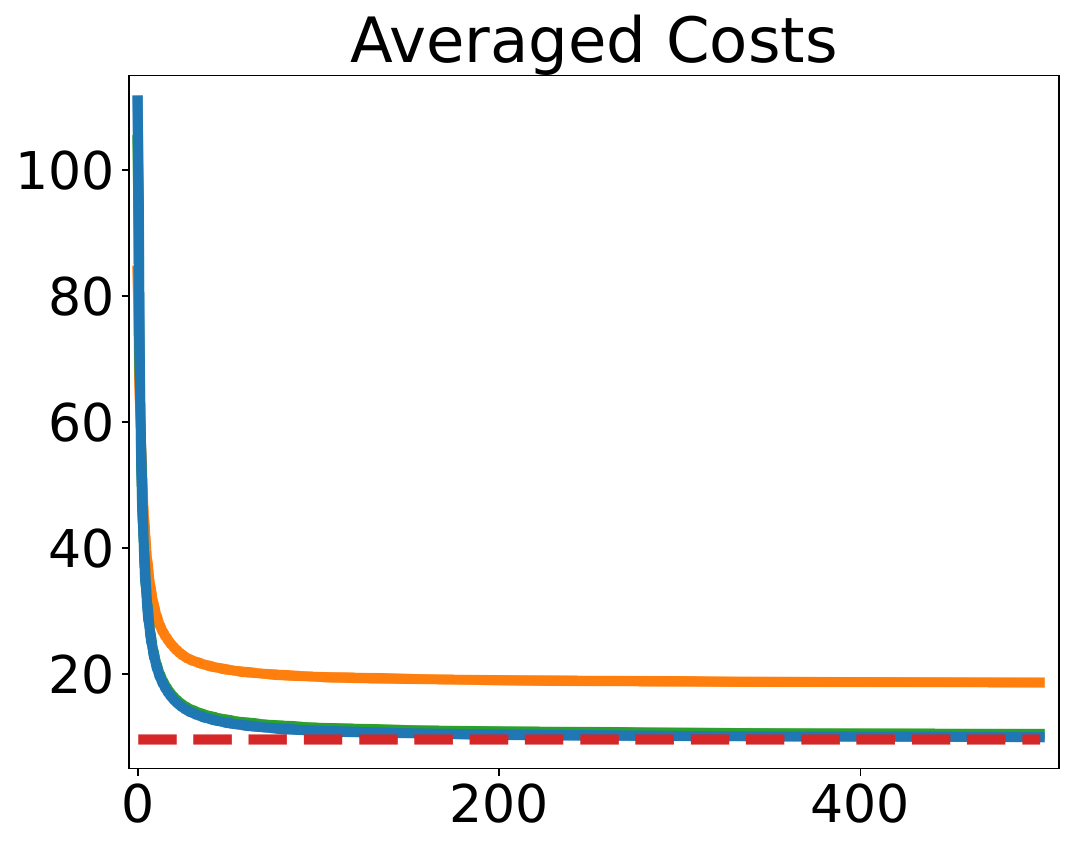}
   \end{minipage}
    \caption{Cumulative closed-loop costs (left) and averaged closed-loop costs (right) for different horizons $N=3$ (orange), $N=4$ (green), and $N=5$ (blue) together with the stationary cost (red).}
    \label{fig:cl_performance}
\end{figure}

To obtain the closed-loop quantities, we run the implementable version of the stochastic MPC algorithm $20{,}000$ times with the initial condition $\mathbf{X}_0 = 3$ up to time $150$.
The optimal control problems arising in each MPC iteration were hereby solved by direct minimization over all possible realization paths using \emph{IPOPT}~\cite{wachter2006ipopt}, which is feasible since the image of the noise process $W(k)$ is finite.

The stability of the MPC closed loop is illustrated in Figure~\ref{fig:cl_distribution} and Figure~\ref{fig:cl_paths}. Figure~\ref{fig:cl_distribution} shows the cumulative distribution function (CDF) of the closed-loop state at time $k=150$ for different prediction horizons $N=3,4,5$, together with the CDF of the stationary process. As can be seen, the CDFs approach the stationary distribution as $N$ increases, indicating stability in distribution of the MPC closed loop.

Similar observations can be made in Figure~\ref{fig:cl_paths}, which shows 15 sample paths of the MPC closed-loop state over time for $N=3$ (left) and $N=5$ (right). 
As can be seen, for increasing $N$ the sample paths increasingly approach the support of the stationary process, indicated by the red lines.

Furthermore, Figure~\ref{fig:cl_performance} shows the cumulative and averaged closed-loop costs for prediction horizons $N=3,4,5$.
In both cases, we observe that the performance improves with increasing prediction horizon $N$, illustrating the near-optimality of the closed loop as a function of the horizon length.
This is consistent with the performance guarantees mentioned in Insight~\ref{pthm:StabilityPerformance_SMPC}.

\end{ex}

\section{Numerical Methods for OCPs and GNEPs}\label{sec:rec_hor_clNE}
\subsection{Numerical Optimal Control}
Numerical optimal control is by now an extremely well-development domain and many powerful open-source tools exist, e.g., \cite{Andersson19a,Kapernick14a,Verschueren22a}. These tools also enable the real-time implementation of MPC for deterministic linear and nonlinear systems, except for extremely large-dimensional settings. In this context, the recent paper \cite{tuhh:stomberg25b} has demonstrated that a state-of-art algorithm for convex quadratic programs (OSQP~\cite{stellato2018osqp}) scales well to very large-dimensional QPs with more than $10^6$ decision variables. These and other results documented in the literature are our motivation to not dive deep into numerical methods for optimal control. Rather we give a snapshot of selected developments for games. 

\subsection{Numerical Methods for Generalized Nash Equilibria} 

In terms of numerical methods, there is a striking difference in difficulty between solving for a general GNEP and solving for the specific subclass of variational equilibria (v-GNE) which are a solution to the VI in~\eqref{eq:VI}. Computing solutions to~\eqref{eq:GNEP} in the general case is particularly difficult as they correspond to quasi-variational inequalities~\cite{facchinei2009generalized}. 
Only few solvers exist which, to the best of our knowledge, have limited or no convergence guarantees~\cite{spica2020real, pustilnik2025generalized, bemporad2025nashopt, dreves2011solution}. 
As a consequence, in the following we focus on solution methods for variational GNEs and present two specific v-GNE seeking algorithms. We consider a special case of GNEP~\eqref{eq:GNEP} with the following quadratic objective
\begin{equation*}
    \ell^v(x,u^v) = \frac{1}{2} \| x\|_{Q^v}^2 + \frac{1}{2} \|u^v\|_{R^v}^2 
\end{equation*}
 In the special case of linear-quadratic dynamic games, the finite-horizon GNEP can be recast as a monotone affine variational inequality (AVI) \cite[Prop. 2]{benenati2026linear}. By stacking the states and control inputs over the horizon, we have
\begin{equation}
   x = \begin{bmatrix} x_1 \\ \vdots \\ x_K \end{bmatrix}, \qquad
    u^v = \begin{bmatrix} u^v_0 \\ \vdots \\ u^v_{K-1} \end{bmatrix},
\end{equation}
and therefore $ x = \Theta \x + \sum_{v\in\mathcal V} \Gamma^v u^v$, where
\begin{equation*}
    \Theta = \begin{bmatrix} A \\ A^2 \\ \vdots \\ A^K \end{bmatrix}, \quad
    \Gamma^v =
    \begin{bmatrix}
    B^v & 0 & \cdots & 0 \\
    AB^v & B^v & \cdots & 0 \\
    \vdots & \vdots & \ddots & \vdots \\
    A^{K-1}B^v & A^{K-2}B^v & \cdots & B^v
    \end{bmatrix}.
\end{equation*}
Let $\bar R^v = I_K \otimes R^v$ and $\bar Q^v = \mathrm{blkdiag}(I_{K-1} \otimes Q^v, {P}^v)$, where $(P^v)_{v \in \mathcal{V}}$ should be the (possibly asymmetric) solutions of the infinite-horizon Riccati--Stein equations. Then, the partial gradient of the objective function of agent $v$ is
\begin{equation} \label{eq:nabla_Ji}
    \nabla_{u^v} J^v(u) = \bar R^v u^v + (\Gamma^v)^\top \bar Q^v \Bigl( \Theta \x + \sum_{v\in\mathcal V} \Gamma^v u^v \Bigr).
\end{equation}

It follows from \cite[Prop. 2]{benenati2026linear} that the open-loop GNEP with quadratic stage-cost is equivalent to an AVI with monotone (strongly monotone if $\lambda_{\min}(R^v)$ is large enough) affine operator $F$ \cite[Def. 2.3.1]{facchinei2003finite}:
\begin{subequations}
\begin{align}
    F(u) & := M u + q \label{eq:F} \\
    M &:= \text{blkdiag}(\bar{R}^v)_{v\in \mc V} \nonumber\\
    &\quad+ \text{blkmat}((\Gamma^v)^{\top}\bar{Q}^v \Gamma^v)_{(v,j)\in\mc V \times \mathcal{V}} \label{eq:F-M}\\ 
    q &:= \col{(\Gamma^v)^{\top}\bar{Q}^v \Theta \x}_{v\in\mc V}.
\end{align}
\end{subequations}

\begin{rmk}
Due to the asymmetry of the matrices $(P^v)_{v \in \mathcal{V}}$, $M$ in \eqref{eq:F-M} is asymmetric in general. Thus, there do not exist $M$ objective functions such that the operator $F$ in \eqref{eq:F} corresponds to their pseudogradient operator, namely, the stacked partial gradients. Consequently, there exists no finite-horizon game (in the original state space) whose OLNE corresponds to an infinite-horizon OLNE.
\end{rmk}
 
The formulation as an AVI is convenient because numerical solvers are available in the presence of convex constraints that arise from state-input constrained  dynamics. In the following, we assume that the affine operator $F(u)=Mu+q$ is strongly monotone and that the feasible set associated with stacked state-input constraints is nonempty and polyhedral, $\mathcal C:=\{v\mid Dv+d\le0\}$. Then$\mathrm{AVI}(\mathcal C,M,q)$ admits a unique solution~\cite[Thm.~2.3.3]{facchinei2003finite}.

\subsection{Douglas--Rachford Algorithms for AVIs}

Let us first present a Douglas--Rachford splitting method for solving the considered variational inequality $\mathrm{AVI}(\mathcal C,M,q)$, where the matrix $M$ is split as $M = (M_1-H) + (M_2-H)$, with $M_1 = M_1^{\top} \succcurlyeq 0, \, M_2 \succ 0$ and $H = H^{\top} \succ 0$.

Consider the following Douglas--Rachford splitting-like method \cite[Eqs. 30, 31]{eckstein1998operator} to solve $\mathrm{AVI}(\mathcal{C}, M, q), \, k \in \mathbb{N}$ where the superscript $l$ now denotes the iteration number:
\begin{subequations}\label{DR-affineVI}
\begin{align}
    y^{l} &= \mathrm{sol}(\mathcal{C}, H+M_1, q+(M_2 - H)u^l), \label{DR-affineVI-1}\\
    u^{l+1} &= (H+M_2)^{-1}\big(H(2\lambda_l y^{l} \nonumber\\&\quad + (1-2\lambda_l)u^l) + M_2u^l\big), \label{DR-affineVI-2}
\end{align}
\end{subequations}
where $\mathrm{sol}(\mathcal{C}, M, q)$ denotes the solution to $\mathrm{AVI}(\mathcal{C}, M, q)$. We note that, unlike $M$,  $H+M_1$ is a symmetric operator, hence the step in \eqref{DR-affineVI-1} is equivalent to solving an optimization problem \cite[\S 1.3.1]{facchinei2003finite}, specifically, a quadratic program (QP), since $\mc C$ is a polyhedron. Under the stated splitting assumptions, this iteration is globally convergent and, under strong monotonicity, converges linearly to the unique AVI solution \cite[Prop. 6, 8]{eckstein1998operator}.

In \cite{baghbadorani:benenati:2026}, the authors choose $H, M_1, M_2$ as follows:
\begin{subequations}\label{eq:example_matrix_splitting}
\begin{align}
H &= (1-\gamma)(M + M^\top) + \varepsilon I, \\
M_1 &= \gamma(M + M^\top), \label{eq:example_matrix_splitting:1} \\
M_2 &=  (M - M^\top) + (1-\gamma)(M + M^\top), \label{eq:example_matrix_splitting:2}
\end{align}
\end{subequations}
for some tuning parameter $\gamma\in(0,1)$. The design above has the following motivation: in the case of $M=M^\top$, with constant step $\lambda=\frac{1}{2}$, the iteration in \eqref{DR-affineVI} becomes
$u^{l+1} =  \mathrm{sol}(\mathcal{C}, \varepsilon I + M, q - \varepsilon u^k)$, which is equivalent to
$    u^{l+1} = \arg\min_{u\in\mc C} \frac{1}{2} \|u + (M+\varepsilon I)^{-1}(q - \varepsilon u^l) \|^2_{M+\varepsilon I}$.
The latter, for $\varepsilon \simeq 0$, resembles a full-step towards the unconstrained solution, $-M^{-1}q$, projected onto the constraint set $\mathcal{C}$. 

\subsection{Smoothing Newton Algorithms for AVIs}

Let us come back to the $\mathrm{AVI}(\mathcal{C},M,q)$ and to its Karush--Kuhn--Tucker (KKT) conditions \cite[Prop.~1.2.1]{facchinei2003finite}:
\begin{subequations}\label{eq:KKT-newton}
\begin{align}
Mu+q-D^\top\lambda &= 0 \label{eq:KKT-newton-a}\\
Du+d &\le 0 \label{eq:KKT-newton-b}\\
\lambda &\ge 0 \label{eq:KKT-newton-c}\\
\lambda^\top(Du+d) &= 0. \label{eq:KKT-newton-d}
\end{align}
\end{subequations}

The complementarity conditions  \eqref{eq:KKT-newton-b}--\eqref{eq:KKT-newton-d} are nonsmooth, so classical smooth Newton methods cannot be applied directly. We therefore use the smoothed Fischer--Burmeister function $\varphi_\mu$~\cite{baghbadorani2026fastnewtonmethodslinearquadratic}, which yields a continuously differentiable approximation of the KKT system.
Specifically, we use $\varphi_\mu(a,b) := \sqrt{a^2+b^2+\mu^2} - a - b$, for small $\mu>0$, which is continuously differentiable, and derive a smoothed KKT system:
\begin{align}\label{eq: NCP-function}
\Phi_\mu(u,\lambda) :=
\begin{bmatrix}
Mu+q-D^\top \lambda\\[1mm]
\varphi_\mu(Du+d,\lambda)
\end{bmatrix} = 0,
\end{align}
with Jacobian 
\begin{align}\label{eq: diff-NCP-function}
\nabla \Phi_\mu(u,\lambda) =
\begin{bmatrix}
M & -D^\top\\
G_\mu D & H_\mu
\end{bmatrix},
\end{align}
for properly derived partial derivatives $G_\mu$ and $H_\mu$ of $\varphi_\mu(Du+d,\lambda)$ \cite[Sec. III]{baghbadorani2026fastnewtonmethodslinearquadratic}. 

Given an iterate $(u^l,\lambda^l)$, the Newton descent direction
$d^l = (\Delta u^l, \Delta \lambda^l)$ is then computed by solving the linear system of equations
\begin{equation}\label{eq:newton-step}
\nabla \Phi_\mu(u^l,\lambda^l)
\begin{bmatrix}
\Delta u^l\\
\Delta \lambda^l
\end{bmatrix} 
= -\Phi_\mu(u^l,\lambda^l),
\end{equation}

where $\nabla\Phi_\mu$ is nonsingular due to
the smoothing of $G_\mu$ and $H_\mu$, and the strong monotonicity of \(M\).

 In this setting, local superlinear convergence to a zero of $\nabla \Phi_{\mu}$ holds for the following smoothing Newton algorithm ($k \in \mathbb{N}$) with step size sequence $(\alpha_k)_{k\in \mathbb{N}}$ \cite[Th. 1]{baghbadorani2026fastnewtonmethodslinearquadratic}:
\begin{subequations}\label{eq:Newton_iteration}
\begin{align}
    d^l &= - \Big(\nabla \Phi_\mu(u^l,\lambda^l)\Big)^{-1}\Phi_\mu(u^l,\lambda^l) \\
    \begin{bmatrix} u^{l+1}\\ \lambda^{l+1}\end{bmatrix} & = \begin{bmatrix} u^{l}\\ \lambda^{l}\end{bmatrix} + \alpha_l d^l.
\end{align}
\end{subequations}

Furthermore, \cite[Th. 2]{baghbadorani2026fastnewtonmethodslinearquadratic} shows a linesearch method to design the step sequence $(\alpha_l)_{l}$ in \eqref{eq:Newton_iteration}, with guaranteed global convergence. 
For receding-horizon implementation, a natural warm start is the equilibrium computed at the previous sampling instant. Finally, we refer to \cite[Sec. IV]{baghbadorani2026fastnewtonmethodslinearquadratic} for numerical experiments that illustrate fast convergence of the proposed smoothing Newton method in \eqref{eq:Newton_iteration} for linear models of longitudinal vehicle dynamics engaged in linear-quadratic dynamic games. These experiments show typical convergence and computational performance on two problem instances, and in turn showcase that our method outperforms forward-backward (FB) and Douglas--Rachford (DR) methods in terms of convergence rate and computational time. All these computational methods are centralized by construction; distributed implementations would require additional decompositions that exploit sparsity in the game structure.

\section{Open Problems and Conclusions} \label{sec:conclusion}

This paper has provided an overview on commonalities and crucial differences between dissipativity-based approaches to optimal control and dynamic games, \edit{as summarized in Table~\ref{tab:multi-agent_mpc_comparison}}. We have also touched upon the consideration of non-Gaussian uncertainties and the development of tailored numerical methods. However, there remain a number of crucial open problems which we summarize next.

\begin{table*}[tb]
\edit{
\centering
\caption{Comparison of receding-horizon control approaches and the role of dissipativity}
\label{tab:multi-agent_mpc_comparison}
\renewcommand{\arraystretch}{1.4}
\newcolumntype{Y}{>{\raggedright\arraybackslash\hsize=1.6\hsize}X}
\newcolumntype{Z}{>{\raggedright\arraybackslash\hsize=1.0\hsize}X}
\newcolumntype{L}{>{\raggedright\arraybackslash\hsize=0.4\hsize}X}
\begin{tabularx}{\textwidth}{@{}L ZY@{}}
\toprule
& OCPs \& model predictive control& GNEPs \& game-theoretic MPC / receding horizon games \\
\midrule
\textbf{Interaction} & cooperative & non-cooperative (self-interested) \\ \addlinespace[.25ex]
\textbf{Objective} & common / shared (separates if costs decoupled) & one per agent, conflicting \\ \addlinespace[.25ex]
\textbf{Solution concept} & social optimum (minimizer) & GT equilibrium (Nash equilibrium  or generalized Nash eq.) \\ \addlinespace[.25ex]
\textbf{Computation} & centralized or distributed & centralized or distributed \\\addlinespace[.25ex]
\textbf{Stability ingredients} &  \vspace{-2mm}\begin{itemize}
    \item terminal costs and constraints
    \item sufficiently long horizons
    \item strict dissipativity
\end{itemize}
or combinations thereof
&
\vspace{-2mm}\begin{itemize}
    \item potential game assumptions
    \item monotonicity of pseudo-gradient map%
    \item strict dissipativity along GNE solutions
\end{itemize}
\\\addlinespace[.25ex]
\textbf{Enabling role of strict dissipativity}&
OCPs:
\vspace{-.5mm}\begin{itemize}
    \item finite-horizon turnpike analysis and characterization of turnpike object
    \item asymptotics of infinite horizon OCPs
    \item deterministic and stochastic settings
\end{itemize}
\, MPC:
\vspace{-0.5mm}\begin{itemize}
    \item closed-loop analysis for generalized stage costs, with and without terminal ingredients
    \item deterministic and stochastic settings
\end{itemize}
&
GNEPs:
\vspace{-0.5mm}\begin{itemize}
    \item finite-horizon turnpike analysis and characterization of turnpike object
    \item infinite-horizon analysis is still open
    \item so far deterministic settings only
\end{itemize}
RHG:
\vspace{-0.5mm}\begin{itemize}
    \item closed-loop analysis with and without terminal ingredients
    \item so far deterministic settings only
\end{itemize}
\\
\bottomrule
\end{tabularx}
}
\end{table*}

\subsection{Open Problems}

\subsubsection*{Deterministic receding horizon games}
The closed-loop stability analysis of RHG only began around 2020-21 and is therefore, compared to its optimization-based MPC counterpart with 30+ years of history, still very much in its infancy. In particular, the fundamental tool of terminal ingredients (terminal constraint sets and penalties) has received little attention. Beyond this, deriving candidate Lyapunov  functions for various classes of nonlinear RHGs would be extremely valuable. Related is the question of how, in nonlinear RHGs where the set of equilibria is non-unique and possibly infinite, one can select and track a specific equilibrium so as to avoid discontinuous jumps in the RHG feedback law, which is essential for providing closed-loop guarantees. So far, all stability analyses of RHG have assumed perfect model knowledge and forecasts. Yet, it is well-known that MPC is not necessarily robust to plant-model mismatch and imperfect forecasts, especially in terms of closed-loop performance. Providing recursive feasibility and closed-loop stability guarantees under noise, plant-model mismatch, and disturbances is thus an essential next step toward real-world applications.

\subsubsection*{Stochastic receding horizon games}
When it comes to the consideration of uncertainty, one can go also beyond mere robustness against noise on the feedback channel. Indeed the integration of uncertainty into receding horizon dynamic games \edit{remains} largely unexplored. Stochastic linear-quadratic (for instance, LQG) dynamic games have been extensively studied, and both open-loop Nash equilibria (OLNE) and feedback Nash equilibria (FNE) can be characterized via stochastic dynamic programming and coupled Riccati--Stein equations, thereby providing a theoretical foundation for unconstrained stochastic dynamic games \cite{SUN2019381}. However, the constrained setting for open-loop and receding horizon games is still nascent. Also the consideration of non-Gaussian exogenous disturbances has not been touched upon. Extending the available deterministic formulations of receding horizon games to stochastic systems may additionally require the integration of risk-sensitive objective functions, or distributionally robust formulations, while simultaneously studying recursive feasibility, closed-loop stability, and performance guarantees. The development of such a stochastic frameworks represents a promising avenue for future research.

\subsubsection*{Stochastic model predictive control}
When it comes to the consideration of non-Gaussian uncertainties and despite the recent progress on stochastic MPC, several research challenges remain open.
A key gap is the mismatch between the analytical framework used to establish turnpike and dissipativity properties and practical stochastic MPC implementations. 
As discussed in Section~\ref{sec:uncertainty}, most existing SOCP analyses are formulated on the level of random variables or probability distributions, whereas practical controllers typically have access to state measurements. 
Extending the presented theory to more general information structures therefore remains an important open problem.
Another challenge concerns computational tractability. 
Throughout Section~\ref{sec:uncertainty}, it was implicitly assumed that the stochastic optimal control problems arising within the MPC scheme are solved exactly. 
While substantial progress on numerical methods for stochastic OCPs has been achieved, e.g.~\cite{ou25polyocp}, for nonlinear systems the computation of exact solutions remains challenging.
Practical stochastic MPC schemes therefore rely on approximations, e.g., by restricting the admissible feedback policies or reducing the number of disturbance scenarios considered during optimization.
A natural question is whether turnpike phenomena persist for such approximate solutions and how dissipativity-based closed-loop analyses can be extended to account for the resulting suboptimality. 
Addressing these questions may lead to a broader turnpike and dissipativity insights that also cover approximate stochastic MPC formulations.

\subsubsection*{Numerical methods}
While Section~\ref{sec:rec_hor_clNE} has presented some of the recent progress on numerical methods for dynamic games, the state of the art is clearly less developed than the OCP/MPC counterparts. Hence, to render RHGs real-time-applicable significant efforts are needed. At the same time, given that the methods presented in Section~\ref{sec:rec_hor_clNE} rely on Douglas-Rachford splitting, one may wonder whether the recent progress on numerical methods for distributed nonconvex optimization for control~\cite{Houska2016,tuhh:stomberg25b} and for optimization problems involving complementarity constraints, e.g.~\cite{pozharskiy2026ccopt}, also allows pushing the frontiers for numerical game-theoretic control concepts. 

\subsection{Conclusions}
This paper provided a tutorial introduction and overview of receding-horizon control across the deterministic, stochastic, multi-agent, and game-theoretic settings. It compared minimizers of optimal control problems with the equilibria of generalized Nash games in terms of cost and constraint handling. We argued that dissipativity theory and the turnpike property are a unifying thread and system-theoretic backbone across open-loop OCPs, open-loop GNEPs and their receding-horizon counterparts. While this paper is the first to give such an overview of results, it also summarizes fundamental insights in terms of parallels and conceptual differences. In conclusion we believe that this overview provides multiple starting points for further research.

\bibliography{DissipativityOCPGames}

% Generated by IEEEtran.bst, version: 1.14 (2015/08/26)
\begin{thebibliography}{100}
\providecommand{\url}[1]{#1}
\csname url@samestyle\endcsname
\providecommand{\newblock}{\relax}
\providecommand{\bibinfo}[2]{#2}
\providecommand{\BIBentrySTDinterwordspacing}{\spaceskip=0pt\relax}
\providecommand{\BIBentryALTinterwordstretchfactor}{4}
\providecommand{\BIBentryALTinterwordspacing}{\spaceskip=\fontdimen2\font plus
\BIBentryALTinterwordstretchfactor\fontdimen3\font minus
  \fontdimen4\font\relax}
\providecommand{\BIBforeignlanguage}[2]{{%
\expandafter\ifx\csname l@#1\endcsname\relax
\typeout{** WARNING: IEEEtran.bst: No hyphenation pattern has been}%
\typeout{** loaded for the language `#1'. Using the pattern for}%
\typeout{** the default language instead.}%
\else
\language=\csname l@#1\endcsname
\fi
#2}}
\providecommand{\BIBdecl}{\relax}
\BIBdecl

\bibitem{Willems72a}
J.~Willems, ``Dissipative dynamical systems part i: General theory,''
  \emph{Archive for Rational Mechanics and Analysis}, vol.~45, no.~5, pp.
  321--351, 1972.

\bibitem{Willems71a}
------, ``Least squares stationary optimal control and the algebraic {Riccati}
  equation,'' \emph{IEEE Transactions on Automatic Control}, vol.~16, no.~6,
  pp. 621--634, 1971.

\bibitem{PMP56}
V.~Boltyanskii, R.~Gamkrelidze, and L.~Pontryagin, ``On the theory of optimal
  processes,'' \emph{Doklady Akademii Nauk SSSR}, vol. 110, pp. 7--10, 1956.

\bibitem{Bellman54a}
R.~Bellman, ``The theory of dynamic programming,'' \emph{Bulletin of the
  American Mathematical Society}, vol.~60, no.~6, pp. 503--515, 1954.

\bibitem{Kalman60a}
R.~Kalman, ``Contributions to the theory of optimal control,'' \emph{Bol. Soc.
  Mat. Mexicana}, vol.~5, no.~2, pp. 102--119, 1960.

\bibitem{Mayne00a}
D.~Mayne, J.~Rawlings, C.~Rao, and P.~Scokaert, ``Constrained model predictive
  control: Stability and optimality,'' \emph{Automatica}, vol.~36, no.~6, pp.
  789--814, 2000.

\bibitem{Andersson19a}
J.~Andersson, J.~Gillis, G.~Horn, J.~Rawlings, and M.~Diehl, ``{CasADi}: a
  software framework for nonlinear optimization and optimal control,''
  \emph{Mathematical Programming Computation}, vol.~11, no.~1, pp. 1--36, 2019.

\bibitem{Kapernick14a}
B.~K{\"a}pernick and K.~Graichen, ``The gradient based nonlinear model
  predictive control software {GRAMPC},'' in \emph{2014 European Control
  Conference ({ECC})}.\hskip 1em plus 0.5em minus 0.4em\relax IEEE, 2014, pp.
  1170--1175.

\bibitem{Verschueren22a}
R.~Verschueren, G.~Frison, D.~Kouzoupis, J.~Frey, N.~v. Duijkeren, A.~Zanelli,
  B.~Novoselnik, T.~Albin, R.~Quirynen, and M.~Diehl, ``{acados} -- a modular
  open-source framework for fast embedded optimal control,'' \emph{Mathematical
  Programming Computation}, vol.~14, no.~1, pp. 147--183, 2022.

\bibitem{Rawlings09b}
J.~Rawlings and R.~Amrit, ``Optimizing process economic performance using model
  predictive control,'' in \emph{Nonlinear Model Predictive Control - Towards
  New Challenging Applications}, ser. Lecture Notes in Control and Information
  Sciences, L.~Magni, D.~Raimondo, and F.~Allg{\"o}wer, Eds.\hskip 1em plus
  0.5em minus 0.4em\relax Springer Berlin, 2009, vol. 384, pp. 119--138.

\bibitem{Diehl11a}
M.~Diehl, R.~Amrit, and J.~Rawlings, ``A {Lyapunov} function for economic
  optimizing model predictive control,'' \emph{IEEE Transactions on Automatic
  Control}, vol.~56, no.~3, pp. 703--707, 2011.

\bibitem{Angeli12a}
D.~Angeli, R.~Amrit, and J.~Rawlings, ``On average performance and stability of
  economic model predictive control,'' \emph{IEEE Transactions on Automatic
  Control}, vol.~57, no.~7, pp. 1615--1626, 2012.

\bibitem{Ramsey28}
F.~P. Ramsey, ``A mathematical theory of saving,'' \emph{The Economic Journal},
  vol.~38, no. 152, pp. 543--559, 1928.

\bibitem{vonNeumann38}
J.~von Neumann, ``{\"U}ber ein \"o{}konomisches {Gleichungssystem} und eine
  {Verallgemeinerung} des {Brouwerschen} {Fixpunktsatzes},'' in
  \emph{{Ergebnisse} eines {Mathematischen} {Seminars}}, K.~Menger, Ed.,
  Vienna, Austria, 1938, vol.~8, pp. 73--83.

\bibitem{Dorfman58}
R.~Dorfman, P.~Samuelson, and R.~Solow, \emph{Linear Programming and Economic
  Analysis}.\hskip 1em plus 0.5em minus 0.4em\relax New York: McGraw-Hill,
  1958.

\bibitem{Wilde72a}
R.~Wilde and P.~Kokotovic, ``A dichotomy in linear control theory,'' \emph{IEEE
  Transactions on Automatic Control}, vol.~17, no.~3, pp. 382--383, 1972.

\bibitem{Wuerth09}
L.~W{\"u}rth, J.~Rawlings, and W.~Marquardt, ``Economic dynamic real-time
  optimization and nonlinear model-predictive control on infinite horizons,''
  in \emph{7th {IFAC} International Symposium on Advanced Control of Chemical
  Processes}, July 2009, pp. 219--224.

\bibitem{Gruene13a}
L.~Gr{\"u}ne, ``Economic receding horizon control without terminal
  constraints,'' \emph{Automatica}, vol.~49, no.~3, pp. 725--734, 2013.

\bibitem{vonNeumann28}
J.~von Neumann, ``Zur {T}heorie der {G}esellschaftsspiele,''
  \emph{Mathematische Annalen}, vol. 100, no.~1, pp. 295--320, 1928.

\bibitem{vonNeumann44}
J.~von Neumann and O.~Morgenstern, \emph{Theory of Games and Economic
  Behavior}.\hskip 1em plus 0.5em minus 0.4em\relax Princeton University Press,
  1944.

\bibitem{Nash50}
J.~F. Nash~Jr, ``Equilibrium points in n-person games,'' \emph{Proceedings of
  the National Academy of Sciences (PNAS)}, vol.~36, no.~1, pp. 48--49, 1950.

\bibitem{Basar08}
T.~Ba{\c{s}}ar and P.~Bernhard, \emph{$H_\infty$ Optimal Control and Related
  Minimax Design Problems: A Dynamic Game Approach}, 2nd~ed.\hskip 1em plus
  0.5em minus 0.4em\relax Birkh{\"a}user Boston, 2008.

\bibitem{fershtman1986turnpike}
C.~Fershtman and E.~Muller, ``Turnpike properties of capital accumulation
  games,'' \emph{Journal of Economic Theory}, vol.~38, no.~1, pp. 167--177,
  Feb. 1986.

\bibitem{carlson1995turnpike}
D.~Carlson and A.~Haurie, \emph{A Turnpike Theory for Infinite Horizon
  Open-Loop Differential Games with Decoupled Controls}.\hskip 1em plus 0.5em
  minus 0.4em\relax Birkhäuser Boston, 1995, pp. 353--376.

\bibitem{carlson1996turnpike}
------, ``A turnpike theory for infinite-horizon open-loop competitive
  processes,'' \emph{SIAM Journal on Control and Optimization}, vol.~34, no.~4,
  pp. 1405--1419, Jul. 1996.

\bibitem{li2025turnpike}
X.~Li, F.~Wu, and X.~Zhang, ``Turnpike properties for zero-sum stochastic
  linear quadratic differential games of {Markovian} regime switching system,''
  Sep. 2025.

\bibitem{cohen2025turnpike}
A.~Cohen and J.~Jian, ``Turnpike properties in linear quadratic {Gaussian}
  {N}-player differential games,'' \emph{ESAIM: Control, Optimisation and
  Calculus of Variations}, vol.~32, p.~47, 2026.

\bibitem{cirant2021long}
M.~Cirant and A.~Porretta, ``Long time behavior and turnpike solutions in
  mildly non-monotone mean field games,'' \emph{ESAIM: Control, Optimisation
  and Calculus of Variations}, vol.~27, p.~86, 2021.

\bibitem{carmona2024leveraging}
R.~A. Carmona and C.~Zeng, ``Leveraging the turnpike effect for mean field
  games numerics,'' \emph{IEEE Open Journal of Control Systems}, vol.~3, pp.
  389--404, 2024.

\bibitem{ersland2025long}
O.~Ersland, E.~R. Jakobsen, and A.~Porretta, ``Long time behaviour of mean
  field games with fractional diffusion,'' May 2025.

\bibitem{fedorov2025studying}
F.~A. Fedorov, ``Studying the well-posedness of the boundary value problem for
  a system of {Riccati} type equations based on the concept of mean field
  games,'' \emph{Moscow University Computational Mathematics and Cybernetics},
  vol.~49, no.~2, pp. 150--164, Jun. 2025.

\bibitem{stephens2015game}
E.~R. Stephens, D.~B. Smith, and A.~Mahanti, ``Game theoretic model predictive
  control for distributed energy demand-side management,'' \emph{IEEE
  Transactions on Smart Grid}, vol.~6, no.~3, pp. 1394--1402, 2015.

\bibitem{lecleach2022algames}
\BIBentryALTinterwordspacing
S.~Le~Cleac'h, M.~Schwager, and Z.~Manchester, ``Algames: a fast augmented
  {Lagrangian} solver for constrained dynamic games,'' \emph{Autonomous
  Robots}, vol.~46, no.~1, pp. 201--215, 2022. [Online]. Available:
  \url{https://doi.org/10.1007/s10514-021-10024-7}
\BIBentrySTDinterwordspacing

\bibitem{benenati2024probabilistic}
E.~Benenati and S.~Grammatico, ``Probabilistic game-theoretic traffic
  routing,'' \emph{IEEE Transactions on Intelligent Transportation Systems},
  vol.~25, no.~10, pp. 13\,080--13\,090, 2024.

\bibitem{hall2024receding}
S.~Hall, L.~Guerrini, F.~D{\"o}rfler, and D.~Liao-McPherson, ``Receding horizon
  games for modeling competitive supply chains,'' \emph{IFAC-PapersOnLine},
  vol.~58, no.~18, pp. 8--14, 2024.

\bibitem{hall2026stability}
S.~Hall, G.~Belgioioso, F.~D{\"o}rfler, and D.~Liao-McPherson, ``Stability
  certificates for receding-horizon games,'' \emph{IEEE Transactions on
  Automatic Control}, vol.~71, no.~5, pp. 3511--3518, 2026.

\bibitem{benenati2026linear}
E.~Benenati and S.~Grammatico, ``Linear-quadratic dynamic games as
  receding-horizon variational inequalities,'' \emph{IEEE Transactions on
  Automatic Control}, vol.~71, no.~4, pp. 2404--2417, 2026.

\bibitem{hall2026towards}
S.~Hall, F.~Dörfler, and T.~Faulwasser, ``Towards closed-loop stability of
  nonlinear receding horizon games,'' 2026, accepted for the 2026 65th {IEEE}
  Conference on Decision and Control ({CDC}), arXiv:2605.12467.

\bibitem{Skogestad05}
S.~Skogestad and I.~Postlethwaite, \emph{Multivariable feedback control:
  analysis and design}, 2nd~ed.\hskip 1em plus 0.5em minus 0.4em\relax
  Chichester: John Wiley \& Sons, 2005.

\bibitem{Boltyanskii60a}
V.~Boltyanskii, R.~Gamkrelidze, and L.~Pontryagin, ``Theory of optimal
  processes. i. the maximum principle,'' \emph{Izvestiya Rossiiskoi Akademii
  Nauk. Seriya Matematicheskaya}, vol.~24, no.~1, pp. 3--42, 1960.

\bibitem{Scattolini09a}
R.~Scattolini, ``Architectures for distributed and hierarchical model
  predictive control--a review,'' \emph{Journal of Process Control}, vol.~19,
  no.~5, pp. 723--731, 2009.

\bibitem{muller2017economic}
M.~A. M{\"u}ller and F.~Allg{\"o}wer, ``Economic and distributed model
  predictive control: Recent developments in optimization-based control,''
  \emph{SICE Journal of Control, Measurement, and System Integration}, vol.~10,
  no.~2, pp. 39--52, 2017.

\bibitem{Bryson69a}
A.~E. Bryson, Jr. and Y.-C. Ho, \emph{Applied Optimal Control: Optimization,
  Estimation, and Control}.\hskip 1em plus 0.5em minus 0.4em\relax Waltham, MA:
  Blaisdell Publishing, 1969.

\bibitem{Bryson99a}
A.~E. Bryson, \emph{Dynamic Optimization}.\hskip 1em plus 0.5em minus
  0.4em\relax Menlo Park, CA: Addison Wesley Longman, 1999.

\bibitem{Sussmann97}
H.~Sussmann and J.~Willems, ``300 years of optimal control: from the
  brachystochrone to the maximum principle,'' \emph{IEEE Control Systems},
  vol.~17, no.~3, pp. 32--44, 1997.

\bibitem{Kalman63}
R.~E. Kalman, ``The theory of optimal control and the calculus of variations,''
  in \emph{Mathematical Optimization Techniques}, R.~Bellman, Ed.\hskip 1em
  plus 0.5em minus 0.4em\relax Berkeley: University of California Press, 1963,
  ch.~16, pp. 309--331.

\bibitem{Bertsekas19a}
D.~Bertsekas, \emph{Reinforcement learning and optimal control}.\hskip 1em plus
  0.5em minus 0.4em\relax Belmont, MA: Athena Scientific, 2019.

\bibitem{pavel2012game}
L.~Pavel, \emph{Game Theory for Control of Optical Networks}, ser. Static and
  Dynamic Game Theory: Foundations and Applications.\hskip 1em plus 0.5em minus
  0.4em\relax Birkhäuser Boston, 2012.

\bibitem{atzeni2013noncooperative}
I.~Atzeni, L.~G. Ordonez, G.~Scutari, D.~P. Palomar, and J.~R. Fonollosa,
  ``Noncooperative and cooperative optimization of distributed energy
  generation and storage in the demand-side of the smart grid,'' \emph{IEEE
  Transactions on Signal Processing}, vol.~61, no.~10, pp. 2454--2472, May
  2013.

\bibitem{krawczyk2005coupled}
J.~B. Krawczyk, ``Coupled constraint {Nash} equilibria in environmental
  games,'' \emph{Resource and Energy Economics}, vol.~27, no.~2, pp. 157--181,
  Jun. 2005.

\bibitem{bahn2008class}
O.~Bahn and A.~Haurie, ``A class of games with couples constraints to model
  international {GHG} emission agreements,'' \emph{International Game Theory
  Review}, vol.~10, no.~04, pp. 337--362, Dec. 2008.

\bibitem{facchinei2009generalized}
F.~Facchinei, A.~Fischer, and V.~Piccialli, ``Generalized {Nash} equilibrium
  problems and {Newton} methods,'' \emph{Mathematical Programming}, vol. 117,
  no.~1, pp. 163--194, Mar. 2009.

\bibitem{bauschke2017convex}
H.~H. Bauschke and P.~L. Combettes, \emph{Convex Analysis and Monotone Operator
  Theory in {Hilbert} Spaces}, 2nd~ed., ser. {CMS} Books in Mathematics.\hskip
  1em plus 0.5em minus 0.4em\relax Springer, 2017.

\bibitem{facchinei2009nash}
F.~Facchinei and J.-S. Pang, ``{Nash} equilibria: the variational approach,''
  in \emph{Convex Optimization in Signal Processing and Communications}, D.~P.
  Palomar and Y.~C. Eldar, Eds.\hskip 1em plus 0.5em minus 0.4em\relax
  Cambridge: Cambridge University Press, dec 2009, ch.~12, pp. 443--493.

\bibitem{gruene2013economic}
L.~Gr{\"u}ne, ``Economic receding horizon control without terminal
  constraints,'' \emph{Automatica}, vol.~49, no.~3, pp. 725--734, mar 2013.

\bibitem{monderer1996potential}
D.~Monderer and L.~S. Shapley, ``Potential games,'' \emph{Games and Economic
  Behavior}, vol.~14, no.~1, pp. 124--143, May 1996.

\bibitem{hall2025system}
S.~Hall, F.~D{\"o}rfler, and T.~Faulwasser, ``System-theoretic analysis of
  dynamic generalized {Nash} equilibria -- turnpikes and dissipativity,'' Oct.
  2025, arXiv:2510.21556.

\bibitem{Kalman60b}
R.~Kalman, ``A new approach to linear filtering and prediction problems,''
  \emph{Journal of Basic Engineering}, vol.~82, pp. 35--45, 1960.

\bibitem{Dreyfus60}
S.~E. Dreyfus, ``Dynamic programming and the calculus of variations,''
  \emph{Journal of Mathematical Analysis and Applications}, vol.~1, no.~2, pp.
  228--239, 1960.

\bibitem{Dreyfus77}
S.~E. Dreyfus and A.~M. Law, \emph{The Art and Theory of Dynamic Programming},
  ser. Mathematics in Science and Engineering.\hskip 1em plus 0.5em minus
  0.4em\relax New York: Academic Press, 1977, vol. 130.

\bibitem{Lee67}
E.~Lee and L.~Markus, \emph{Foundations of Optimal Control Theory}, ser. The
  {SIAM} Series in Applied Mathematics.\hskip 1em plus 0.5em minus 0.4em\relax
  New York: John Wiley \& Sons, 1967.

\bibitem{Propoi63}
A.~Propoi, ``Application of linear programming methods for the synthesis of
  automatic sampled-data systems,'' \emph{Avtomat. i Telemekh}, vol.~24, no.~7,
  pp. 912--920, 1963.

\bibitem{Keerthi88}
S.~Keerthi and E.~Gilbert, ``Optimal infinite-horizon feedback laws for a
  general class of constrained discrete-time systems: Stability and
  moving-horizon approximations,'' \emph{Journal of Optimization Theory and
  Applications}, vol.~57, no.~2, pp. 265--293, 1988.

\bibitem{Maciejowski02a}
J.~Maciejowski, \emph{Predictive control: with constraints}.\hskip 1em plus
  0.5em minus 0.4em\relax Pearson Education Limited, 2002.

\bibitem{kouvaritakis2016mpc}
B.~Kouvaritakis and M.~Cannon, \emph{Model Predictive Control: Classical,
  Robust and Stochastic}, 1st~ed., ser. Advanced Textbooks in Control and
  Signal Processing.\hskip 1em plus 0.5em minus 0.4em\relax Cham: Springer,
  2016.

\bibitem{Rawlings17}
J.~Rawlings, D.~Mayne, and M.~Diehl, \emph{Model Predictive Control: Theory,
  Computation, and Design}, 2nd~ed.\hskip 1em plus 0.5em minus 0.4em\relax
  Madison, WI: Nob Hill Publishing, 2017.

\bibitem{Gruene17a}
L.~Gr{\"u}ne and J.~Pannek, \emph{Nonlinear Model Predictive Control: Theory
  and Algorithms}, 2nd~ed., ser. Communications and Control Engineering.\hskip
  1em plus 0.5em minus 0.4em\relax Cham: Springer, 2017.

\bibitem{tudo:faulwasser24b}
T.~Faulwasser, E.~Kerrigan, F.~Logist, S.~Lucia, M.~M{\"o}nnigmann, A.~Parisio,
  and M.~Schulze~Darup, ``Teaching model predictive control: What, when, where,
  why, who, and how?'' \emph{IEEE Control Systems Magazine}, vol.~44, no.~4,
  pp. 47--65, Aug. 2024.

\bibitem{Kellett14}
C.~Kellett, ``A compendium of comparison function results,'' \emph{Mathematics
  of Control, Signals, and Systems}, vol.~26, no.~3, pp. 339--374, 2014.

\bibitem{Chen98}
H.~Chen and F.~Allg{\"o}wer, ``A quasi-infinite horizon nonlinear model
  predictive control scheme with guaranteed stability,'' \emph{Automatica},
  vol.~34, no.~10, pp. 1205--1217, 1998.

\bibitem{lincoln2006relaxing}
B.~Lincoln and A.~Rantzer, ``Relaxing dynamic programming,'' \emph{IEEE
  Transactions on Automatic Control}, vol.~51, no.~8, pp. 1249--1260, 2006.

\bibitem{Gruene09a}
L.~Gr{\"u}ne, ``Analysis and design of unconstrained nonlinear mpc schemes for
  finite and infinite dimensional systems,'' \emph{SIAM Journal on Control and
  Optimization}, vol.~48, no.~2, pp. 1206--1228, 2009.

\bibitem{Gruene10a}
L.~Gr{\"u}ne, J.~Pannek, M.~Seehafer, and K.~Worthmann, ``Analysis of
  unconstrained nonlinear mpc schemes with time varying control horizon,''
  \emph{SIAM Journal on Control and Optimization}, vol.~48, no.~8, pp.
  4938--4962, 2010.

\bibitem{Boccia14a}
A.~Boccia, L.~Gr{\"u}ne, and K.~Worthmann, ``Stability and feasibility of state
  constrained {MPC} without stabilizing terminal constraints,'' \emph{Systems
  \& Control Letters}, vol.~72, pp. 14--21, 2014.

\bibitem{keviczky2006decentralized}
T.~Keviczky, F.~Borrelli, and G.~J. Balas, ``Decentralized receding horizon
  control for large scale dynamically decoupled systems,'' \emph{Automatica},
  vol.~42, no.~12, pp. 2105--2115, 2006.

\bibitem{camponogara2002distributed}
E.~Camponogara, D.~Jia, B.~H. Krogh, and S.~Talukdar, ``Distributed model
  predictive control,'' \emph{IEEE Control Systems Magazine}, vol.~22, no.~1,
  pp. 44--52, 2002.

\bibitem{Boyd2011}
S.~Boyd, N.~Parikh, E.~Chu, B.~Peleato, and J.~Eckstein, ``Distributed
  optimization and statistical learning via the alternating direction method of
  multipliers,'' \emph{Foundations and Trends in Machine Learning}, vol.~3,
  no.~1, pp. 1--122, 2011.

\bibitem{Houska2016}
B.~Houska, J.~Frasch, and M.~Diehl, ``An augmented {Lagrangian} based algorithm
  for distributed nonconvex optimization,'' \emph{SIAM Journal on
  Optimization}, vol.~26, no.~2, pp. 1101--1127, 2016.

\bibitem{Stomberg2025b}
G.~Stomberg, A.~Engelmann, M.~Diehl, and T.~Faulwasser, ``Decentralized
  real-time iterations for distributed {NMPC},'' \emph{IEEE Transactions on
  Automatic Control}, vol.~71, no.~3, pp. 1584--1599, 2026.

\bibitem{dunbar2006distributed}
W.~B. Dunbar and R.~M. Murray, ``Distributed receding horizon control for
  multi-vehicle formation stabilization,'' \emph{Automatica}, vol.~42, no.~4,
  pp. 549--558, 2006.

\bibitem{Venkat08a}
A.~Venkat, I.~Hiskens, J.~Rawlings, and S.~Wright, ``Distributed {MPC}
  strategies with application to power system automatic generation control,''
  \emph{IEEE Transactions on Control Systems Technology}, vol.~16, no.~6, pp.
  1192--1206, 2008.

\bibitem{tuhh:stomberg25b}
G.~Stomberg, M.~Raetsch, A.~Engelmann, and T.~Faulwasser, ``Large problems are
  not necessarily hard: {A} case study on distributed {NMPC} paying off,'' in
  \emph{2025 European Control Conference ({ECC})}.\hskip 1em plus 0.5em minus
  0.4em\relax IEEE, 2025, pp. 2630--2637.

\bibitem{tudo:stomberg22a}
G.~Stomberg, A.~Engelmann, and T.~Faulwasser, ``A compendium of optimization
  algorithms for distributed linear-quadratic {MPC},'' \emph{at -
  Automatisierungstechnik}, vol.~70, no.~4, pp. 317--330, 2022.

\bibitem{gu2008differential}
D.~Gu, ``A differential game approach to formation control,'' \emph{IEEE Trans.
  Control Syst. Technol.}, vol.~16, no.~1, pp. 85--93, Jan. 2008.

\bibitem{liniger2020noncooperative}
A.~Liniger and J.~Lygeros, ``A noncooperative game approach to autonomous
  racing,'' \emph{{IEEE} Transactions on Control Systems Technology}, vol.~28,
  no.~3, pp. 884--897, May 2020.

\bibitem{wang2021game}
M.~Wang, Z.~Wang, J.~Talbot, J.~C. Gerdes, and M.~Schwager, ``Game-theoretic
  planning for self-driving cars in multivehicle competitive scenarios,''
  \emph{{IEEE} Transactions on Robotics}, vol.~37, no.~4, pp. 1313--1325, 2021.

\bibitem{mignoni2023distributed}
N.~Mignoni, R.~Carli, and M.~Dotoli, ``Distributed noncooperative {MPC} for
  energy scheduling of charging and trading electric vehicles in energy
  communities,'' \emph{IEEE Transactions on Control Systems Technology},
  vol.~31, no.~5, pp. 2159--2172, 2023.

\bibitem{paola2018distributed}
A.~D. Paola, F.~Fele, D.~Angeli, and G.~Strbac, ``Distributed coordination of
  price-responsive electric loads: A receding horizon approach,'' in \emph{2018
  {IEEE} Conference on Decision and Control ({CDC})}.\hskip 1em plus 0.5em
  minus 0.4em\relax IEEE, dec 2018, pp. 6033--6040.

\bibitem{hall2022receding}
S.~Hall, G.~Belgioioso, D.~Liao-McPherson, and F.~D{\"o}rfler, ``Receding
  horizon games with coupling constraints for demand-side management,'' in
  \emph{2022 {IEEE} 61st Conference on Decision and Control ({CDC})}, 2022, pp.
  3795--3800.

\bibitem{benenati2023optimal}
E.~Benenati, W.~Ananduta, and S.~Grammatico, ``Optimal selection and tracking
  of generalized {Nash} equilibria in monotone games,'' \emph{IEEE Transactions
  on Automatic Control}, vol.~68, no.~12, pp. 7644--7659, 2023.

\bibitem{hall2025limits}
S.~Hall, F.~D{\"o}rfler, H.~H. Nax, and S.~Bolognani, ``The limits of
  ``fairness'' of the variational generalized {Nash} equilibrium,'' in
  \emph{2025 {IEEE} 64th Conference on Decision and Control ({CDC})}.\hskip 1em
  plus 0.5em minus 0.4em\relax IEEE, 2025, pp. 5354--5360.

\bibitem{hall2026solving}
S.~Hall and A.~Bemporad, ``Solving multiparametric generalized {Nash}
  equilibrium problems and explicit game-theoretic model predictive control,''
  Jul. 2026.

\bibitem{gruene2014asymptotic}
L.~Gr{\"u}ne and M.~Stieler, ``Asymptotic stability and transient optimality of
  economic {MPC} without terminal conditions,'' \emph{Journal of Process
  Control}, vol.~24, no.~8, pp. 1187--1196, aug 2014.

\bibitem{bemporad2025nashopt}
A.~Bemporad, ``{NashOpt}: A {Python} library for computing generalized {Nash}
  equilibria and game design,'' 2025,
  \url{https://github.com/bemporad/nashopt}.

\bibitem{Willems07a}
J.~Willems, ``Dissipative dynamical systems,'' \emph{European Journal of
  Control}, vol.~13, no. 2-3, pp. 134--151, 2007.

\bibitem{Byrnes94}
C.~I. Byrnes and W.~Lin, ``Losslessness, feedback equivalence, and the global
  stabilization of discrete-time nonlinear systems,'' \emph{IEEE Transactions
  on Automatic Control}, vol.~39, no.~1, pp. 83--98, 1994.

\bibitem{IEEECSM02a}
R.~Sepulchre, ``50 years of dissipativity theory, part i,'' \emph{IEEE Control
  Systems Magazine}, vol.~42, no.~2, pp. 6--9, 2022.

\bibitem{IEEECSM02b}
------, ``50 years of dissipativity theory, part ii,'' \emph{IEEE Control
  Systems Magazine}, vol.~42, no.~3, pp. 5--7, 2022.

\bibitem{Gruene16a}
L.~Gr{\"u}ne and M.~M{\"u}ller, ``On the relation between strict dissipativity
  and turnpike properties,'' \emph{Systems \& Control Letters}, vol.~90, pp.
  45--53, 2016.

\bibitem{tudo:faulwasser22b}
T.~Faulwasser and C.~M. Kellett, ``Dissipativity in infinite-horizon optimal
  control: {Willems}---1971 paper revisited,'' \emph{IFAC-PapersOnLine},
  vol.~55, no.~30, pp. 49--54, 2022, 25th {IFAC} Symposium on Mathematical
  Theory of Networks and Systems MTNS 2022.

\bibitem{zanon2026rethinking}
M.~Zanon, ``Rethinking strict dissipativity for economic mpc,'' \emph{arXiv
  preprint arXiv:2603.08535}, 2026.

\bibitem{tudo:faulwasser21a}
T.~Faulwasser and C.~Kellett, ``On continuous-time infinite horizon optimal
  control -- {Dissipativity}, stability and transversality,''
  \emph{Automatica}, vol. 134, p. 109907, 2021.

\bibitem{epfl:faulwasser15h}
T.~Faulwasser, M.~Korda, C.~Jones, and D.~Bonvin, ``On turnpike and
  dissipativity properties of continuous-time optimal control problems,''
  \emph{Automatica}, vol.~81, pp. 297--304, 2017.

\bibitem{Stieler14a}
T.~Damm, L.~Gr{\"u}ne, M.~Stieler, and K.~Worthmann, ``An exponential turnpike
  theorem for dissipative optimal control problems,'' \emph{SIAM Journal on
  Control and Optimization}, vol.~52, no.~3, pp. 1935--1957, 2014.

\bibitem{epfl:faulwasser14e}
T.~Faulwasser, M.~Korda, C.~Jones, and D.~Bonvin, ``Turnpike and dissipativity
  properties in dynamic real-time optimization and economic {MPC},'' in
  \emph{Proc. of the 53rd {IEEE} Conference on Decision and Control}, Los
  Angeles, California, USA, 2014, pp. 2734--2739.

\bibitem{tudo:faulwasser22a}
T.~Faulwasser and L.~Gr\"une, ``Turnpike properties in optimal control: An
  overview of discrete-time and continuous-time results,'' in \emph{Handbook of
  Numerical Analysis}, E.~Zuazua and E.~Tr\'elat, Eds.\hskip 1em plus 0.5em
  minus 0.4em\relax Elsevier, Jan. 2022, vol.~23, ch.~11, pp. 367--400.

\bibitem{Carlson91}
D.~A. Carlson, A.~B. Haurie, and A.~Leizarowitz, \emph{Infinite Horizon Optimal
  Control: Deterministic and Stochastic Systems}, 2nd~ed.\hskip 1em plus 0.5em
  minus 0.4em\relax Berlin, Heidelberg: Springer, 1991.

\bibitem{Trelat15a}
E.~Tr{\'e}lat and E.~Zuazua, ``The turnpike property in finite-dimensional
  nonlinear optimal control,'' \emph{Journal of Differential Equations}, vol.
  258, no.~1, pp. 81--114, 2015.

\bibitem{carlson2000infinite}
D.~A. Carlson and A.~B. Haurie, \emph{Infinite Horizon Dynamic Games with
  Coupled State Constraints}.\hskip 1em plus 0.5em minus 0.4em\relax
  Birkhäuser Boston, 2000, pp. 195--212.

\bibitem{koutsoupias1999worst}
E.~Koutsoupias and C.~Papadimitriou, ``Worst-case equilibria,'' in \emph{Annual
  symposium on theoretical aspects of computer science}.\hskip 1em plus 0.5em
  minus 0.4em\relax Springer, 1999, pp. 404--413.

\bibitem{papadimitriou2001algorithms}
C.~Papadimitriou, ``Algorithms, games, and the internet,'' in \emph{Proceedings
  of the thirty-third annual ACM symposium on Theory of computing}, 2001, pp.
  749--753.

\bibitem{kit:faulwasser18c}
T.~Faulwasser, L.~Gr\"une, and M.~M{\"u}ller, ``Economic nonlinear model
  predictive control: Stability, optimality and performance,''
  \emph{Foundations and Trends in Systems and Control}, vol.~5, no.~1, pp.
  1--98, 2018.

\bibitem{grunekellett2017}
L.~Gr{\"u}ne, C.~M. Kellett, and S.~R. Weller, ``On the relation between
  turnpike properties for finite and infinite horizon optimal control
  problems,'' \emph{Journal of Optimization Theory and Applications}, vol. 173,
  no.~3, pp. 727--745, 2017.

\bibitem{kit:faulwasser18e_2}
T.~Faulwasser and M.~Zanon, ``Asymptotic stability of economic {NMPC}: {The}
  importance of adjoints,'' \emph{IFAC-PapersOnLine}, vol.~51, no.~20, pp.
  157--168, 2018.

\bibitem{kit:zanon18a}
M.~Zanon and T.~Faulwasser, ``Economic {MPC} without terminal constraints:
  Gradient-correcting end penalties enforce stability,'' \emph{Journal of
  Process Control}, vol.~63, pp. 1--14, Mar. 2018.

\bibitem{mesbah2016stochastic}
A.~Mesbah, ``Stochastic model predictive control: An overview and perspectives
  for future research,'' \emph{IEEE Control Systems Magazine}, vol.~36, no.~6,
  pp. 30--44, 2016.

\bibitem{fristedt1997modern}
B.~Fristedt and L.~Gray, \emph{A Modern Approach to Probability Theory}, ser.
  Probability and Its Applications.\hskip 1em plus 0.5em minus 0.4em\relax
  Boston, MA: Birkh{\"a}user, 1997.

\bibitem{Protter2005stochastic}
P.~E. Protter, \emph{Stochastic Integration and Differential Equations},
  2nd~ed., ser. Stochastic Modelling and Applied Probability.\hskip 1em plus
  0.5em minus 0.4em\relax Springer Berlin Heidelberg, 2005, vol.~21.

\bibitem{Kolokoltsov2012}
V.~Kolokoltsov and W.~Yang, ``Turnpike theorems for {Markov} games,''
  \emph{Dynamic Games and Applications}, vol.~2, no.~3, pp. 294--312, may 2012.

\bibitem{Marimon1989}
R.~Marimon, ``Stochastic turnpike property and stationary equilibrium,''
  \emph{Journal of Economic Theory}, vol.~47, no.~2, pp. 282--306, apr 1989.

\bibitem{Sun2022}
J.~Sun, H.~Wang, and J.~Yong, ``Turnpike properties for stochastic
  linear-quadratic optimal control problems,'' \emph{Chinese Annals of
  Mathematics, Series B}, vol.~43, no.~6, pp. 999--1022, nov 2022.

\bibitem{schiessl2023pathwise}
J.~Schie{\ss}l, R.~Ou, T.~Faulwasser, M.~H. Baumann, and L.~Gr{\"u}ne,
  ``Pathwise turnpike and dissipativity results for discrete-time stochastic
  linear-quadratic optimal control problems,'' in \emph{2023 62nd {IEEE}
  Conference on Decision and Control ({CDC})}.\hskip 1em plus 0.5em minus
  0.4em\relax IEEE, 2023, pp. 2790--2795.

\bibitem{ou2021simulation}
R.~Ou, M.~H. Baumann, L.~Gr{\"u}ne, and T.~Faulwasser, ``A simulation study on
  turnpikes in stochastic {LQ} optimal control,'' \emph{IFAC-PapersOnLine},
  vol.~54, no.~3, pp. 516--521, 2021, 16th {IFAC} Symposium on Advanced Control
  of Chemical Processes {ADCHEM} 2021.

\bibitem{sun2023stochturnpikepaths}
J.~Sun and H.~Wu, ``Long-time behavior of stochastic linear-quadratic optimal
  control problems,'' in \emph{2023 62nd {IEEE} Conference on Decision and
  Control ({CDC})}.\hskip 1em plus 0.5em minus 0.4em\relax IEEE, 2023, pp.
  2796--2802.

\bibitem{ou25polyocp}
R.~Ou, L.~Januzi, J.~Schie{\ss}l, M.~H. Baumann, L.~Gr{\"u}ne, and
  T.~Faulwasser, ``Polyocp. jl -- a julia package for stochastic ocps and
  mpc,'' \emph{European Journal of Control}, p. 101617, 2026.

\bibitem{muehlpfordt20polychaos}
T.~M{\"u}hlpfordt, F.~Zahn, V.~Hagenmeyer, and T.~Faulwasser, ``{PolyChaos.jl}
  -- a {Julia} package for polynomial chaos in systems and control,''
  \emph{IFAC-PapersOnLine}, vol.~53, no.~2, pp. 7210--7216, 2020, 21st {IFAC}
  World Congress.

\bibitem{Doob1953}
J.~Doob, \emph{Stochastic Processes}, ser. Probability and Statistics
  Series.\hskip 1em plus 0.5em minus 0.4em\relax Wiley, 1953.

\bibitem{schiessl2024relationship}
J.~Schie{\ss}l, M.~H. Baumann, T.~Faulwasser, and L.~Gr{\"u}ne, ``On the
  relationship between stochastic turnpike and dissipativity notions,''
  \emph{IEEE Transactions on Automatic Control}, vol.~70, no.~6, pp.
  3527--3539, 2025.

\bibitem{gros2022stochDissi}
S.~Gros and M.~Zanon, ``Economic {MPC} of {Markov} decision processes:
  {Dissipativity} in undiscounted infinite-horizon optimal control,''
  \emph{Automatica}, vol. 146, p. 110602, 2022.

\bibitem{schiessl2025turnpikeLQP}
J.~Schie{\ss}l, R.~Ou, T.~Faulwasser, M.~H. Baumann, and L.~Gr{\"u}ne,
  ``Turnpike and dissipativity in generalized discrete-time stochastic
  linear-quadratic optimal control,'' \emph{SIAM Journal on Control and
  Optimization}, vol.~63, no.~2, pp. 1432--1457, 2025.

\bibitem{schiessl26stability}
J.~Schie{\ss}l, H.~Selder, R.~Ou, M.~H. Baumann, T.~Faulwasser, and
  L.~Gr{\"u}ne, ``Stability and performance of stochastic economic {MPC} --
  stochastic characterization of the closed-loop asymptotics,'' Jun. 2026,
  accepted for the Special Issue for Prof. Jean-Michel Coron, Chinese Annals of
  Mathematics, Series B.

\bibitem{lucia2020stability}
S.~Lucia, S.~Subramanian, D.~Limon, and S.~Engell, ``Stability properties of
  multi-stage nonlinear model predictive control,'' \emph{Systems \& Control
  Letters}, vol. 143, p. 104743, 2020.

\bibitem{mcallister2022stochMPC}
R.~D. McAllister and J.~B. Rawlings, ``Nonlinear stochastic model predictive
  control: {Existence}, measurability, and stochastic asymptotic stability,''
  \emph{IEEE Transactions on Automatic Control}, vol.~68, no.~3, pp.
  1524--1536, 2023.

\bibitem{chatterjee2014stability}
D.~Chatterjee and J.~Lygeros, ``On stability and performance of stochastic
  predictive control techniques,'' \emph{IEEE Transactions on Automatic
  Control}, vol.~60, no.~2, pp. 509--514, 2015.

\bibitem{hewing2018stochMPC}
L.~Hewing and M.~N. Zeilinger, ``Stochastic model predictive control for linear
  systems using probabilistic reachable sets,'' in \emph{2018 {IEEE} 57th
  Conference on Decision and Control ({CDC}) : Proceedings}, 2018, pp.
  5182--5188.

\bibitem{hewing2020stochMPC}
\BIBentryALTinterwordspacing
L.~Hewing, K.~P. Wabersich, and M.~N. Zeilinger, ``Recursively feasible
  stochastic model predictive control using indirect feedback,''
  \emph{Automatica}, vol. 119, p. 109095, 2020. [Online]. Available:
  \url{https://www.sciencedirect.com/science/article/pii/S0005109820302934}
\BIBentrySTDinterwordspacing

\bibitem{koehler2025stochMPC}
J.~K{\"o}hler and M.~N. Zeilinger, ``Predictive control for nonlinear
  stochastic systems: {Closed}-loop guarantees with unbounded noise,''
  \emph{IEEE Transactions on Automatic Control}, vol.~70, no.~11, pp.
  7382--7397, 2025.

\bibitem{kordabad2022MDP}
A.~B. Kordabad and S.~Gros, ``Functional stability of discounted {Markov}
  decision processes using economic {MPC} dissipativity theory,'' in \emph{2022
  European Control Conference ({ECC})}.\hskip 1em plus 0.5em minus 0.4em\relax
  IEEE, 2022, pp. 1858--1863.

\bibitem{gale1967optimal}
D.~Gale, ``On optimal development in a multi-sector economy,'' \emph{The Review
  of Economic Studies}, vol.~34, no.~1, pp. 1--18, 1967.

\bibitem{schiessl2024nearOptimal}
J.~Schie{\ss}l, R.~Ou, T.~Faulwasser, M.~H. Baumann, and L.~Gr{\"u}ne,
  ``Near-optimal performance of stochastic economic {MPC},'' in \emph{2024
  {IEEE} 63rd Conference on Decision and Control ({CDC})}.\hskip 1em plus 0.5em
  minus 0.4em\relax IEEE, 2024, pp. 2565--2571.

\bibitem{bertsekas1996stochasticDPP}
D.~P. Bertsekas and S.~E. Shreve, \emph{Stochastic Optimal Control: The
  Discrete-Time Case}, ser. Optimization and Neural Computation Series.\hskip
  1em plus 0.5em minus 0.4em\relax Belmont, MA: Athena Scientific, 1996,
  vol.~5, originally published by Academic Press, 1978.

\bibitem{schiessl2025riskcost}
J.~Schie{\ss}l, R.~Ou, M.~H. Baumann, T.~Faulwasser, and L.~Gr{\"u}ne,
  ``Towards turnpike-based performance analysis of risk-averse stochastic
  predictive control,'' in \emph{2025 {IEEE} 64th Conference on Decision and
  Control ({CDC})}.\hskip 1em plus 0.5em minus 0.4em\relax IEEE, 2025, pp.
  329--335.

\bibitem{schluter2022stochastic}
H.~Schl{\"u}ter and F.~Allg{\"o}wer, ``Stochastic model predictive control
  using initial state optimization,'' \emph{IFAC-PapersOnLine}, vol.~55,
  no.~30, pp. 454--459, 2022, 25th International Symposium on Mathematical
  Theory of Networks and Systems {MTNS} 2022.

\bibitem{schluter2023stochastic}
------, ``Stochastic model predictive control using initial state and variance
  interpolation,'' in \emph{2023 62nd {IEEE} Conference on Decision and Control
  ({CDC})}.\hskip 1em plus 0.5em minus 0.4em\relax IEEE, 2023, pp. 6700--6706.

\bibitem{schiessl2026riskconstraints}
J.~Schie{\ss}l, R.~Ou, M.~H. Baumann, T.~Faulwasser, and L.~Gr{\"u}ne,
  ``Closed-loop analysis of linear stochastic {MPC} with risk-averse
  constraints,'' 2026, accepted for the 2026 65th {IEEE} Conference on Decision
  and Control ({CDC}), arXiv:2604.11183.

\bibitem{wachter2006ipopt}
A.~W{\"a}chter and L.~T. Biegler, ``On the implementation of an interior-point
  filter line-search algorithm for large-scale nonlinear programming,''
  \emph{Mathematical Programming}, vol. 106, no.~1, pp. 25--57, 2006.

\bibitem{stellato2018osqp}
B.~Stellato, G.~Banjac, P.~Goulart, A.~Bemporad, and S.~Boyd, ``{OSQP}: An
  operator splitting solver for quadratic programs,'' in \emph{2018 {UKACC}
  12th International Conference on Control ({CONTROL})}.\hskip 1em plus 0.5em
  minus 0.4em\relax IEEE, 2018, p. 339.

\bibitem{spica2020real}
R.~Spica, E.~Cristofalo, Z.~Wang, E.~Montijano, and M.~Schwager, ``A real-time
  game theoretic planner for autonomous two-player drone racing,'' \emph{IEEE
  Transactions on Robotics}, vol.~36, no.~5, pp. 1389--1403, Oct. 2020.

\bibitem{pustilnik2025generalized}
M.~Pustilnik and F.~Borrelli, ``Generalized {Nash} equilibrium solutions in
  dynamic games with shared constraints,'' Feb. 2025, arXiv:2502.19569.

\bibitem{dreves2011solution}
A.~Dreves, F.~Facchinei, C.~Kanzow, and S.~Sagratella, ``On the solution of the
  {KKT} conditions of generalized {Nash} equilibrium problems,'' \emph{SIAM
  Journal on Optimization}, vol.~21, no.~3, pp. 1082--1108, 2011.

\bibitem{facchinei2003finite}
F.~Facchinei and J.-S. Pang, \emph{Finite-Dimensional Variational Inequalities
  and Complementarity Problems}, ser. Springer Series in Operations Research
  and Financial Engineering.\hskip 1em plus 0.5em minus 0.4em\relax New York,
  NY: Springer, 2003.

\bibitem{eckstein1998operator}
J.~Eckstein and M.~C. Ferris, ``Operator-splitting methods for monotone affine
  variational inequalities, with a parallel application to optimal control,''
  \emph{{INFORMS} Journal on Computing}, vol.~10, no.~2, pp. 218--235, 1998.

\bibitem{baghbadorani:benenati:2026}
R.~R. Baghbadorani, E.~Benenati, and S.~Grammatico, ``A {Douglas--Rachford}
  splitting method for solving monotone variational inequalities in
  linear-quadratic dynamic games,'' in \emph{IFAC World Congress}, 2026,
  arXiv:2504.05757.

\bibitem{baghbadorani2026fastnewtonmethodslinearquadratic}
R.~R. Baghbadorani and S.~Grammatico, ``Fast {Newton} methods for
  linear-quadratic dynamic games with application to autonomous vehicle
  platooning and intersection crossing,'' in \emph{{IEEE} Intelligent
  Transportation Systems Conference}, 2026, arXiv:2605.01898.

\bibitem{SUN2019381}
J.~Sun and J.~Yong, ``Linear–quadratic stochastic two-person nonzero-sum
  differential games: Open-loop and closed-loop {Nash} equilibria,''
  \emph{Stochastic Processes and their Applications}, vol. 129, no.~2, pp.
  381--418, 2019.

\bibitem{pozharskiy2026ccopt}
A.~Pozharskiy, F.~Pacaud, M.~Diehl, and A.~Nurkanovi{\'c}, ``{CCOpt}: an
  open-source solver for large-scale mathematical programs with complementarity
  constraints,'' Apr. 2026, arXiv:2604.18726.

\end{thebibliography}

\end{document}